# THE QUANTUM MECHANICS OF COSMOLOGY

James B. Hartle
Department of Physics
University of California
Santa Barbara, CA 93106 USA

TABLE OF CONTENTS

## III. GENERALIZED QUANTUM MECHANICS

## IV. TIME IN QUANTUM MECHANICS

## V. THE QUANTUM MECHANICS OF SPACETIME

## VI. PRACTICAL QUANTUM COSMOLOGY

## ACKNOWLEDGEMENTS

## REFERENCES

## APPENDIX: BUZZWORDS

## I. INTRODUCTION

It is an inescapable inference from the physics of the last sixty years that we live in a quantum mechanical universe — a world in which the basic laws of physics conform to that framework for prediction we call quantum mechanics. If this inference is correct, then there must be a description of the universe as a whole and everything in it in quantum mechanical terms. The nature of this description and its observable consequences are the subject of quantum cosmology.

Our observations of the present universe on the largest scales are crude and a classical description of them is entirely adequate. Providing a quantum mechanical description of these observations alone might be an interesting intellectual challenge, but it would be unlikely to yield testable predictions differing from those of classical physics. Today, however, we have a more ambitious aim. We aim, in quantum cosmology, to provide a theory of the initial condition of the universe which will predict testable correlations among observations today. There are no realistic predictions of any kind that do not depend on this initial condition, if only very weakly. Predictions of certain observations may be testably sensitive to its details. These include the large scale homogeneity and isotropy of the universe, its approximate spatial flatness, the spectrum of density fluctuations that produced the galaxies, the homogeneity of the thermodynamic arrow of time, and the existence of classical spacetime. Further, one of the main topics of this school is the question of whether the coupling constants of the effective interactions of the elementary particles at accessible energy scales may depend, in part, on the initial condition of the universe. It is for such reasons that the search for a theory of the initial condition of the universe is just as necessary and just as fundamental as the search for a theory of the dynamics of the elementary particles.*

The physics of the very early universe is likely to be quantum mechanical in an essential way. The singularity theorems of classical general relativity† suggest that an early era preceded ours in which even the geometry of spacetime exhibited significant quantum fluctuations. It is for a theory of the initial condition that describes this era, and all later ones, that we need a quantum mechanics of cosmology. That quantum mechanics is the subject of these lectures.

The "Copenhagen" frameworks for quantum mechanics, as they were formulated in the 1930's and '40's and as they exist in most textbooks today,‡ are inadequate for quantum cosmology on at least two counts. First, these formulations characteristically assumed a possible division of the world into "obsever" and "observed", assumed that "measurements" are the primary focus of scientific statements and, in effect, posited the existence of an external "classical domain". However, in a theory of the whole thing there can be no fundamental division into observer and observed. Measurements and observers cannot be fundamental notions in a theory that seeks to describe the early universe when neither existed. In a basic formulation of quantum mechanics there is no reason in general for there to be any variables that exhibit classical behavior in all circumstances. Copenhagen quantum mechanics thus needs to be generalized to provide a quantum framework for cosmology.

---

* For reviews of quantum cosmology see lectures by Halliwell in this volume and Hartle (1988c, 1990a). For a bibliography of papers on the subject through 1989 see Halliwell (1990).

† For a review of the singularity theorems of classical general relativity see Geroch and Horowitz (1979). For the specific application to cosmology see Hawking and Ellis (1968).

‡ There are various "Copenhagen" formulations. For a classic exposition of one of them see London and Bauer (1939).

The second count on which the familiar formulations of quantum mechanics are inadequate for cosmology concerns their central use of a preferred time. Time is a special observable in Hamiltonian quantum mechanics. Probabilities are predicted for observations "at one moment of time". Time is the only observable for which there are no interfering alternatives as a measurement of momentum is an interfering alternative to a measurement of position. Time is the sole observable not represented by an operator in the familiar quantum mechanical formalism, but rather enters the theory as a parameter describing evolution.

Were there a fixed geometry for spacetime, that background geometry would give physical meaning to the preferred time of Hamiltonian quantum mechanics. Thus, the Newtonian time of non-relativistic spacetime is unambiguously taken over as the preferred time of non-relativistic quantum mechanics. The spacetime of special relativity has many timelike directions. Any one of them could supply the preferred time of a quantum theory of relativistic fields. It makes no difference which is used for all such quantum theories are unitarily equivalent because of relativistic causality. However, spacetime is not fixed fundamentally. In a quantum theory of gravity, in the domain of the very early universe, it will fluctuate and be without definite value. If spacetime is quantum mechanically variable, there is no one fixed spacetime to supply a unique notion of "timelike, "spacelike" or "spacelike surface". In a quantum theory of spacetime there is no one background geometry to give meaning to the preferred time of Hamiltonian quantum mechanics. There is thus a conflict between the general covariance of the physics of gravitation and the preferred time of Hamiltonian quantum mechanics. This is the "problem of time" in quantum gravity.* Again the familiar framework of quantum mechanics needs to be generalized.

These lectures discuss the generalizations of Copenhagen quantum mechanics necessary to deal with cosmology and with quantum cosmological spacetimes. Their purpose is to sketch a coherent framework for the process of quantum mechanical prediction in general, and for extracting the predictions of theories of the initial condition in particular. Throughout I shall assume quantum mechanics and I shall assume spacetime. Some hold the view that "quantum mechanics is obviously absurd but not obviously wrong" in its application to the macroscopic domain. I hope to show that it is not obviously absurd. Some believe that spacetime is not fundamental. If so, then a quantum mechanical framework at least as general as that sketched here will be needed to discuss the effective physics of spacetime on all scales above the Planck length and revisions of the familiar framework at least as radical as those suggested here will be needed below it.

The strategy of these lectures is to discuss the generalizations needed for a quantum mechanics of cosmology in two steps. I shall begin by assuming in Section II a fixed background spacetime that supplies a preferred family of timelike directions for quantum mechanics. This of course, is an excellent approximation on accessible scales for times later than $10^{-43}$ sec after the big bang. The familiar apparatus of Hilbert space, states, Hamiltonian and other operators then may be applied to process of prediction. Indeed, in this context the quantum mechanics of cosmology is in no way distinguished from the quantum mechanics of a large isolated box containing both observers and observed.*

---

\* For classic reviews of this problem from the perspective of canonical quantum gravity see Wheeler (1979) and Kuchař (1981).

The quantum framework for cosmology I shall describe in Section II has its origins in the work of Everett and has been developed by many.† Especially taking into account its recent developments, notably the work of Zeh (1971), Zurek (1981, 1982), Joos and Zeh (1985), Griffiths (1984), Omnès (1988abc, 1989) and others, it is sometimes called the post-Everett interpretation of quantum mechanics. I shall follow the development in Gell-Mann and Hartle (1990).

Everett's idea was to take quantum mechanics seriously and apply it to the universe as a whole. He showed how an observer could be considered part of this system and how its activities — measuring, recording, calculating probabilities, etc. — could be described within quantum mechanics. Yet the Everett analysis was not complete. It did not adequately describe within quantum mechanics the origin of the "classical domain" of familiar experience or, in an observer independent way, the meaning of the "branching" that replaced the notion of measurement. It did not distinguish from among the vast number of choices of quantum mechanical observables that are in principle available to an observer, the particular choices that, in fact, describe the classical domain.‡ It did not sufficiently address the construction of history, so important for cosmology, independently of the memory of an observer.

The post-Everett framework stresses the importance of histories for quantum mechanics. It stresses the consistency of probability sum rules as the primary criterion for assigning probabilities to histories rather than any notion of "measurement". It stresses the initial condition of the universe as the ultimate origin within quantum mechanics of the classical domain. There are many problems in this approach to quantum mechanics yet to be solved, but, as I hope to show, post-Everett quantum mechanics provides a framework that is sufficiently general for cosmology and sufficiently detailed that the remaining questions can be attacked.

Having obtained in Section II an understanding of how to generalize familiar quantum mechanics to make predictions for closed systems, the remainder of the lectures will be concerned with the generalizations needed to accomodate quantum spacetime. In Section III a broad framework for quantum mechanical theories will be introduced. Hamiltonian quantum mechanics is one instance of such a quantum mechanics but not the only possible one. This will be illustrated in Section IV where, motivated by the problem of time, various examples of generalized quantum mechanics that have no equivalent Hamiltonian formulations will be discussed. In Section V a generalized sum-over-histories quantum mechanics for closed cosmo-logical spacetimes will be described that is free from the problem of time.* This quantum mechanics too has no obvious equivalent Hamiltonian formulation. Finally, in Section VI we shall review the rules by which semiclassical predictions are extracted from a wave function of the universe.

---

* Thus, those who find it unsettling for some reason to consider the universe as a whole may substitute the words "closed system" for "cosmology" and "universe" without any change in meaning in Sections II-IV. As we shall see in Section II.6, however, as far as physical mechanisms which insure the emergence of a classical domain, the only "closed system" of interest *is* essentially the universe as a whole.

† The original reference is Everett (1957). The idea was developed by many, among them Wheeler (1957), DeWitt (1970), Geroch (1984), and Mukhanov (1985) and independently arrived at by others, e.g. Gell-Mann (1963) and Cooper and Van Vechten (1969). There is a useful collection of early papers in DeWitt and Graham (1973).

‡ This is sometimes called the "basis problem".

Any generalization of the familiar framework of quantum mechanics has the obligation to recover that framework in suitable limiting circumstances. For the generalizations discussed here those circumstances concern the existence of a "classical domain" and, in particular, the existence of classical spacetime. Classical behavior is not a consequence of all states in quantum theory; it is a property of particular states. It will be a constant theme of these lectures that, in the generalizations discussed, the familiar formulations of quantum mechanics are recovered as limiting cases in circumstances defined by the particular state the universe does have. That is, most fundamentally, they are recovered because of this universe's particular quantum initial condition. The "classical domain" of the Copenhagen interpretations is not a general feature of a quantum theory of the universe but it may be a feature of its particular initial conditions and dynamics at late times. In a similar way Hamiltonian quantum mechanics, with its preferred time, may not be the most general formulation of quantum mechanics, but it may be an approximation to a yet more general sum-over-histories framework appropriate in the late universe where a nearly classical background spacetime is realized because of a specific initial condition. In this way, the Copenhagen formulations of quantum mechanics can be seen as approximations in which certain approximate classical features of the universe are idealized as exact — approximations that are not generally applicable in quantum theory, but made appropriate by a specific initial condition. From this perspective, the "classical domain" with its classical spacetime are "excess baggage" in the fundamental theory of a kind that is seen elsewhere in the development of physics. (See, e.g. Hartle, 1990b) They are true features of the late epoch of this universe perceived to be fundamental because of the limited character of our observations. They may be more successfully viewed as but one possibility out of many in a yet more general theory.

## II. POST-EVERETT QUANTUM MECHANICS†

### II.1. Probability

#### II.1.1. *Probabilities in general*

Even apart from quantum mechanics, there is no certainty in this world and therefore physics deals in probabilities.‡ It deals most generally with the probabilities for alternative time histories of the universe. From these, conditional probabilities appropriate when information about our specific history is known may be constructed.

---

* Several authors have suggested, in various ways, that sum-over-histories quantum mechanics might be a fruitful approach to a generally covariant quantum mechanics of cosmological spacetime, among them recently Teitelboim (1983abc), Sorkin (1989) and the author (Hartle, 1986b, 1988ab, 1989b). The latter approach is described in Section V.

† Most of the material in this Section is an abridgement or amplification of Gell-Mann and Hartle (1990).

‡ For a lively review of the use of probability in physics most of whose viewpoints are compatible with those expressed here see Deutsch (1991).

To understand what these probabilities mean, it is best to understand how they are used. We deal, first of all, with probabilities for *single* events of the *single* system that is the universe as a whole. When these probabilities become sufficiently close to zero or one there is a definite prediction on which we may act. How sufficiently close to 0 or 1 the probabilities must be depends on the circumstances in which they are applied. There is no certainty that the sun will come up tomorrow at the time printed in our daily newspapers. The sun may be destroyed by a neutron star now racing across the galaxy at near light speed. The earth's rotation rate could undergo a quantum fluctuation. An error could have been made in the computer that extrapolates the motion of the earth. The printer could have made a mistake in setting the type. Our eyes may deceive us in reading the time. We watch the sunrise at the appointed time because we compute, however imperfectly, that the probability of these things happening is sufficiently low.

Various strategies can be employed to identify situations where probabilities are near zero or one. Acquiring information and considering the conditional probabilities based on it is one such strategy. Current theories of the initial condition of the universe predict almost no probabilities near zero or one without further conditions. The "no boundary" wave function of the universe, for example, does not predict the present position of the sun on the sky. It will predict, however, that the conditional probability for the sun to be at the position predicted by classical celestial mechanics given a few previous positions is a number very near unity.

Another strategy to isolate probabilities near 0 or 1 is to consider ensembles of repeated observations of identical subsystems. There are no genuinely infinite ensembles in the world so we are necessarily concerned with the probabilities for deviation of a finite ensemble from the expected behavior of an infinite one. These are probabilities for a single feature (the deviation) of a single system (the whole ensemble). To give a quantum mechanical example, consider an ensemble of $N$ spins each in a state $|\psi\rangle$. Suppose we measure whether the spin is up or down for each spin. The predicted relative frequency of finding $n_\uparrow$ spin-ups is

$$f_N^\uparrow = \frac{n_\uparrow}{N} = |<\uparrow|\psi>|^2 \, , \tag{II.1.1}$$

where $|\uparrow\rangle$ is state with the spin definitely up. Of course, there is no certainty that we will get this result but as $N$ becomes large we expect the probability of significant deviations away from this value to be very small.

In the quantum mechanics of the whole ensemble this prediction would be phrased as follows: There is an observable $f_N^\uparrow$ corresponding to the relative frequency of spin up. Its operator is easily defined on the basis in which all the spins are either up or down as

$$f_N^\uparrow = \sum_{s_1\cdots s_N} |s_1\rangle \cdots |s_N\rangle \left( \sum_i \frac{\delta_{s_i}\uparrow}{N} \right) <s_n|\cdots<s_1| \, . \tag{II.1.2}$$

Here, $|s\rangle$, with $s=\uparrow$ or $\downarrow$, are the spin eigenstates in the measured direction. The eigenvalue in brackets is just the number of spin ups in the state $|s_1\rangle \cdots |s_N\rangle$. The operator $f_N^\uparrow$ thus has the discrete spectrum $1/N, 2/N, \cdots, 1$. We can now calculate the probability that $f_N^\uparrow$ has one of these possible values in the state

$$|\Psi\rangle = |\psi\rangle \cdots |\psi\rangle \quad (N \text{ times}) \, , \tag{II.1.3}$$

which describes $N$ independent subsystems each in the state $|\psi>$. The result is simply a binomial distribution. The probability of finding relative frequency $f$ is

$$p(f) = \binom{N}{fN} p_\uparrow^{fN} p_\downarrow^{N(1-f)} \qquad (\text{II.1.4})$$

where $p_\uparrow = |<\uparrow|\psi>|^2$ and $p_\downarrow = 1 - p_\uparrow$. As $N$ becomes large this approaches a continuum normal distribution that is sharply peaked about $f = p_\uparrow$. The width becomes arbitrarily small with large $N$ as $N^{-\frac{1}{2}}$. Thus, the probability for finding $f$ in some range about $p_\uparrow$ can be made close to one by choosing $N$ sufficiently large yielding a definite prediction for the relative frequency. In a given experiment how large does $N$ have to be before the prediction is counted as definite? It must be large enough so the probability of error is sufficiently small to isolate a result of significance given the status of competing theories, competing groups, the consequences of a lowered reputation if wrong, the limitations of resources, etc.

The existence of large ensembles of repeated observations in identical circumstances and their ubiquity in laboratory science should not obscure the fact that in the last analysis physics must predict probabilities for the single system which is the ensemble as a whole. Whether it is the probability of a successful marriage, the probability of the present galaxy-galaxy correlation function, or the probability of the fluctuations in an ensemble of repeated observations, we must deal with the probabilities of single events in single systems. In geology, astronomy, history, and cosmology, most predictions of interest have this character. For some it is easier to discuss such probabilities by employing the fiction that they are definite predictions of the relative frequencies in an imaginary infinite ensemble of repeated indentical universes.* Here, I shall deal directly with the individual events.

The goal of physical theory is, therefore, most generally to predict the probabilities of histories of single events of a single system. Such probabilities are, of course, not measurable quantities. The success of a theory is to be judged by whether its definite predictions (probabilities sufficiently close to 0 or 1) are confirmed by observation or not.

Probabilities need be assigned to histories by physical theory only up to the accuracy they are used. Two theories that predict probabilities for the sun not rising tomorrow at its classically calculated time that are both well beneath the standard on which we act are equivalent for all practical purposes as far as this prediction is concerned. For example, a model of the Earth's rotation that includes the gravitational effects of Sirius gives different probabilities from one which does not, but ones which are equivalent for all practical purposes to those of the model in which this effect is neglected.

The probabilities assigned by physical theory must conform to the standard rules of probability theory: The probability for both of two exclusive events is the sum of the probabilities for each. The probabilities of an exhaustive set of alternatives must sum to unity. The probability of the empty alternative is zero. Because probabilities are meaningful only up to the standard by which they are used, it is useful to consider *approximate probabilities* which need satisfy the rules of probability theory only up to the same standard. A theory which assigns approximate

---

* For developments of this point of view in the quantum mechanical context see Finkelstein (1963), Hartle (1968), Graham (1970), Farhi, Goldstone and Gutmann (1989).

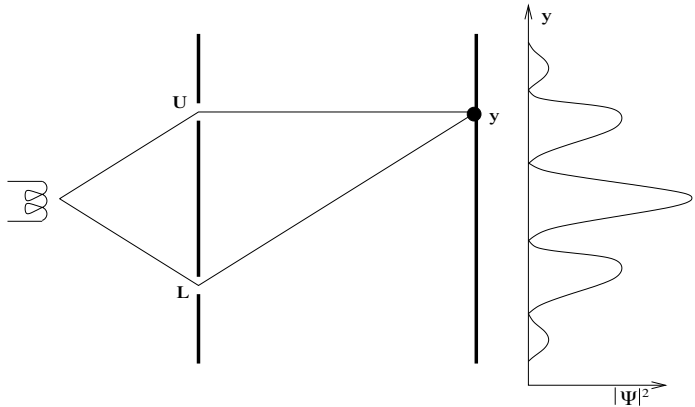

**Fig. 1:** *The two-slit experiment. An electron gun at right emits an electron traveling towards a screen with two slits, its progress in space recapitulating its evolution in time. When precise detections are made of an ensemble of such electrons at the screen it is not possible, because of interference, to assign a probability to the alternatives of whether an individual electron went through the upper slit or the lower slit. However, if the electron interacts with apparatus that measures which slit it passed through, then these alternatives decohere and probabilities can be assigned*

probabilities in this sense could always be augmented by a prescription for renormalizing the probabilities so that the rules are exactly obeyed without changing their values in any relevent sense. As we shall see, it is only through the use of such approximate probabilities that quantum mechanics can assign probabilities to interesting time histories at all. We shall return to issues connected with the use of approximate probabilities in Section II.11.

### II.1.2. *Probabilities in Quantum Mechanics*

The characteristic feature of a quantum mechanical theory is that not every history that can be described can be assigned a probability. Nowhere is this more clearly illustrated than in the two slit experiment. In the usual "Copenhagen" discussion if we have not measured which of the two slits the electron passed through on its way to being detected at the screen, then we are not permitted to assign probabilities to these alternative histories. It would be inconsistent to do so since the correct probability sum rule would not be satisfied. Because of interference, the probability to arrive at $y$ is not the sum of the probabilities to arrive at $y$ going through the upper or lower slit:

$$p(y) \neq p_U(y) + p_L(y) \tag{II.1.5}$$

because

$$|\psi_L(y) + \psi_U(y)|^2 \neq |\psi_L(y)|^2 + |\psi_U(y)|^2 \quad . \tag{II.1.6}$$

If we *have* measured which slit the electron went through, then the interference is destroyed, the sum rule obeyed, and we *can* meaningfully assign probabilities to these alternative histories.

We cannot have such a rule in quantum cosmology because there is not a fundamental notion of "measurement". There is no fundamental division into observer and observed and no fundamental reason for the existence of classically behaving measuring apparatus. In particular, in the early universe none of these concepts seem relevant. We need an observer-independent, measurement-independent rule for which histories can be assigned probabilities and which cannot. It is to this rule that I now turn.

## II.2. Decoherent Histories

### II.2.1. *Fine and Coarse Grained Histories*

I shall now describe the rules that specify which histories of the universe may be assigned approximate probabilities. They are essentially those put forward by Griffiths (1984), developed by Omnès (1988abc, 1989), and independently but later arrived at by Gell-Mann and the author (Gell-Mann and Hartle, 1990). Since histories are our concern, it's convenient to begin with Feynman's sum-over-histories formulation of quantum mechanics. A completely fine-grained history is then specified by giving a set of generalized coördinates $q^i(t)$ as functions of time. These might be the values of fundamental fields at different points of space

Completely fine-grained histories cannot be assigned probabilities; only suitable coarse-grained histories can. Examples of coarse grainings are: (1) Specifying the $q^i$ not at all times but at a discrete set of times. (2) Specifying not all the $q^i$ at any one time but only some of them. (3) Specifying not definite values of these $q^i$ but only ranges of values. An exhaustive set of ranges at any one time consists of regions $\{\Delta_\alpha\}$ that make up the whole space spanned by the $q^i$ as $\alpha$ passes over all values. An exhaustive set of coarse-grained histories is then defined by exhaustive sets of ranges $\{\Delta_\alpha^i\}$ at times $t_i$, $i = 1, \cdots, n$.

### II.2.2. *Decohering Sets of Coarse Grained Histories*

The important theoretical construct for giving the rule that determines whether probabilities may be assigned to a given set of alternative histories, and what these probabilities are, is the decoherence functional $D$ [(history)', (history)]. This a complex functional on any pair of histories in a coarse-grained set. It is most transparently defined in the sum-over-histories framework for completely fine-grained history segments between an initial time $t_0$ and a final time $t_f$, as follows:

$$D\left[q'^i(t), q^i(t)\right] = \delta\left(q'^i_f - q^i_f\right)\ \exp\left\{i\left(S[q'^i(t)] - S[q^i(t)]\right)/\hbar\right\}\rho(q'^i_0, q^i_0)\ . \quad \text{(II.2.1)}$$

Here, $\rho$ is the initial density matrix of the universe in the $q^i$ representation, $q'^i_0$ and $q^i_0$ are the initial values of the complete set of variables, and $q'^i_f$ and $q^i_f$ are the final values. The decoherence functional for coarse-grained histories is obtained from $(2.1)^*$ according to the principle of superposition by summing over all that is not specified by the coarse graining. Thus,

$$D\left([\Delta_{\alpha'}], [\Delta_\alpha]\right) = \int_{[\Delta_{\alpha'}]} \delta q' \int_{[\Delta_\alpha]} \delta q \delta(q'^i_f - q^i_f)\ e^{i\left\{\left(S[q'^i] - S[q^i]\right)/\hbar\right\}}\rho(q'^i_0, q^i_0)\quad . \quad \text{(II.2.2)}$$

---

* (2.1) refers to eq.(II.2.1). Section numbers are omitted when referring to equations within a given section.

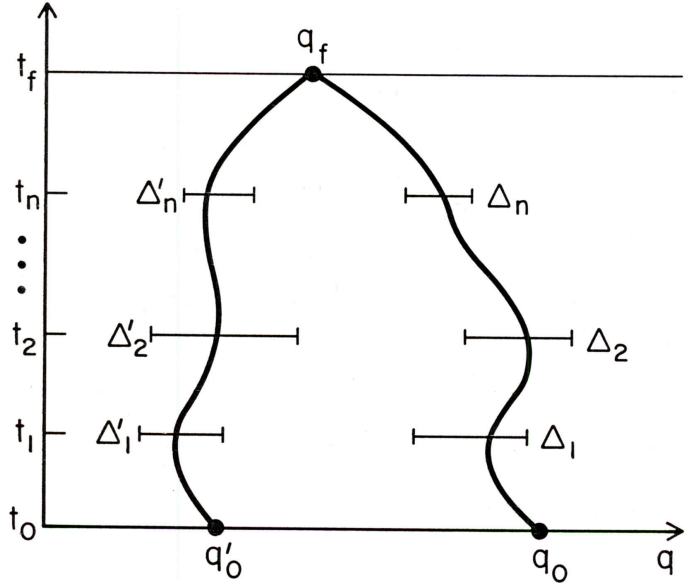

**Fig. 2:** *The sum-over-histories construction of the decoherence functional.*

More precisely, the sum is as follows (Fig. 2): It is over all histories $q'^i(t)$, $q^i(t)$ that begin at $q'^i_0$, $q^i_0$ respectively, pass through the ranges $[\Delta_{\alpha'}]$ and $[\Delta_\alpha]$ respectively, and wind up at a common point $q^i_f$ at any time $t_f > t_n$. It is completed by summing over $q'^i_0$, $q^i_0$, and $q^i_f$ (Fig. 2). The result is independent of $t_f$. The three forms of information necessary for prediction — initial condition, action, and specific history are manifest in this formula as $\rho$, $S$, and $[\Delta_\alpha]$ respectively.

The connection between coarse-grained histories and completely fine-grained ones is transparent in the sum-over-histories formulation of quantum mechanics. However, the sum-over-histories formulation does not allow us to consider coarse-grained histories of the most general type directly. For the most general histories one needs to exploit the transformation theory of quantum mechanics and for this the Heisenberg picture is convenient.* In the Heisenberg picture $D$ can be written

$$D\Big([P_{\alpha'}],[P_\alpha]\Big) = Tr\left[P^n_{\alpha_n}(t_n)\cdots P^1_{\alpha'_1}(t_1)\rho P^1_{\alpha_1}(t_1)\cdots P^n_{\alpha_n}(t_n)\right] \ , \qquad \text{(II.2.3)}$$

where the $P^k_\alpha(t)$ are a set of projection operators corresponding to an exhaustive set of alternatives at one time. These satisfy

$$\sum_\alpha P^k_\alpha(t) = 1 \ , \qquad P^k_\alpha(t)P^k_\beta(t) = \delta_{\alpha\beta}P^k_\beta(t) \ . \qquad \text{(II.2.4)}$$

Here, $k$ labels the set, $\alpha$ the alternative, and $t$ the time. The operators representing the same alternatives at different times are connected by

$$P^k_\alpha(t) = e^{iHt/\hbar}P_\alpha(0)e^{-iHt/\hbar} \ . \qquad \text{(II.2.5)}$$

---

* The utility of this Heisenberg formulation of quantum mechanics has been stressed by many authors, among them Groenewold (1952) Wigner (1963), Aharonov, Bergmann, and Lebovitz (1964), Unruh (1986), and Gell-Mann (1987).

A set of alternative histories, $[P_\alpha]$, is represented by a set of exhaustive projections $(P^1_{\alpha_1}(t_1),\ P^2_{\alpha_2}(t_2),\cdots,P^n_{\alpha_n}(t_n))$ as $\alpha_1,\cdots,\alpha_n$ range over all values. An individual history in the set is a particular set of values $\alpha_1,\cdots,\alpha_n$. In the Heisenberg picture a completely fine-grained set of histories is defined by giving a *complete* set of projections (one dimensional ones) at each and every time. Every possible set of alternative histories may then be obtained by coarse graining the various fine-grained sets, that is, by using $P$'s in the coarser grained sets which are *sums* of those in the finer grained sets. Thus, if $[\overline{P_\beta}]$ is a coarse graining of the set of histories $\{[P_\alpha]\}$, we write

$$D\left([\overline{P_{\beta'}}],[\overline{P_\beta}]\right) = \sum_{\substack{\text{all } P'_\alpha \\ \text{not fixed by } [\overline{P_{\beta'}}]}} \sum_{\substack{\text{all } P_\alpha \\ \text{not fixed by } [\overline{P_\beta}]}} D\left([P_{\alpha'}],\ [P_\alpha]\right) \quad . \tag{II.2.6}$$

A set of coarse-grained alternative histories is said to *decohere* when the off-diagonal elements of $D$ are sufficiently small:

$$D\left([P_{\alpha'}],[P_\alpha]\right) \approx 0 \quad , \quad \text{for any } \alpha'_k \neq \alpha_k \quad . \tag{II.2.7}$$

This is a generalization of the condition for the absence of interference in the two-slit experiment (approximate equality of the two sides of (1.6)). It has as a consequence the purely diagonal formula

$$D\left([\overline{P_\beta}],[\overline{P_\beta}]\right) \approx \sum_{\substack{\text{all } P_\alpha \text{ not} \\ \text{fixed by } [P_\beta]}} D\left([P_\alpha],[P_\alpha]\right) \quad . \tag{II.2.8}$$

The rule for when approximate probabilities can be assigned to a set of histories of the universe is then this: To the extent that a *set* of alternative histories decoheres, probabilities can be assigned to its individual members. The probabilities are the *diagonal* elements of $D$. Thus,

$$\begin{aligned}
p([P_\alpha]) &= D([P_\alpha],[P_\alpha]) \\
&= Tr\left[P^n_{\alpha_n}(t_n)\cdots P^1_{\alpha_1}(t_1)\rho P^1_{\alpha_1}(t_1)\cdots P^n_{\alpha_n}(t_n)\right]
\end{aligned} \tag{II.2.9}$$

when the set decoheres. We shall frequently write $p(\alpha_n t_n,\cdots\alpha_1 t_1)$ for these probabilities, suppressing the labels of the sets.

The probabilities defined by (2.9) obey the rules of probability theory as a consequence of decoherence. The principal requirement is that the probabilities be additive on "disjoint sets of the sample space". For histories this gives the sum rule

$$p\left([\overline{P_\beta}]\right) \approx \sum_{\substack{\text{all } P_\alpha \text{ not} \\ \text{fixed by } [\overline{P_\beta}]}} p\left([P_\alpha]\right) \quad . \tag{II.2.10}$$

These relate the probabilities for a set of histories to the probabilities for *all* coarser grained sets that can be constructed from it. For example, the sum rule eliminating all projections at one time is

$$\begin{aligned}
\sum_{\alpha_k} p(\alpha_n t_n,&\cdots\alpha_{k+1}t_{k+1},\alpha_k t_k,\alpha_{k-1}t_{k-1},\cdots,\alpha_1 t_1) \\
&\approx p(\alpha_n t_n,\cdots\alpha_{k+1}t_{k+1},\alpha_{k-1}t_{k-1},\cdots,\alpha_1 t_1) \quad . \tag{II.2.11}
\end{aligned}$$

The $p([P_\alpha])$ are *approximate* probabilities in the sense of Section II.1 which approximately obey the probability sum rules. If a given standard by which these sum rules are satisfied is required, it can be met by coarse graining at the requisite level. It is possible to demand exact decoherence. For example, sets of histories consisting of alternatives at a single time exactly decohere because of the cyclic property of the trace in (2.3) and (2.4). Once a standard is met, further coarse graining of a decoherent set of alternative histories produces a set of decoherent histories since the probability sum rules continue to be satisfied. (Those for the coarser grained set are contained among those for the finer grained set.) Further fine graining can result in the loss of decoherence.

Given this discussion, the *fundamental formula* of quantum mechanics may be reasonably taken to be

$$D\left([P_{\alpha'}], [P_\alpha]\right) \approx \delta_{\alpha_1' \alpha_1} \cdots \delta_{\alpha_n' \alpha_n} p([P_\alpha]) \qquad \text{(II.2.12)}$$

for all $[P_\alpha]$ in a set of alternative histories. Vanishing of the off-diagonal elements of $D$ gives the rule for when probabilities may be consistently assigned. The diagonal elements give their values.

We could have used a weaker condition than (2.7) as the definition of decoherence. Eq. (2.7) is sufficient. To understand the necessary condition, consider the weakest coarse graining in which just two projections, $P_a(t)$ and $P_b(t)$ in an exhaustive set of alternatives at one time $t$ are lumped together in a single alternative $(P_a(t) + P_b(t))$ in the coarser grained set. (This corresponds to the logical operation "or".) The probability sum rules (2.10) require

$$D\left(\cdots (P_a(t) + P_b(t)) \cdots \rho \cdots (P_a(t) + P_b(t)) \cdots\right) =$$

$$D\left(\cdots P_a(t) \cdots \rho \cdots P_a(t) \cdots\right) + D\left(\cdots P_b(t) \cdots \rho \cdots P_b(t) \cdots\right) , \qquad \text{(II.2.13)}$$

or equivalently

$$D\left(\cdots P_a(t) \cdots \rho \cdots P_b(t) \cdots\right) + D\left(\cdots P_b(t) \cdots \rho \cdots P_a(t) \cdots\right) = 0 . \qquad \text{(II.2.14)}$$

Considering all such cases the *necessary* and sufficient condition for the validity of the sum rules (2.10) of probability theory is:

$$D\left([P_\alpha], [P_{\alpha'}]\right) + D\left([P_{\alpha'}], [P_\alpha]\right) \approx 0 \qquad \text{(II.2.15)}$$

for any $\alpha_k' \neq \alpha_k$, or equivalently

$$Re\left\{D\left([P_\alpha], [P_{\alpha'}]\right)\right\} \approx 0 \quad . \qquad \text{(II.2.16)}$$

This is the condition used by Griffiths (1984) as the requirement for "consistent histories". However, while, as we shall see, it is easy to identify physical situations in which the off-diagonal elements of $D$ approximately vanish as the result of coarse graining, it is hard to think of a general mechanism that suppresses only their real parts. In the usual analysis of measurement (as in the two-slit experiment, cf. (1.6)) the off-diagonal parts of $D$ approximately vanish. We shall, therefore, explore the stronger condition (2.7) in what follows.

### II.2.3. *No Moment by Moment Definition of Decoherence*

Decoherence is a property of coarse-grained sets of alternative histories of the universe. The decoherence of alternatives in a given coarse-grained set in the past can be affected by further fine graining in the future. The further fine graining produces a *different* coarse-grained set of histories that may or may not decohere.

Consider by way of example, a Stern-Gerlach experiment in which an atomic beam divides in an inhomogeneous magnetic field according to a spin component, $s_z$, of the atoms, and later to recombines under the action of a further appropriate inhomogeneous magnetic field. In a coarse graining that concerns only $s_z$ at moments when the beams were separated, the alternative values of this variable would decohere because they are correlated with orthogonal trajectories of the beams. In a coarse graining that, in addition, includes $s_z$ at later moments when the beams are recombined, the alternative values of $s_z$ when the beams are separated would not decohere. The interference destroyed by separating the beams has been restored by recombining them.

Thus, generally, decoherence cannot be viewed as an evolving phenomenon in which certain alternatives decohere and remain so. Decoherence is a property of sets of alternative histories, not of any summary of the system at a moment of time. Further fine grain the set and it may no longer decohere. Having made this general point it should also be noted that I shall later argue (Section II.7) that the decoherence of coarse-grained histories constructed from certain kinds of variables associated with the classical domain of familiar experience are insensitive to further fine graining by the same kinds of variables. Even here, however, as we shall see, there is always *some* fine graining in the future which will destroy the decoherence of alternatives in the past.*

### II.3. Prediction, Retrodiction, and History

### II.3.1. *Prediction and Retrodiction*

Decoherent histories are what we utilize in the process of prediction in quantum mechanics, for they may be assigned probabilities. Decoherence thus generalizes and replaces the notion of "measurement", which served this role in the Copenhagen interpretations. Decoherence is a more precise, more objective, more observer-independent idea. For example, if their associated histories decohere, we may assign probabilities to various values of reasonable scale density fluctuations in the early universe whether or not anything like a "measurement" was carried out on them and certainly whether or not there was an "observer" to do it. We shall return to a specific discussion of typical measurement situations in Section II.9.

The joint probabilities $p(\alpha_n t_n, \cdots, \alpha_1 t_1)$ for the individual histories in a decohering set are the raw material for prediction and retrodiction in quantum cosmology. From them, relevant conditional probabilities may be computed. The conditional probability of one subset, $\{\alpha_i t_i\}$, given the rest, $\overline{\{\alpha_i t_i\}}$, is generally

$$p\Big(\{\alpha_i t_i\} | \overline{\{\alpha_i t_i\}}\Big) = \frac{p(\alpha_n t_n, \cdots, \alpha_1 t_1)}{p\Big(\overline{\{\alpha_i t_i\}}\Big)} \quad . \tag{II.3.1}$$

---

\* This has been forcefully stated by Bell (1975).

For example, the probability for *predicting* alternatives $\alpha_{k+1}, \cdots, \alpha_n$, given that the alternatives $\alpha_1, \cdots, \alpha_k$ have already happened, is

$$p(\alpha_n t_n, \cdots, \alpha_{k+1} t_{k+1} | \alpha_k t_k, \cdots, \alpha_1 t_1) = \frac{p(\alpha_n t_n, \cdots, \alpha_1 t_1)}{p(\alpha_k t_k, \cdots, \alpha_1 t_1)} \quad . \tag{II.3.2}$$

The probability that $\alpha_{n-1}, \cdots, \alpha_1$ happened in the *past*, given an alternative $\alpha_n$ at the present time $t_n$, is

$$p(\alpha_{n-1} t_{n-1}, \cdots, \alpha_1 t_1 | \alpha_n t_n) = \frac{p(\alpha_n t_n, \cdots, \alpha_1 t_1)}{p(\alpha_n t_n)} \quad . \tag{II.3.3}$$

Decoherence ensures that the probabilities defined by $(3.1) - (3.3)$ will approximately add to unity when summed over all remaining alternatives, because of the probatility sum rules (2.10).

Despite the similarity between (3.2) and (3.3), there are differences between prediction and retrodiction. Future predictions can all be obtained from an effective density matrix summarizing information about what has happened. If $\rho_{\text{eff}}$ is defined by

$$\rho_{\text{eff}}(t_k) = \frac{P^k_{\alpha_k}(t_k) \cdots P^1_{\alpha_1}(t_1) \rho P^1_{\alpha_1}(t_1) \cdots P^k_{\alpha_k}(t_k)}{Tr[P^k_{\alpha_k}(t_k) \cdots P^1_{\alpha_1}(t_1) \rho P^1_{\alpha_1}(t_1) \cdots P^k_{\alpha_k}(t_k)]} \quad , \tag{II.3.4}$$

then

$$\begin{aligned} p(\alpha_n t_n, \cdots, &\alpha_{k+1} t_{k+1} | \alpha_k t_k, \cdots, \alpha_1 t_1) \\ &= Tr[P^n_{\alpha_n}(t_n) \cdots P^{k+1}_{\alpha_{k+1}}(t_{k+1}) \rho_{\text{eff}}(t_k) P^{k+1}_{\alpha_{k+1}}(t_{k+1}) \cdots P^n_{\alpha_n}(t_n)] \quad . \end{aligned} \tag{II.3.5}$$

The density matrix $\rho_{\text{eff}}(t_k)$ represents the usual notion of "state of the system at time $t_k$". It is given here in the Heisenberg picture and is constant between $t_k$ and $t_{k+1}$ after which a new $\rho_{\text{eff}}(t_{k+1})$ must be used for future prediction. Its Schrödinger picture representative which evolves with time, would be given by

$$e^{-iH(t-t_k)/\hbar} \rho_{\text{eff}}(t_k) e^{iH(t-t_k)/\hbar} \tag{II.3.6}$$

for $t_k < t < t_{k+1}$. In contrast to prediction, there is no effective density matrix representing present information from which probabilities for the past can be derived. As (3.3) shows, history requires knowledge of *both* present records *and* the initial condition of the universe.

Prediction and retrodiction differ in another way. Because of the cyclic property of the trace in (2.3), *any* final alternative decoheres and a probability can be predicted for it. By contrast we expect only certain variables to decohere in the past, appropriate to present data and the initial $\rho$.

These differences between prediction and retrodiction are aspects of the arrow of time in quantum mechanics. Mathematically they are consequences of the time ordering in the decoherence functional (2.3). The theory can be rewritten with the opposite time ordering. Field theory is invariant under CPT. Performing a CPT transformation on (2.3) or (2.9) results in an equivalent expression in which the CPT transformed $\rho$ is assigned to the far future and the CPT-transformed projections are anti-time-ordered. (See Section IV.2 for more details) Either time ordering can, therefore, be used; the important point is that there is a knowable Heisenberg $\rho$ from which probabilities can be predicted. It is by convention that we think of it as an "initial condition", with the projections in increasing time order from the inside out in (2.3) and (2.9). The words "prediction" and "retrodiction" are used in this paper in the context of this convention.

While the formalism of quantum mechanics allows the universe to be discussed with either time ordering, the physics of the universe is time asymmetric, with a simple condition in what we call "the past." For example, the present homogeneity of the thermodynamic arrow of time can be traced to the near homogeneity of the "early" universe implied by $\rho$ and the implication that the progenitors of approximately isolated subsystems started out far from equilibrium at "early" times.

## II.3.2. *The Reconstruction of History*

In classical physics reconstructing the past history of the universe, or any subsystem of it, is most honestly viewed as the process of assigning probabilities to alternatives in the past given present records. We assign the date 55BC to the Roman conquest of Britain on the basis of present textual records. We use present observations of the position of the sun and moon on the sky to reconstruct their past trajectories. We use the fossil record to estimate that the probability is high that dinosaurs roamed the earth from 230-65 million years ago. We believe that matter and radiation were in thermal equilibrium some 12 billion years ago on the basis of records such as the present values of the Hubble constant, the mean mass density, and the temperature of the cosmic background radiation. History becomes predictive and testable when we predict that *further* present records will be consistent with those already found. Texts yet to be discovered are predicted to be consistent with the story of Caesar. Present records of past eclipses are predicted to be consistent with our past extrapolations of the position of the sun and moon. Fossils yet to be unearthed are predicted to lie in appropriate strata. New measures of the age of the universe (such as that provided by the evolution of globular clusters) are predicted to be consistent with those obtained from other sources. In such ways history becomes a predictive science.

Most fundamentally, the reconstruction of history, including the classical process described above, must be seen in the context of the process quantum mechanical retrodiction discussed in the preceding subsection. Certain features of the quantum mechanical process deserve to be stressed to contrast them with the classical process.

First, quantum mechanics does not allow probabilities to be assigned to arbitrary sets of alternative histories; the set must decohere. As the two slit example shows the reconstruction of history *generally* is forbidden in quantum mechanics. Second, for interesting sets of alternatives that do decohere, the decoherence and the assigned probabilities will be approximate. It is unlikely, for example, that the initial state of the universe is such that the interference is *exactly* zero between two past positions of the sun on the sky. (See Section II.10 for further discussion.) Third, the decoherence of a set of histories as well as the probabilities for the individual histories in the set depend on the initial condition of the universe as well as on present data. Eq.(3.3) gives the conditional probability for a string of alternatives $\alpha_1, \cdots, \alpha_{n-1}$ in the past, given alternatives $\alpha_n$ representing the values of present records. This depends on $\rho$ as well as the $\alpha_n$ and, therefore, there is no present effective density matrix for retrodiction as there is for prediction. The reconstruction of history on the basis of present data alone is not possible in quantum mechanics in general. The classical reconstruction of history from present data alone is possible only for sets of histories that exhibit high levels of classical correlation in time. This will be discussed in Section II.7.

In classical physics new and better present data lead to new and more accurate probabilities for the past. It was the vision of classical physical physics that probabilities were the result of ignorance and sufficient fine graining would establish a unique past. This is not the case in quantum mechanics. Arbitrarily fine-grained sets of histories do not decohere. Sets sufficiently coarse-grained to be assigned probabilities will generally have alternative pasts with probabilities neither zero or one. However, in quantum mechanics there is not even a unique set of alternative histories. Alteration of the coarse graining in the future can change the possibilities for retrodiction of the past as discussed in Section II.2.3. Consider the Schrödinger cat experiment carried out at a certain time. In future coarse grainings confined to the quantities of classical physics the cat can be said to have been either alive or dead with certain probabilities at the time of the experiment. However, in a coarse graining that, at a later time, involves operators sensitive to the interference between configurations in which the cat is alive or dead it will not, in general, be possible even to assign probbabilities to these past alternatives.

## II.4. Branches (Illustrated by a Pure $\rho$)

Decohering sets of alternative histories give a definite meaning to Everett's branches. For each such set of histories, the exhaustive set of $P_{\alpha_k}^k$ at each time $t_k$ corresponds to a branching. To illustrate this even more explicitly, consider an initial density matrix that is a pure state, as in typical proposals for the wave function of the universe:

$$\rho = |\Psi><\Psi| \quad . \tag{II.4.1}$$

The initial state may be decomposed according to the projection operators that define the set of alternative histories

$$|\Psi> = \sum_{\alpha_1\cdots\alpha_n} P_{\alpha_n}^n(t_n)\cdots P_{\alpha_1}^1(t_1)|\Psi> \tag{II.4.2}$$

$$\equiv \sum_{\alpha_1\cdots\alpha_n} |[P_\alpha],\Psi> \quad . \tag{II.4.3}$$

The states $|[P_\alpha],\Psi>$ are approximately orthogonal as a consequence of their decoherence

$$<[P_{\alpha'}],\Psi|[P_\alpha],\Psi> \approx 0, \quad \text{for any} \quad \alpha_k' \neq \alpha_k \quad . \tag{II.4.4}$$

Eq.(4.4) is just a reëxpression of the definition of decoherence (2.7), given (4.1).

When the initial density matrix is pure, it is easily seen that some coarse graining in the present is always needed to achieve decoherence in the past. If the $P_{\alpha_n}^n(t_n)$ for the last time $t_n$ in the decoherence functional (2.3) were projections onto a complete set of states, $D$ would factor and could never satisfy the condition for decoherence (2.7) except for trivial histories composed of projections that are *exactly* correlated with the $P_{\alpha_n}^n(t_n)$. Similarly, it is not difficult to show that some coarse graining is required at any time in order to have decoherence of previous alternatives with the same kind of exceptions.

After normalization, the states $|[P_\alpha],\Psi>$ represent the individual histories or individual branches in the decohering set. We may, as for the effective density matrix of Section II.3.1, summarize present information for prediction just by giving one of these wave functions with projections up to the present.

## II.5. Sets of Histories with the Same Probabilities

If the projections $P$ are not restricted to a particular class (such as projections onto ranges of $q^i$ variables), so that coarse-grained histories consist of arbitrary exhaustive families of projection operators, then the problem of exhibiting the decohering sets of strings of projections arising from a given $\rho$ is a purely algebraic one. Assume, for example, that the initial condition is known to be a pure state as in (4.1). The problem of finding ordered strings of exhaustive sets of projections $[P_\alpha]$ so that the histories $P_{\alpha_n}^n \cdots P_{\alpha_1}^1 |\Psi>$ decohere according to (4.4) is purely algebraic and involves just subspaces of Hilbert space. The problem is the same for one vector $|\Psi>$ as for any other. Indeed, using subspaces that are *exactly* orthogonal, we may identify sequences that *exactly* decohere.

However, it is clear that the solution of the mathematical problem of enumerating the sets of decohering histories of a given Hilbert space has no physical content by itself. No description of the histories has been given. No reference has been made to a theory of the fundamental interactions. No distinction has been made between one vector in Hilbert space as a theory of the initial condition and any other. The resulting probabilities are merely abstract numbers.

We obtain a description of the sets of alternative histories of the universe when the operators corresponding to the fundamental fields are identified. We make contact with the theory of the fundamental interactions if the evolution of these fields is given by a fundamental Hamiltonian. Different initial vectors in Hilbert space will then give rise to decohering sets having different descriptions in terms of the fundamental fields. The probabilities acquire physical meaning.

Two different simple operations allow us to construct from one set of histories another set with a *different description* but the *same probabilities*.* First consider unitary transformations of the $P$'s that are constant in time and leave the initial $\rho$ fixed

$$\rho = U\rho U^{-1} \quad , \tag{II.5.1}$$

$$\tilde{P}_\alpha^k(t) = U P_\alpha^k(t) U^{-1} \quad . \tag{II.5.2}$$

If $\rho$ is pure there will be very many such transformations; the Hilbert space is large and only a single vector is fixed. The sets of histories made up from the $\{\tilde{P}_\alpha^k\}$ will have an identical decoherence functional to the sets constructed from the corresponding $\{P_\alpha^k\}$. If one set decoheres, the other will and the probabilities for the individual histories will be the same.

In a similar way, decoherence and probabilities are invariant under arbitrary reassignments of the times in a string of $P$'s (as long as they continue to be ordered), with the projection operators at the altered times unchanged as operators in Hilbert space. This is because in the Heisenberg picture every projection is, at *any* time, a projection operator for *some* quantity.

The histories arising from constant unitary transformations or from reassignment of times of a given set of $P$'s will, in general, have very different descriptions in terms of fundamental fields from that of the original set. We are considering transformations such as (5.2) in an active sense so that the field operators and Hamiltonian are unchanged. (The passive transformations, in which these are transformed, are easily understood.)    A set of projections onto the ranges of field values in a spatial region is generally transformed by (5.2) or by any reassignment of the times into an extraordinarily complicated combination of all fields and all momenta at all positions in the universe! Histories consisting of projections onto values of similar quantities at different times can thus become histories of very different quantities at various other times.

---

* Discussions with R. Penrose were useful on this point.

In ordinary presentations of quantum mechanics, two histories with different descriptions can correspond to physically distinct situations because it is presumed that the various different Hermitian combinations of field operators are potentially measurable by different kinds of external apparatus. In quantum cosmology, however, apparatus and system are considered together and the notion of physically distinct situations may have a different character.

## II.6. The Origins of Decoherence in Our Universe

### II.6.1. *On What Does Decoherence Depend?*

What are the features of coarse-grained sets of histories that decohere in our universe? In seeking to answer this question it is important to keep in mind the basic aspects of the theoretical framework on which decoherence depends. Decoherence of a set of alternative histories is not a property of their operators *alone*. It depends on the relations of those operators to the density matrix $\rho$, the Hamiltonian $H$, and the fundamental fields. Given these, we could, in principle, *compute* which sets of alternative histories decohere.

We are not likely to carry out a computation of all decohering sets of alternative histories for the universe, described in terms of the fundamental fields, anytime in the near future, if ever. However, if we focus attention on coarse grainings of particular variables, we can exhibit widely occurring mechanisms by which they decohere in the presence of the actual $\rho$ of the universe. We have mentioned that decoherence is automatic if the projection operators $P$ refer only to one time; the same would be true even for different times if all the $P$'s commuted with one another. In cases of interest, each $P$ typically factors into commuting projection operators, and the factors of $P$'s for different times often fail to commute with one another, for example factors that are projections onto related ranges of values of the same Heisenberg operator at different times. However, these non-commuting factors may be correlated, given $\rho$, with other projection factors that do commute or, at least, effectively commute inside the trace with the density matrix $\rho$ in eq.(2.3) for the decoherence functional. In fact, these other projection factors may commute with all the subsequent $P$'s and thus allow themselves to be moved to the outside of the trace formula. When all the non-commuting factors are correlated in this manner with effectively commuting ones, then the off-diagonal terms in the decoherence functional vanish, in other words, decoherence results. Of course, all this behavior may be approximate, resulting in approximate decoherence.

This type of situation is fundamental in the interpretation of quantum mechanics. Non-commuting quantities, say at different times, may be correlated with commuting or effectively commuting quantities because of the character of $\rho$ and $H$, and thus produce decoherence of strings of $P$'s despite their non-commutation. For a pure $\rho$, for example, the behavior of the effectively commuting variables leads to the orthogonality of the branches of the state $|\Psi>$, as defined in (4.4). Correlations of this character are central to understanding historical records (Section II.3.2) and measurement situations (Section II.9).

Specific models of this kind of decoherence have been discussed by many authors, among them Joos and Zeh (1985), Zurek (1984), and Caldeira and Leggett (1983), and Unruh and Zurek (1989). We shall now discuss two examples.

## II.6.2. A Two Slit Model

Let us begin with a very simple model due to Joos and Zeh (1985) in its essential features. We consider the two slit example again but this time suppose that in the neighborhood of the slits there is a gas of photons or other light particles colliding with the electrons (Fig. 3). Physically it is easy to see what happens, the random uncorrelated collisions can carry away delicate phase correlations between the beams even if they do not affect the trajectories of the electrons very much. The interference pattern will then be destroyed and it will be possible to assign probabilities to whether the electron went through the upper slit or the lower slit. Let us see how this picture is reflected in mathematics. Initially, suppose the state of the entire system is a state of the electron $|\psi>$ and $N$ distinguishable "photons" in states $|\varphi_1>$, $|\varphi_2>$, etc., viz.

$$|\Psi> = |\psi>|\varphi_1>|\varphi_2> \cdots |\varphi_N> \ . \tag{II.6.1}$$

$|\psi>$ is a coherent superposition of a state in which the electron passes through the upper slit $|U>$ and the lower slit $|L>$. Explicitly:

$$|\psi> = \alpha|U> + \beta|L> \ . \tag{II.6.2}$$

The wave functions of both states are confined to moving wave packets in the $x$-direction so that position in $x$ recapitulates history in time. We now ask whether for the initial condition (6.1) of this "universe", the history where the electron passes through the upper slit and arrives at a detector at point $y$ on the screen decoheres from that in which it passes through the lower slit and arrives at point $y$. That is, as in Section II.4, we ask whether the two vectors

$$P_y P_U|\Psi> \quad , \quad P_y P_L|\Psi> \tag{II.6.3}$$

are nearly orthogonal, the times of the projections being those for the nearly classical motion in $x$. The overlap can be worked out in the Schrödinger picture where the initial state evolves and the projections on the electron's position are applied to it at the appropriate times. Collisions occur, but the states $|U>$ and $|L>$ are left more or less undisturbed. The states of the "photons", of course, are significantly affected. If the photons are dilute enough to be scattered once by the electron in its time to traverse the gas the two states of (6.3) will be approximately

$$\alpha P_y|U> S_U|\varphi_1> S_U|\varphi_2> \cdots S_U|\varphi_N> \ , \tag{II.6.4a}$$

and

$$\beta \ P_y|L> S_L|\varphi_1> S_L|\varphi_2> \cdots S_L|\varphi_N> \ . \tag{II.6.4b}$$

Here, $S_U$ and $S_L$ are the scattering matrices from an electron in the vicinity of the upper slit and the lower slit respectively. The two branches in (6.4) decohere because the states of the "photons" are nearly orthogonal. The overlap is proportional to

$$<\varphi_1|S_U^\dagger S_L|\varphi_1><\varphi_2|S_U^\dagger S_L|\varphi_2> \cdots <\varphi_N|S_U^\dagger S_L \ |\varphi_N> \ . \tag{II.6.5}$$

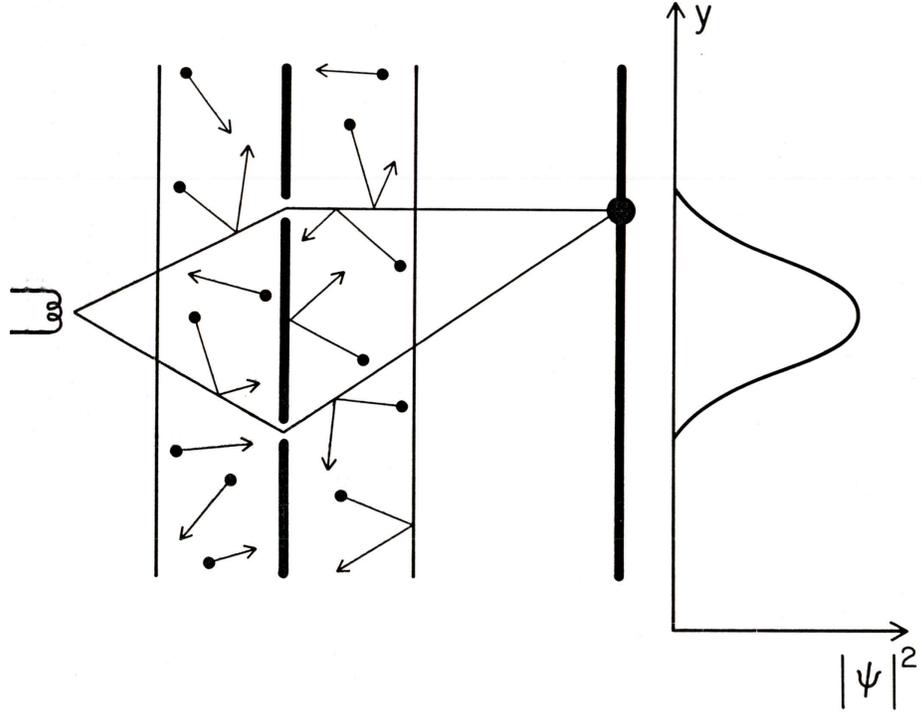

**Fig. 3:** *The two slit experiment with an interacting gas. Near the slits light particles of a gas collide with the electrons. Even if the collisions do not affect the trajectories of the electrons very much they can still carry away the phase correlations between the histories in which the electron arrived at point y on the screen by passing through the upper slit and that in which it arrived at the same point by passing through the lower slit. A coarse graining that consisted only of these two alternative histories of the electron would approximately decohere as a consequence of the interactions with the gas given adequate density, cross-section, etc. Interference is destroyed and probabilities can be assigned to these alternative histories of the electron in a way that they could not be if the gas were not present (cf. Fig. 1). The lost phase information is still available in correlations between states of the gas and states of the electron. The alternative histories of the electron would not decohere in a coarse graining that included both the histories of the electron and operators that were sensitive to the correlations between the electrons and the gas.*

*This model illustrates a widely occuring mechanism by which certain types of coarse-grained sets of alternative histories decohere in the universe.*

Now the $S$-matrices for scattering off the upper position or the lower position can be connected to that of an electron at the orgin by a translation

$$S_U = e^{-i\mathbf{k}\cdot\mathbf{x}_U}\, S\, e^{+i\mathbf{k}\cdot\mathbf{x}_U} \ , \tag{II.6.6a}$$

$$S_L = e^{-i\mathbf{k}\cdot\mathbf{x}_L} S\, e^{+i\mathbf{k}\cdot\mathbf{x}_L} \ . \tag{II.6.6b}$$

Here, $\hbar\mathbf{k}$ is the momentum of a photon, $\mathbf{x}_U$ and $\mathbf{x}_L$ are the positions of the slits and $S$ is the scattering matrix from an electron at the origin.

$$< \mathbf{k}'|S|\mathbf{k} >= \delta^{(3)}\left(\mathbf{k} - \mathbf{k}'\right) + \frac{i}{2\pi\omega_{\mathbf{k}}} f\left(\mathbf{k}, \mathbf{k}'\right)\delta\left(\omega_k - \omega_k'\right) \ , \qquad (\text{II}.6.7)$$

where $f$ is the scattering amplitude and $\omega_k = |\vec{k}|$.

Consider the case where all the photons are in plane wave states in an interaction volume $V$, all having the same energy $\hbar\omega$, but with random orientations for their momenta. Suppose further that the energy is low so that the electron is not much disturbed by a scattering and low enough so the wavelength is much longer than the separation between the slits, $k|\mathbf{x}_U - \mathbf{x}_L| << 1$. It is then possible to work out the overlap. The answer according to Joos and Zeh (1985) is

$$\left(1 - \frac{(k|\mathbf{x}_U - \mathbf{x}_L|)^2}{8\pi^2 V^{2/3}}\sigma\right)^N \qquad (\text{II}.6.8)$$

where $\sigma$ is the effective scattering cross section. Even if $\sigma$ is small, as $N$ becomes large this tends to zero. The characteristic time for this loss of coherence is that for which the number of collisions times the second term in the argument of (6.8) is near unity. That is,

$$t_{\text{decoherence}} \sim \frac{V^{2/3}\tau}{(k|\mathbf{x}_U - \mathbf{x}_L|)^2\sigma} \ , \qquad (\text{II}.6.9)$$

where $\tau$ is the collision time. In this way decoherence becomes a quantitative phenomenon.

### II.6.3. The Caldeira-Leggett Oscillator Model

A more sophisticated model has been studied by Caldeira and Leggett (1983), Zurek (1984), and others. The model consists of a distinguished oscillator in one dimension, interacting linearly with a large number of other oscillators, and a coarse graining which involves only the coördinate of the distinguished oscillator. Let $x$ be the coördinate of the distinguished oscillator and $X_k$ the coördinates of the rest. The Lagrangians for the distinguished oscillators by themselves are

$$H_{\text{free}}(p, x) = \frac{1}{2M}(p^2 + \omega^2 x^2) \qquad (\text{II}.6.10)$$

and

$$H_{\text{osc}}(p_k, X_k)) = \frac{1}{2m}\sum_k\left(P_k^2 + \omega_k^2 X_k^2\right) \quad . \qquad (\text{II}.6.11)$$

The interaction is linear,

$$H_{\text{int}}(X, X_k) = x\sum_k C_k X_k \ , \qquad (\text{II}.6.12)$$

defining couplings $C_k$. Consider the special case where the initial density matrix of the whole system factors into a density matrix $\bar{\rho}(x', x)$ for the particle and a density matrix describing a thermal bath at temperature $T = 1/\beta k$ for the oscillators:

$$< x'X_k'|\rho|xX_k >= \bar{\rho}(x', x)\prod_k \rho_k(X_k', X_k) \ . \qquad (\text{II}.6.13)$$

The density matrix for each oscillator of the bath is

$$\rho_k(X_k', X_k) = <X_k'|e^{-\beta H_{\text{osc}}}|X_k> / Tr(e^{-\beta H}) = \left[\frac{m\omega_k}{\pi\hbar}\tanh\left(\frac{\hbar\omega_k\beta}{2}\right)\right]^{\frac{1}{2}} \quad \text{(II.6.14)}$$

$$\times \exp\left[-\left\{\frac{m\omega_k}{2\hbar\sinh(\hbar\beta\omega_k)}\left[\left(X_k'^2+X_k^2\right)\cosh(\hbar\beta\omega_k)-2X_k'X_k\right]\right\}\right] . \quad \text{(II.6.15)}$$

Now, imagine constructing the decoherence functional following only intervals of the position of the particle, $[\Delta_k]$, and completely integrating out the coördinates of the bath. It should be clear that the integrals can be done because everything about the bath is the exponential of a quadratic form. All integrals are Gaussian path integrals. The result has the general form

$$D\left([\Delta_{\alpha'}], [\Delta_\alpha]\right) = \int_{[\Delta_{\alpha'}]}\delta x' \int_{[\Delta_\alpha]}\delta x\,\delta(x_f'-x_f)\exp\Bigg\{i\bigg(S_{\text{free}}[x'(t)]-S_{\text{free}}[x(t)]$$

$$+W[x'(t),x(t)]\bigg)/\hbar\Bigg\}\bar\rho(x_0',x_0) \quad , \quad \text{(II.6.16)}$$

where $S_{\text{free}}$ is the free action of the distinguished oscillator with frequency renormalized by the interaction to $\omega_R$. The intervals $[\Delta_\alpha]$ refer only to the variables of the distinguished particle. The sum over the rest of the oscillators has been carried out and is summarized by the Feynman-Vernon influence functional $\exp(iW[x'(t),x(t)])$. The remaining sum over $x'(t)$ and $x(t)$ is as in (2.2).

$W[x'(t),x(t)]$ will be quadratic in the paths of the special particle. I shall not quote its general form given in Caldeira and Leggett (1983), but just that of a simple case. This is a cutoff continuum of oscillators with couplings

$$\rho_D(\omega)C^2(\omega) = \begin{cases} \frac{4Mm\gamma\omega^2}{\pi} & \omega < \Omega \\ 0 & \omega > \Omega \end{cases} \quad \text{(II.6.17)}$$

where $\rho_D(\omega)$ is the density of oscillators with frequency $\omega$. Then in the further Fokker-Planck limit where $kT >> \hbar\Omega >> \hbar\omega_R$

$$W[x'(t),x(t)] = -M\gamma\int dt\,[x'\dot x'-x\dot x+x'\dot x-x\dot x']$$

$$+i\frac{2M\gamma kT}{\hbar}\int dt\,[x'(t)-x(t)]^2 \quad , \quad \text{(II.6.18)}$$

where $\gamma$ summarizes the interaction strengths of the distinguished oscillator with the bath. The real part of $W$ contributes dissipation to the equations of motion. The imaginary part squeezes the trajectories $x(t)$ and $x'(t)$ together, thereby providing approximate decoherence. Very roughly, primed and unprimed position intervals separated by distances $d$ in (6.16) will decohere when spaced in time by intervals

$$t_{\text{decoherence}} \gtrsim \frac{1}{\gamma}\left[\left(\frac{\hbar}{\sqrt{2MkT}}\right)\cdot\left(\frac{1}{d}\right)\right]^2 \quad . \quad \text{(II.6.19)}$$

As stressed by Zurek (1984), for typical macroscopic parameters this minimum time for decoherence can be many orders of magnitude smaller than a characteristic dynamical time, say the damping time $1/\gamma$. (The ratio is around $10^{-40}$(!) for $M \sim$ gm, $T \sim 300°K, d \sim$ cm.)

What the above models convincingly show is that decoherence will be widespread in the universe for certain familiar "classical" variables. Alternative histories of the position of a mm. size dust grain, initially in a coherent superposition of two different positions separated by similar dimensions, decohere, if for no other reason, by the interaction of the grain with the 3° cosmic background radiation if the successive localizations are spaced by more than a nanosecond (Joos and Zeh, 1985).*

## II.6.4. The Evolution of Reduced Density Matrices

The coarse graining in the Caldeira-Leggett model focusses on one variable at a succession of times. The probabilities for this variable at any one time can be computed from a reduced density matrix on the Hilbert space of the distinguished oscillator

$$\tilde{\rho}_{\text{eff}}(t) = Sp[\rho_{\text{eff}}(t)] \ . \tag{II.6.20}$$

Here, $\rho_{\text{eff}}(t)$ is the effective density matrix for the whole system at the time $t$ as introduced in Section II.3 and $Sp$ denotes a trace over the Hilbert spaces of all the other oscillators. The mechanism for decoherence that has been described for this example also leads to interesting time behavior of this reduced density matrix.

In the position representation there is a convenient path integral summary* of the evolution of $\tilde{\rho}_{\text{eff}}$.

$$< x'|\tilde{\rho}_{\text{eff}}(t)|x> = \int_{[\Delta_{\beta'}]} \delta x' \int_{[\Delta_{\beta}]} \delta x$$

$$\times \exp\left\{ i\Big( S_{\text{free}}[x'(t)] - S_{\text{free}}[x(t)] + W[x'(t),x(t)]\Big)/\hbar \right\} \bar{\rho}(x'_0, x_0) \ . \tag{II.6.21}$$

The integral over $x(t)$ is over paths that begin at $x_0$, pass through all intervals $[\Delta_{\alpha}]$ that are at times *before* $t$, and end at time $t$ at the position $x$ specified by the matrix element $< x'|\tilde{\rho}_{\text{eff}}(t)|x>$. The integration over $x'(t)$ is analogous. Eq.(6.21) is similar to (6.16) except that the class of paths integrated over is different. In particular the paths in (6.21) do not end at a common value that is then integrated over as they do in (6.16). The similarity is enough, however, to show that the same imaginary part of $W$ that squeezes the coarse-grained histories $x(t)$ and $x'(t)$ together will cause the reduced density matrix $\tilde{\rho}_{\text{eff}}$ to evolve to near diagonal form in the position representation on the decoherence time scale (6.19).

The approach to diagonal form of a reduced density matrix has often been discussed in connection with mechanisms that effect decoherence. However, the approach to diagonal form of a reduced density matrix cannot be taken to be the *definition* of decoherence. First, a reduced density matrix is suitable only for limited kinds of coarse grainings — those which distinguish particular variables and the same variables at each time. (More precisely it is appropriate only coarse grainings defined as sequences of projections that operate on a fixed number of factors of a tensor product Hilbert space.) There are many more general and more realistic kinds of coarse graining. Second, and more importantly, as discussed in II.2.3, decoherence is a property of sets of alternative histories and therefore cannot be described by an effective density matrix at a moment of time. Even if the off-diagonal elements vanish at one moment of time there is nothing to guarantee that at a later moment they may not become non-vanishing again.†

---

* It should be clear from examples such as this that the only realistic closed systems in which many widespread mechanisms of decoherence operate are of cosmological dimensions — the cosmological event horizon if one is discussing the history of the universe and even light hours in the discussion of laboratory experiments.

* A slight generalization of that of Feynman and Vernon (1963).

## II.7. Towards a Classical Domain

As observers of the universe, we deal with coarse grainings that reflect our own limited sensory perceptions, extended by instruments, communication and records but in the end characterized by a large amount of ignorance. Yet, we have the impression that the universe exhibits a finer graining, independent of us, defining an always decohering "classical domain", to which our senses are adapted, but deal with only a small part of. Setting out for a journey to a distant, unseen part of the universe we do not imagine that we need to equip ourselves with spacesuits having receptors sensitive, say, to coherent superpositrons of familiar "classical variables". We expect the finer graining resulting from adjoining sufficiently coarse-grained "classical variables" in the new region to continue to decohere and to exhibit correlations in time for the most part conforming to classical dynamical laws.

To what should we attribute the existence of our "classical domain". Fundamentally there are three elements of the framework of quantum mechanics under discussion — the initial condition $\rho$, the Hamiltonian describing evolution, and the projection operators defining the possible alternative coarse-grained histories. There are no sets of operators comprising a coarse graining that define a classical domain in every circumstance. Rather, like decoherence itself, a classical domain can only be a property of the initial condition of the universe and the Hamiltonain describing evolution. Given the Hamiltonian of the elementary particles, there may be a wide range of initial conditions that give rise to a classical domain even though most do not. The existence of a classical domain would then not be much of a test of a theory of the initial condition. Yet, given the Hamiltonian it is still interesting to ask, What is the class of initial conditions that give rise to classical domains? Are the familiar variables of classical physics uniquely singled out to describe such coarse grainings or are there other possibilities? Does the initial condition of our universe define one or more classical domains? To answer this kind of question we need more precise criteria for what kinds of coarse grainings constitute a classical domain. Such criteria would apply both to the probabilities of the individual histories in the classical domain and to their descriptions in terms of fundamental fields. No completely satisfactory criteria have yet been given* but some attributes of a successful definition can at least be sketched in words if not yet fully in equations:

A classical domain would be a *set* of coarse-grained, alternative, decohering histories with at least the following properties:

---

† There are interesting examples of this. See e.g. Leggett, et al. (1987).
* For a fuller discussion see Gell-Mann and Hartle (1990).

(1)   A classical domain should be maximally refined consistent with decoherence so that it is a property of the universe and not the choice of any particular observer.   However, it should not contain trivial refinements such as would be obtained, for example, by mindlessly interpolating projections on the particular branch at every time

(2)   A classical domain should be made up of histories that consist, for the most part, of the same variables at different times.  That is, they should be made up of habitually decohering variables.  However, the histories cannot consist entirely of such variables because, as we shall see, in a measurement situation there may be very different variables that decohere, not habitually, but only by virtue of their correlation with a habitually decohering one.

(3)   The histories of a classical domain should exhibit, as much as possible, patterns of classical correlation among the habitually decohering variables.  That is, successive projections onto related ranges of habitually decohering variables should follow roughly classical orbits with probabilities as near to unity as possible.  However, this pattern of classical correlation cannot be exact or otherwise we would never know quantum mechanics!  The pattern of classical correlation may be disturbed by inclusion, in the set of projection operators, of other variables neither habitually decohering nor normally classically correlated as in a quantum measurement situation.  The pattern may also be disturbed by quantum spreading and by quantum and classical fluctuations.

Thus we can, at best, deal with quasiclassical sets of alternative decohering histories with trajectories that split and fan out.  There are no classical domains only quasiclassical ones.  We shall refer to the operators that habitually define them as "quasiclassical operators".

We can understand the origin of at least some quasiclassical operators in reasonably general terms as follows:  In the earliest instants of the universe the operators defining spacetime on scales well above the Planck scale emerge from the quantum fog as quasiclassical.  Any theory of the initial condition that does not imply this is simply inconsistent with observation in a manifest way.  A background spacetime is thus defined and conservation laws arising from spacetime symmetries have meaning.  Then, where there are suitable conditions of low temperature, etc., various sorts of hydrodynamic variables may emerge as quasiclassical operators.  These are integrals over suitable small volumes of densities of conserved or nearly conserved quantities.  Examples are densities of energy, momentum, baryon number, and, in later epochs, nuclei, and even chemical species.  The sizes of the volumes are limited above by the requirement that the histories be refined as much as possible consistent with decoherence.  They are limited below by classicality because they require sufficient "inertia" to enable them to resist deviations from predictability caused by their interactions with one another, by quantum spreading, and by the quantum and statistical fluctuations summed over to produce decoherence.  Suitable integrals of densities of approximately conserved quantities are thus candidates for habitually decohering quasiclassical operators.  Field theory is local, and it is an interesting question whether that locality somehow picks out local densities as the source of habitually decohering quantities.  It is hardly necessary to note that such hydrodynamic variables are among the principal variables of classical physics.

In the case of densities of conserved quantities, the integrals would not change at all if the volumes were infinite.  For smaller volumes we expect approximate persistence.  When, as in hydrodynamics, the rates of change of the integrals form part of an approximately closed system of equations of motion, the resulting evolution is just as classical as in the case of persistence.

It would be a striking and deeply important fact of the universe if among the decoherent sets of alternative histories there were one roughly equivalent group with much higher classicalities than all the others. That would then be *the* quasi-classical domain, completely independent of any subjective criterion, and realized within quantum mechanics by utilizing only the initial condition of the universe and the Hamiltonian of the elementary particles. It would have the form of *alternative histories*, constantly branching and fanning out. Supplemented by the specific information gained from observation, which restricts the branches, it would be *the* arena for prediction in quantum mechanics.

It might seem at first sight that in such a picture the complementarity of quantum mechanics would be lost. In a given situation, for example, *either* a momentum *or* a coördinate could be measured, leading to different kinds of histories. That impression is illusory. The history in which an observer, as part of the universe, measures $p$ and the history in which that observer measures $x$ are two decohering alternatives. In each of these branches, numerous variables referring to things like the $3°K$ photons are integrated over. (These variables are not necessarily the same for all branches, so that some aspects of the $3°K$ background radiation, for example, may belong to one branch of the quasiclassical domain but not to another.) The important point is that the decoherent histories of a quasiclassical domain contain all possible choices that might be made by all possible observers that might exist, now, in the past, or in the future.

The EPR or EPRB situation is no more mysterious. There, a choice of measurements, say, $\sigma_x$ or $\sigma_y$ for a given electron, is correlated with the behavior of $\sigma_x$ or $\sigma_y$ for another electron because the two together are in a singlet spin state even though widely separated. Again, the two measurement situations (for $\sigma_x$ and $\sigma_y$) decohere from each other, but here, in each, there is also a correlation between the information obtained about one spin and the information that can be obtained about the other.

## II.8. The Branch Dependence of Decoherence

As the discussion in Sections II.6 and II.7 shows, physically interesting mechanisms for decoherence will operate differently in different alternative histories for the universe. For example, hydrodynamic variables defined by a relatively small set of volumes may decohere at certain locations in spacetime in those branches where a gravitationally condensed body (e.g. the earth) actually exists, and may not decohere in other branches where no such condensed body exists at that location. In the latter branch there simply may be not enough "inertia" for densities defined with too small volumes to resist deviations from predictability. Similarly, alternative spin directions associated with Stern-Gerlach beams may decohere for those branches on which a photographic plate detects their beams and not in a branch where they recombine coherently instead. There are no variables that are expected to decohere universally. Even the mechanisms causing spacetime geometry at a given location to decohere on scales far above the Planck length cannot necessarily be expected to operate in the same way on a branch where the location is the center of a black hole as on those branches where there is no black hole nearby.

How is such "branch dependence" described in the formalism we have elaborated? It is not described by considering histories where the *set* of alternatives at one time (the $k$ in a set of $P_\alpha^k$) depends on *specific* alternatives (the $\alpha$'s) of sets of earlier times. Such dependence would destroy the derivation of the probability sum rules from the fundamental formula. However, there is no such obstacle to the set of alternatives at one time depending on the *sets* of alternatives at all previous times. It is by exploiting this possibility, together with the possibility of present records of past events, that we can correctly describe the sense in which there is branch dependence of decoherence, as we shall now discuss.

A record is a present alternative that is, with high probability, correlated with an alternative in the past. The construction of the relevant probabilities was discussed in Section II.3, including their dependence on the initial condition of the universe (or at least on information that effectively bears on that initial condition). Even non-commuting alternatives such as a position and its momentum at different, even nearby times may be stored in presently commuting record variables.

The branch dependence of histories becomes explicit when sets of alternatives are considered that include records of specific events in the past. To illustrate this, consider the example above, where different sorts of hydrodynamic variables might decohere or not depending on whether there was a gravitational condensation. The set of alternatives that decohere must refer both to the records of the condensation *and* to hydrodynamic variables. Hydrodynamic variables with smaller volumes would be part of the subset with the record that the condensation took place and vice versa.

The branch dependence of decoherence provides the most direct argument against the position that a classical domain should simply be *defined* in terms of a certain set of variables (e.g. values of spacetime averages of the fields in the classical action). There are unlikely to be any physically interesting variables that decohere independent of circumstance.

## II.9. Measurement

When a correlation exists between the ranges of values of two operators of a quasiclassical domain, there is a *measurement situation*. From a knowledge of the value of one, the value of the other can be deduced because they are correlated with probability near unity. Any such correlation exists in some branches of the universe and not in others; for example, measurements in a laboratory exist only in those branches where the laboratory was actually constructed!

I use the term "measurement situation" rather than "measurement" for such correlations to stress that nothing as sophisticated as an "observer" need be present for them to exist. If there are many significantly different quasiclassical domains, different measurement situations may be exhibited by each one.

When the correlation we are discussing is between the ranges of values of two quasiclassical operators, of the kind discussed in Section II.7, which *habitually* decohere, we have a measurement situation of a familiar classical kind. However, besides the quasiclassical operators, the highly classical sets of alternative histories of a quasiclassical domain may include *other* operators whose ranges of values are highly correlated with the quasiclassical ones at particular times. Such operators, not normally decohering, are, in fact, only included among the decohering set by virtue of their correlation with a habitually decohering one. In this case we have a measurement situation of the kind usually discussed in quantum mechanics. Suppose, for example, in the inevitable Stern-Gerlach experiment, that $\sigma_z$ of a spin-$1/2$ particle is correlated with the orbit of an atom in an inhomogeneous magnetic field. If the two directions decohere because of interaction with something else that is summed over, (the atomic excitations in a photographic plate for example), then the spin direction will be included in the set of decoherent histories, fully correlated with the decohering orbital directions. The spin direction is thus measured.

The recovery of the Copenhagen rule for when probabilities may be assigned is immediate. Measured quantities are correlated with decohering histories. Decohering histories can be assigned probabilities. Thus in the two-slit experiment (Fig. 1), when the electron interacts with an apparatus that determines which slit it passed through, it is the decoherence of the alternative configurations of the apparatus that enables probabilities to be assigned for the electron.

Correlation between the ranges of values of operators of a quasiclassical domain is the *only* defining property of a measurement situation. Conventionally, measurements have been characterized in other ways. Essential features have been seen to be irreversibility, amplification beyond a certain level of signal-to-noise, association with a macroscopic variable, the possibility of further association with a long chain of such variables, and the formation of enduring records. Efforts have been made to attach some degree of precision to words like "irreversible", "macroscopic", and "record", and to discuss what level of "amplification" needs to be achieved. Such characterizations of measurement are difficult to define precisely.

An example of this occurs in the case of "null measurements" discussed by Renninger (1960), Dicke (1981), and others. An atom decays at the center of a spherical cavity. A detector that covers all but a small opening in the sphere does *not* register. We conclude that we have measured the direction of the decay photon to an accuracy set by the solid angle subtended by the opening. Certainly there is an interaction of the electromagnetic field with the detector, but did the escaping photon suffer an "irreversible act of amplification"? The point in the present approach is that the *set* of alternatives, detected and not detected, decohere because of the place of the detector in the universe.

Despite the lack of precise measures, characteristics such as irreversibility, amplification, etc. may be seen to follow *roughly* from the present definition in familiar measurement situations as follows:

Correlation of a variable with a quasiclassical domain (actually, inclusion in its set of histories) accomplishes the amplification beyond noise and the association with a macroscopic variable that can be extended to an indefinitely long chain of such variables. The relative predictability of the classical domain is a generalized form of record. The approximate constancy of, say, a mark in a notebook is just a special case; persistence in a classical orbit is just as good.

Irreversibility is more subtle. One measure of it is the cost (in energy, money, etc.) of tracking down the phases specifying coherence and restoring them. This is intuitively large in many typical measurement situations. Another, related measure is the negative of the logarithm of the probability of doing so. If the probability of reversing the phases in any particular measurement situation were significant, then we would not have the necessary amount of decoherence. The correlation could not be inside the set of decohering histories. Thus, this measure of irreversibility is large. Indeed, in many circumstances where the phases are carried off to infinity or lost in photons impossible to catch up with, the probability of recovering them is truly zero and the situation perfectly irreversible — infinitely costly to reverse and with zero probability for reversal!

Defining a measurement situation solely as the existence of correlations in a quasiclassical domain, if a suitable general definition of classicality can be found, would have the advantages of clarity, economy, and generality. Measurement situations occur throughout the universe and without the necessary intervention of anything as sophisticated as an "observer". Thus, by this definition, the production of fission tracks in mica deep in the earth by the decay of a uranium nucleus leads to a measurement situation in a quasiclassical domain in which the tracks directions decohere, whether or not these tracks are ever registered by an observer.

## II.10. The Ideal Measurement Model and the Copenhagen Approximation to Quantum Mechanics

In conventional discussions of measurement in quantum mechanics it is useful to consider ideal models of the measurement process (See, e.g. von Neumann 1932, London and Bauer, 1939, Wigner, 1963 or almost any current text on quantum mechanics). Such models idealize various approximate properties of realistic measurement situations as exact features of the model. For example, configurations of an apparatus corresponding to different results of an experiment are typically represented by *exactly* orthogonal states in these models. This kind of ideal model is useful in isolating the essential features of many laboratory measurement situations in an easily analysable way. Ideal measurement models are useful in quantum cosomology for the same reasons. Beyond that, however, they are useful in indicating how the Copenhagen formulation of quantum theory can be derived as an approximation to the quantum mechanics of the universe described here. I shall describe one such model.

Consider a closed system one part of which is a subsystem to be studied and the rest of which can be organized into various types of measuring apparatus. The latter includes any "observer" that may be present. Corresponding to this division, we assume a Hilbert space that is a tensor product, $\mathcal{H}_s \otimes \mathcal{H}_r$, of a Hilbert space for the subsystem and a Hilbert space for the rest. We assume an "initial condition" that is a product of a density matrix for the subsystem in $\mathcal{H}_s$ and another for the rest in $\mathcal{H}_r$

$$\rho = \rho_s \otimes \rho_r \ . \tag{II.10.1}$$

Various sets of alternatives for the subsystem are represented by exhaustive and exclusive sets of projection operators $\{S_\alpha^k(t)\}$, $\alpha = 1, 2, 3, \cdots$. Their Schrödinger picture representatives are of the form $S_\alpha^k = s_\alpha^k \otimes I_r$ where the $s_\alpha^k$ are a set of projection operators acting on $\mathcal{H}_s$. Of course, since the subsystem and the rest are interacting, the Heisenberg picture representatives $S_\alpha^k(t)$ will not in general have this product form. The various possible configurations of an apparatus which measures the set of alternatives $\{S_\alpha^k(t)\}$ are described by an exhaustive set of alternatives for the rest $\{R_\beta^{(k,\tau)}(t)\}$, $\beta = 1, 2, 3 \cdots$. The operator $R_\beta^{(k,\tau)}(t)$ corresponds to the alternative that the apparatus has recorded the alternative $\beta$ for the subsystem studied in the set $k$ at time $\tau$. We can ask about the value of this record at any time and so $R_\beta^{(k,\tau)}(t)$ itself depends on $t$. For example, we could ask whether the record of the result of the measurement persists. The $S$'s and the $R$'s at the same time are assumed to commute with one another.

Two of the three crucial assumptions defining the ideal measurement model are the following:

i. *Correlation*: The alternatives $\{S_\alpha^k(t)\}$ and $\{R_\beta^{(k,\tau)}(t)\}$ are exactly correlated, that is

$$Tr\Big[R_{\beta_n'}^{(n,t_n)}(t_n)S_{\alpha_n'}^n(t_n)\cdots R_{\beta_1'}^{(1,t_1)}(t_1)S_{\alpha_1'}^1(t_1)\rho S_{\alpha_1}^1(t_1)R_{\beta_1}^{(1,t_1)}(t_1)\cdots S_{\alpha_n}^n(t_n)R_{\beta_n}^{(n,t_n)}(t_n)\Big]$$

$$\propto \delta_{\alpha_n'\beta_n'}\cdots\delta_{\alpha_1'\beta_1'}\delta_{\alpha_1\beta_1}\cdots\delta_{\alpha_n\beta_n} \ . \tag{II.10.2}$$

This is an idealization of the measurement situation correlations discussed in the previous section. The existence of such correlations is not inherent in the properties of the operators $\{S_\alpha^k(t)\}$ and $\{R_\alpha^k(t)\}$. Their existence depends also on the Hamiltonian and on choosing an initial $\rho$ that models an experimental preparation of apparatus and subsystem and which will lead to a measurement situation. We assume in the model that we have a $\rho$ and $H$ of this character.*

ii. *Persistent Records*: We assume that as a consequence of $\rho$ and $H$, distinguishable, persistent, non-interacting *records* are formed of the results of the measurements of the various times $t$. The $R_\beta^{(k,\tau)}(t)$ describe the alternative values of these records at time $t$. More precisely we assume that if $t_2 > t_1$ are any two times later than $\tau$ then these operators have the property

$$Tr\Big[\cdots R_{\beta_2'}^{(k,\tau)}(t_2)\cdots R_{\beta_1'}^{(k,\tau)}(t_1)\cdots \rho \cdots R_{\beta_1}^{(k,\tau)}(t_1)\cdots R_{\beta_2}^{(k,\tau)}(t_2)\cdots\Big] \propto \delta_{\beta_2'\beta_1'}\delta_{\beta_1\beta_1}\ , \tag{II.10.3}$$

where the elipses $(\cdots)$ stand for any combination of $R$'s and $S$'s in the correct time order. That is, the record projections effectively commute with all other projections at time $t > \tau$. Eq. (10.3) is the statement that values of the records at later times are exactly correlated with those of earlier times. An assumption like (10.3) is not needed if only one measurement situation at one time is to be discussed, as is common in models of the measurement process. It is needed for discussions of sequences of measurements, as here, to ensure that subsequent interactions do not reëstablish the coherence of different measurement alternatives.

The questions of interest in this model are whether the set of histories of "measured" alternatives $\{[S_\alpha]\}$ decoheres, and, if so, what their probabilities are. The answers are supplied by analysing the decoherence functional

$$D\left([S_{\alpha'}],[S_\alpha]\right) = Tr\left[S_{\alpha_n'}^n(t_n)\cdots S_{\alpha_1'}^1(t_1)\rho S_{\alpha_1}^1(t_1)\cdots S_{\alpha_n}^n(t_n)\right]\ . \tag{II.10.4}$$

Alongside each $S_{\alpha_k}^k(t_k)$ in the above expression insert a resolution of the identy into record variables

$$\sum_{\beta_k} R_{\beta_k}^{(k,t_k)}(t_k) = 1\ . \tag{II.10.5}$$

Because of the assumption (i) of *exact* correlation between the subsystem alternatives $S_{\alpha_k}^k$ and the records [eq. (10.2)], only the term with $\beta_k = \alpha_k$ is this sum survives.

A consequence of condition (10.3) and the properties of projections (2.4) is that all the inserted $R$'s can be dragged to the outside of the decoherence functional and evaluated at the last time. The decoherence functional is then

$$D\left([S_{\alpha'}],[S_\alpha]\right) = Tr\Big[R_{\alpha_n'}^{(n,t_n)}(t_n)\cdots R_{\alpha_1'}^{(1,t_1)}(t_n)S_{\alpha_n'}^n(t_n)\cdots S_{\alpha_1'}^1(t_1)\rho$$

$$\times S_{\alpha_1}^1(t_1)\cdots S_{\alpha_n}^n(t_n)R_{\alpha_1}^{(1,t_1)}(t_n)\cdots R_{\alpha_n}^{(n,t_n)}(t_n)\Big]\ . \tag{II.10.6}$$

---

* A more realistic model would treat a more general $\rho$ but include as the first set of projections in the string defining a history a set one member of which is the alternative "ready to measure the alternatives in the set $k$". The relevant probabilities defining the correlations of a measurement situation would then be conditioned on this alternative for the apparatus.

Then, since the record variables, $R_\beta^{(k,\tau)}$, are exclusive by construction, we may use the cyclic property of the trace to show that the off-diagonal terms in the $\alpha$'s of (10.6) vanish identically. The records decohere. However, since the records are exactly correlated with measured properties of the system studied according to assumption (i), this decoherence accomplishes the decoherence of the measured alternatives of the system. Thus, as a consequence of the *existence* of alternatives $\{R_\beta^{(k,\tau)}(t)\}$ with the properties (i) and (ii) the decoherence functional (10.4) is exactly diagonal and we can write

$$D\left([S_{\alpha'}], [S_\alpha]\right) = \delta_{\alpha'_n \alpha_n} \cdots \delta_{\alpha'_1 \alpha_1} Tr\left[S_{\alpha_n}^n(t_n) \cdots S_{\alpha_1}^1(t_1) \rho S_{\alpha_1}^1(t_1) \cdots S_{\alpha_n}^n(t_n)\right]$$

$$\text{(II.10.7)}$$

Put differently, we can say that the decoherence of the records in the larger universe has accomplished the exact decoherence of the measured quantity of the subsystem studied.* noindent The third assumption of the ideal measurement model is the following: (iii) *Measured Quantities are Undisturbed.* We assume that the diagonal elements of the decoherence functional (10.5), which give the probabilities of the histories $[S_\alpha]$, are the same as if they were calculated with the operators $s_\alpha^k(t) \otimes I_r$ where $s_\alpha^k(t)$ are the alternatives for the subsystem evolved with its *own* Hamiltonian. Thus,

$$p([S_\alpha]) = tr\left[s_{\alpha_n}^n(t_n) \cdots s_{\alpha_1}^1(t_1) \rho_s s_{\alpha_1}^1(t_1) \cdots s_{\alpha_n}^n(t_n)\right] \quad . \qquad \text{(II.10.8)}$$

where $\rho_s$, the projection operators $\{s_\alpha^k(t)\}$, and the trace $tr$ refer to the Hilbert space $\mathcal{H}_s$. This is the assumption that the measurement interaction instantaneously reduces the off-diagonal elements of the subsystem's decoherence functional to zero while leaving the diagonal elements unchanged. The values of measured quantities are thus left undisturbed.

In idealized models of this kind, the fundamental formula is exact and the rule for assigning probabilities can be restated: *Probabilities can be assigned to histories that have been* **measured** *and the probability is* (10.8). This is the rule of the Copenhagen interpretations for assigning probabilities. Eq. (10.8) may be unfamiliar to those used to working with a state vector that evolves unitarily in between measurements and by reduction of the state vector at a measurement. In fact, it is a compact and efficient expression of these two forms of evolution as has been stressed by Groenewold (1952), Wigner (1963), Aharonov, Bergmann, and Lebovitz (1964), Unruh (1986), and Gell-Mann (1987) among others. I shall demonstrate this equivalence explicitly below but for the moment let us discuss the significance of the ideal measurement model.

The ideal measurement model shows how the Copenhagen rule for assigning probabilities fits into the more general post-Everett framework of quantum cosmology. The rule holds in the model because certain approximate features of some measurement situations have been idealized as exact. Specifically, these idealizations include the *exact* factorization of the initial density matrix $\rho$ [eq. (10.2)], the

---

* A slightly different idealization leading to the same result would be to assume that the correlations expressed in (10.2) are with projections $\{R_\alpha^{(k,\tau)}(t)\}$ that *always* exactly decohere because of the properties of the initial $\rho$ no matter where located in a string of projections in the decoherence functional. Such variables are typically described as "macroscopic". There is some economy in such a sweeping idealization but the model of persistent records suggests a mechanism by which such decoherence might be accomplished. (Cf. the discussion in Section II.6.1)

*exact* correlation between measured system and registering apparatus [eq. (10.2)], and the *exact* persistence and independence of measurement records [eq. (10.3)]. In practice none of these idealizations will be *exactly* true. There are many typical experimental situations involving measurements at a single time, however, where they are true to an excellent approximation. (See Section II.11 for some estimates of the degree of approximation)    The further idealization that measured quantities are undisturbed almost never holds for measurements of microscopic quantities but is typical for measurements of macroscopic ones. For experimental situations where the idealizations of measurement model are approximately true, the Copenhagen rule supplies an *approximation* for the probabilities of the fundamental formula. The fundamental formula, however, applies more generally and precisely, for example, to situations in the early universe where nothing like the idealizations of this measurement model may be appropriate.

I conclude this subsection by returning to the equivalence of eq.(10.8) with the usual picture of a unitarily evolving state vector reduced on measurement. To see the equivalence let us calculate the probability for a sequence of just two measurements at times $t_1$ and $t_2$ according to the usual story in the Heisenberg picture, given and an initial pure $\rho_s = |\psi><\psi|$ at time $t_0$. The state $|\psi>$ is constant from $t_0$ to $t_1$. The probability that the outcome of the first measurement is $\alpha_1$ is

$$p(\alpha_1) = < \psi|s^1_{\alpha_1}(t_1)|\psi > \ . \tag{II.10.9}$$

The normalized state after the measurement is reduced to

$$|\psi_{\alpha_1} > = \frac{s^1_{\alpha_1}(t_1)|\psi >}{\sqrt{< \psi|s^1_{\alpha_1}(t_1)|\psi >}} \ . \tag{II.10.10}$$

The probability of obtaining the result $\alpha_2$ on the next measurement given the result $\alpha_1$ on the first is

$$\begin{aligned} p(\alpha_2|\alpha_1) &= < \psi_{\alpha_1}|s^1_{\alpha_1}(t_2)|\psi_{\alpha_1} > \\ &= \frac{< \psi|s^1_{\alpha_1}(t_1)s^2_{\alpha_2}(t_2)s^1_{\alpha_1}(t_1)|\psi >}{< \psi|s^1_{\alpha_1}(t_1)|\psi >} \ . \end{aligned} \tag{II.10.11}$$

The *joint* probability for $\alpha_2$ followed by $\alpha_1$ is

$$p(\alpha_2, \alpha_1) = p(\alpha_2|\alpha_1)p(\alpha_1) \ . \tag{II.10.12}$$

so that using (10.9) and (10.11) we have

$$p(\alpha_2, \alpha_1) = < \psi|s^1_{\alpha_1}(t_1)s^2_{\alpha_2}(t_2)s^1_{\alpha_1}(t_1)|\psi > \ . \tag{II.10.13}$$

This is just the formula (10.8) for the Copenhagen probabilities for the special case of a history with two times and a pure initial density matrix $\rho_s$.

## II.11. Approximate Probabilities Again

The discussion of the ideal measurement model just given provides a convenient opportunity for reviewing in a more concrete context the notion of approximate probability introduced in Section II.1.

As discussed in Section II.5, given the initial condition $\rho$, it is possible to exhibit the sets of alternative histories of the universe that *exactly* decohere and for which the probability sum rules are exactly satisfied. Among these exactly decohering sets, there may be some that have the correlations of the ideal measurement model perhaps even with an "initial" effective density matrix of the product form (10.1). For these situations the assumptions and consequences of the model would be exact, and the probabilities with which it deals would exactly satisfy the probability sum rules. However, measurement situations of familiar kinds, in which quasiclassical variables participate as records, will not correspond to exact decoherence. Rather, the decoherence and the probabilities of the fundamental formula will be approximate. In interesting situations this approximation will be VERY, VERY good.

Consider, for example, a measurement situation in which one of the participants in the defining correlations is the center of mass position $X$ and momentum $P$ of a massive body (many atoms) such as the often discussed "pointer". Suppose further, that the decoherence of alternative histories is to be effected by the orthogonality or near orthogonality of states of the massive body with sufficiently differing position and momentum much as in the example described at the beginning of Section II.6. More precisely, the states of the body are to be concentrated on cells of phase space of size $\Delta X$ and $\Delta P$ consistent with the uncertainty principle. A wave function cannot be *exactly* concentrated on a phase space cell. If $\psi(X)$ vanishes outside a region of compact support, then

$$\phi(P) = \int dX \exp\left(-iPX/\hbar\right)\psi(X) \tag{II.11.1}$$

will be analytic in $P$ and cannot vanish except at isolated points. There are, therefore, no *exactly* orthogonal states corresponding to phase space cells. However, one can find *approximately* orthogonal states. For example, gaussian wave packets with minimum uncertainties ($\Delta X_{\min}$, $\Delta P_{\min}$) contained within the phase space cells would do the job.* Wave functions concentrated on different cells would at most have overlaps of order

$$\exp\left[-\frac{1}{4}\left(\Delta X/\Delta X_{\min}\right)^2\right] . \tag{II.11.2}$$

Thus, in such a measurement situation, decoherence would be approximate and the resulting probabilities approximately satisfy the sum rules up to a standard set roughly by (11.2).

For accuracies $\Delta X$ only modestly larger than $\Delta X_{\min}$ the violation of probability sum rules suggested by (11.2) will be very small. This, however, is not the whole story, for we know from the discussion in Section II.6 that in realistic situations the center of mass of the massive body will become coupled by collisions with a large number of photons, molecules, and similar variables. As the model at the start of that Section suggests, this coupling acts to *improve* the orthogonality of the states of combined system of massive body and decohering agents corresponding

---

* These have approximate classical evolution as well, (See, e.g. Hepp, 1974). For more on approximate projectors onto phase space cells see Omnès (1989) and the references therein

to different values of $X$ and $P$. Eq.(11.2) is, in effect, multiplied by the overlaps of the coupled many particle states of the form (6.5). Accepting uncritically the estimates of Joos and Zeh (1985), contained essentially in (6.8), one finds that after a milisecond the overlap factor for a pointer of linear dimension 1 cm interacting with molecules of air might be of order

$$10^{-10^{40}} \qquad\qquad (\text{II}.11.3)$$

where the words "of order" refer to the exponent of the exponent! This is a VERY small number. One cannot expect such estimates to be reliable in any usual sense. They suffice, however, to show that the decoherence will be very, very nearly exact in such circumstances and the probability sum rules very, very nearly satisfied.

One standard by which this accuracy might be measured is the probability that we have simply imagined our whole personal histories up until now. The author does not pretend to know how to estimate this in a sensible fashion but a naive guess might be $(p)^N$ where $p$ is an atomic tunneling probability per relevent atom and $N$ is the number of atoms involved. Simple guesses for $N$ and $p$ give numbers that are negligibly small but possibly larger than (11.3). Thus, when I speak of approximate decoherence and approximate probabilities in discussions of quasiclassical domains and measurement situations I mean, decoherence achieved and probability sum rules satisfied beyond any standard that might be conceivably contemplated for the accuracy of prediction and the comparison of theory with experiment.

A theory which uses approximate probabilities conforming to standards such as those described above can always be converted into a theory for which the probability sum rules are exact. Simply augment the theory by an *ad hoc* rule for renormalizing the probabilities of exhaustive sets of alternatives when the failure of the sum rules to be satisfied falls below a certain level. There surely exists some level below which the predictions of such a theory would be indistinguishable from one using approximate probabilities such as described above. At such a standard it seems simpler to employ approximate probabilities.

The small numbers which estimate the failure of decoherence among familiar quasiclassical operators show how excellent the approximation of exact decoherence is in the ideal measurement model of the Copenhagen approximation to quantum mechanics. Other features of the model such as the exact correlations and the exact persistance are likely to be much less accurate approximations. Similarly small numbers measure the minute failure of decoherence of alternative past positions of massive bodies (e.g. the sun) thus justifying in part the approximation of reconstructing their histories classically. Indeed, these numbers are so small that one might reasonably ask whether it is not possible to find sets of histories that are "nearly" the quasiclassical ones but which *exactly* decohere. The main issue in this question is whether operators defining such histories would have a description in terms of repeated quantities for the most part correlated in time according to classical laws of motion.

## II.12. Complex Adaptive Systems

Our picture is of a universe that, as a consequence of a particular initial condition and of the underlying Hamiltonian, exhibits at least one quasiclassical domain made up of sets of alternative histories with as much classicality as possible. The quasiclassical domains are a consequence of the theory and its boundary condition, not an artifact of our construction. How do we characterize our place as a collectivity of observers in the universe?

Both singly and collectively we are examples of the general class of complex adaptive systems. (See, e.g. Gell-Mann, 1990.) When they are considered as portions of the universe, making observations, we refer to such complex adaptive systems as information gathering and utilizing systems ($IGUS$es). From a quantum mechanical point of view the foremost characteristic of an $IGUS$ is that, in some form of approximation, however crude or classical, it employs the fundamental formula, with what amounts to a rudimentary theory of $\rho$, $\tilde{H}$, and quantum mechanics. Probabilities of interest for observations by the $IGUS$ include those for correlations between its memory and its external world up to the present. (Typically these are assumed perfect; not always such a good approximation!) The approximate fundamental formula is used to compute probabilities on the basis of present data, make predictions, control future perceptions on the basis of these predictions (i.e., exhibit behavior), acquire further data, make further predictions, and so on.

To carry on in this way, an $IGUS$ uses probabilities for histories referring both to the future and the past. Typically, an $IGUS$ performs further coarse graining on a quasiclassical domain. Naturally, its coarse graining is very much coarser than that of the quasiclassical domain and utilizes only a few of the variables in the universe.

The reason such systems as $IGUS$es exist, functioning in such a fashion, is to be sought in their evolution within the universe. It is reasonable to suppose that they evolved to make predictions because it is adaptive to do so. The reason, therefore, for their focus on decohering variables is that these are the *only* variables for which predictions can be made. The reason for their focus on the histories of a quasiclassical domain is that these present enough regularity over time to permit prediction by relatively rudimentary, easily evolved algorithms. The reason, specifically, that we do not see Mars spread out in a quantum superposition of different positions is that we have evolved to use, in perception, a coarse graining in which such superpositions rapidly decohere.

If there is essentially only one quasiclassical domain, then naturally $IGUS$es evolve to utilize further coarse grainings of it. If there are many essentially inequivalent quasiclassical domains, then there is an interesting question of which particular domain or set of such domains $IGUS$es evolve to exploit. However they evolve, $IGUS$es, including human beings, occupy no special place and play no preferred role in the laws of physics. They merely utilize the probabilities presented by quantum mechanics in the context of a quasiclassical domain of this universe.

Thus, the most fundamental, assumption free, way of "including the observer in the universe" is to see it as a system that has evolved within the universe. Understanding this evolutionary process would seem an intractable task were it not plausibly divisible into two parts: The first of these is understanding why this quantum mechanical universe exhibits one or more quasiclassical domains. The second part is understanding nuclear, chemical, and biological evolution in a universe that exhibits quasiclassical domains. Providing criteria for a quasiclassical domain is a way of dividing the problem into these two parts. The first of these may be tractable within physics.

## II.13. Open Questions

There are many open questions whose resolutions would help to complete, test, and affirm the view of quantum mechanics adumbrated in this section: The mechanisms of decoherence need to be explored quantitatively in increasingly realistic models especially in regard to the quasiclassical coarse grainings of Section II.7. Sets of alternative decohering histories need to be exhibited explicitly for model initial conditions and Hamiltonians. It is central to complete the definition of a quasiclassical domain by finding the general definition for classicality. Once that is accomplished, the question of how many and what kinds of essentially enequivalent quasiclassical domains follow from $\rho$ and $H$ is a topic for serious theoretical research. So is the question of what kinds of *IGUS*es can exist in the universe exploiting particular quasiclassical domains, or the unique one if there is only one.

Beyond these specific questions, resolution of the problems of interpretation presented by quantum mechanics seems best accomplished not by further intense scrutiny of the subject as it applies to reproducible laboratory situations, but rather through an examination of the origin of the universe and its subsequent history. Quantum mechanics is best and most fundamentally understood in the context of quantum cosmology. The founders of quantum mechanics were right in pointing out that something external to the framework of wave function and Schrödinger equation *is* needed to interpret the theory. But it is not a postulated classical domain to which quantum mechanics does not apply. Rather it is the initial condition of the universe that, together with the action function of the elementary particles and the throws of quantum dice since the beginning, is the likely origin of quasiclassical domain(s) within quantum theory itself.

## III. GENERALIZED QUANTUM MECHANICS

### III.1. General Features

What might we mean, most generally, by quantum mechanics for a closed system such as the universe as a whole? Roughly speaking, by a quantum mechanics we mean a theory that admits a notion of fine and coarse grained histories, the amplitudes for which are connected by the principle of superposition and for which there is a rule (decoherence) for when coarse-grained histories can be assigned probabilities obeying the standard sum rules of probability calculus. More precisely, from the discussion in the preceeding section it is possible to abstract the following three elements of quantum mechanics in general:

1) *Fine-Grained Histories:* These are the sets of fine-grained, exhaustive, alternative histories of the universe $\{f\}$ which are the most refined description to which one can contemplate assigning probabilities. There may be many such sets.

2) *Coarse Graining:* A coarse graining of an exhaustive set of histories is a *partition* of that set into exhaustive and exclusive classes $\{h\}$. The possible coarse-grained sets of alternative histories of the universe are all possible coarse grainings of fine-grained sets.

3) *Decoherence Functional:* The decoherence functional, $D(h, h')$, is defined on each pair of coarse-grained histories in an exhaustive set $\{h\}$. With it a probability $p(h)$ is assigned to individual members of a set of coarse grained histories that decohere according to the fundamental formula.

$$D(h, h') \approx \delta_{hh'} p(h) \ . \tag{III.1.1}$$

The decoherence functional $D$ on the fine-grained histories must satisfy the following properties:

i)   *Hermiticity:*
$$D(f, f') = D^*(f', f) \ , \tag{III.1.2}$$

ii)  *Positivity:*
$$D(f, f) \geq 0 \ , \tag{III.1.3}$$

iii) *Normalization:*
$$\sum_{f, f'} D(f, f') = 1 \ . \tag{III.1.4}$$

The decoherence functional a the coarse-grained set of alternative histories may then be *defined* by the principle of superposition:

iv)  *The principle of superposition:*
$$D(h, h') = \sum_{\substack{\text{all } f \\ \text{in } h}} \sum_{\substack{\text{all } f' \\ \text{in } h'}} D(f, f') \ . \tag{III.1.5}$$

This definition must be consistent; if a set of histories can arise by a coarse graining of two different fine-grained sets the same decoherence functional must result.

As a consequence of its definition, the decoherence functional $D(h, h')$ on coarse-grained sets of histories satisfies four analogous conditions. In particular if $\{h\}$ is an exhaustive set of alternative coarse-grained histories and $\{\bar{h}\}$ is a coarser grained partition of it we have

i)   *Hermiticity:*
$$D(h, h') = D^*(h', h) \ , \tag{III.1.2a}$$

ii)  *Positivity:*
$$D(h, h) \geq 0 \ , \tag{III.1.3a}$$

iii) *Normalization:*
$$\sum_{h, h'} D(h, h') = 1 \ , \tag{III.1.4a}$$

iv)  *The principle of superposition:*
$$D(\bar{h}, \bar{h}') = \sum_{\substack{\text{all } h \\ \text{in } \bar{h}}} \sum_{\substack{\text{all } h' \\ \text{in } \bar{h}'}} D(h, h') \ . \tag{III.1.5a}$$

These conditions are equivalent to the conditions (1.2)-(1.4) if the set of all coarse-grained sets of histories is taken to include the fine-grained sets.

As a consequence of these four conditions, the approximate probabilities for prediction and retrodiction defined by the fundamental formula (1.1) will obey the rules of probability theory to the standard that decoherence is enforced. By virtue of (i), (ii), and (iii) the probabilities $p(h)$ are real numbers lying between 0 and 1. By virtue of (iv) they satisfy the sum rules

$$p(\bar{h}) = \sum_{\substack{\text{all } h \\ \text{in } \bar{h}}} p(h) \tag{III.1.6}$$

for all coarse grainings.

These three elements — fine-grained histories, coarse graining, and decoherence — capture the essential features of quantum mechanical prediction for a closed system in a way that is general enough for the physical situations to be considered later. They do not, however, represent the most general formulation which could be given. For example, following the discussion at the end of Section II.2 a fundamental formula with $ReD$ replacing $D$ in (1.1) is sufficient for the probability sum rules.

Hamiltonian quantum mechanics implements the three elements in a specific way. However, I shall argue that the Hamiltonian framework is not the only way of implementing these elements of quantum mechanics and that alternative implementations that are, in this sense, generalizations of Hamiltonian quantum mechanics may be of interest in connection with quantum mechanical theories of spacetime. In the remainder of this section I shall show how some familiar formulations of quantum mechanics look from this generalized point of view.

### III.2. Hamiltonian Quantum Mechanics

First, we consider Hamiltonian quantum mechanics from this general perspective. In Hamiltonian quantum mechanics sets of histories are represented by time sequences of projections onto exhaustive sets of orthogonal subspaces of a Hilbert space. The three elements of a quantum mechanics are implemented as follows:

1) *Fine-Grained Histories:* These correspond to sequences of sets of projections onto a *complete* set of states, one set at every time. There are thus *many* different sets of fine-grained histories corresponding to the various possible complete sets of states at each and every time.

2) *Coarse Graining:* A set of histories is a coarse graining of a finer set if each projection in the coarser grained set is a sum of projections in the finer grained set. The projections constructed as sums define a partition of the histories in the finer grained set.

By way of example, consider the quantum mechanics of a particle and the completely fine-grained set of histories that consists of specifying the position at every moment of time, that is, specifying the particle's path in configuration space. A coarse graining consisting of projections onto an exhaustive set of ranges of position at, say, three different times defines a partition of the configuration space paths into those that pass through the various possible combinations of ranges at the different times.

3) *Decoherence Functional:* For Hamiltonian quantum mechanics this is (II.2.3). In the present notation $h$ stands for the history corresponding to a particular sequence of projections $[P_\alpha]$. Thus,

$$D(h, h') = Tr\left[P^n_{\alpha_n}(t_n)\cdots P^1_{\alpha_1}(t_1)\rho P^1_{\alpha'_1}(t_1)\cdots P^n_{\alpha'_n}(t_n)\right] , \qquad \text{(III.2.1)}$$

which is easily seen to satisfy properties (i)-(iv) above.

The structure of sets of alternative coarse-grained histories of Hamiltonian quantum mechanics is shown schematically in Fig. 4. The sets of coarse-grained histories form a partially ordered set defining a semi-lattice. For any pair of sets of histories, the least coarse grained set of which they are both fine grainings can be defined. However, there is not, in general, a unique most fine-grained set of which two sets are a coarse graining. There is an operation of "join" but not of "meet". The many possible fine-grained starting points in Hamiltonian quantum mechanics are a reflection of the democracy of transformation theory. No one basis is distinguished from any other.

### III.3. Sum-Over-Histories Quantum Mechanics for Theories with a Time.

The three elements of a sum-over-histories formulation of the quantum mechanics of a theory with a well defined physical time are as follows:

1) *Fine-Grained Histories:* The fine-grained histories are the possible paths in a configuration space of generalized coördinates $q^i$ expressed as *single-valued* functions of the physical time. Only one configuration is possible at each instant. Sum-over-histories quantum mechanics, therefore, starts from a *unique* fine-grained set of alternative histories of the universe in contrast to Hamiltonian quantum mechanics that starts from many.

2) *Coarse Graining:* There are many ways of partitioning the fine-grained paths into exhaustive and exclusive classes. However, the existence of a physical time allows an especially natural coarse graining because paths cross a constant time surface in the extended configuration space $(t, q^i)$ once and only once. Specifying an exhaustive set of regions $\{\Delta_\alpha\}$ of the $q^i$ at one time, therefore, partitions the paths into the class of those that pass through $\Delta_1$ at that time, the class of those that pass through $\Delta_2$ at that time, etc. More generally, different exhaustive sets of regions $\{\Delta_\alpha^k\}$ at times $t_k$, $k = 1$, $\cdots$, $n$ similarly define a partition of the fine-grained histories into exhaustive and exclusive classes. Other kinds of partitions can be contemplated (see e.g. Hartle, 1988a) but these will suffice for our later discussions.

3) *Decoherence Functional:* The decoherence functional for sum-over-histories quantum mechanics for theories with a time is

$$D(h, h') = \int_h \delta q \int_{h'} \delta q' \delta(q_f^i - q_f'^i) \ \exp\left\{ i \left( S[q^i(t)] - S[q'^i(t)] \right) / \hbar \right\} \rho(q_0^i, q_0'^i) \ .$$
(III.3.1)

Here, we consider an interval of time from an initial instant $t_0$ to some final time $t_f$. The first integral is over paths $q^i(t)$ that begin at $q_0^i$, end at $q_f^i$, and lie in the partition $h$. The integral includes an integration over $q_0^i$ and $q_f^i$. The second integral over paths $q'^i(t)$ is similarly defined. If $\rho(q^i, q'^i)$ is a density matrix, then it is easy to verify that $D$ defined by (3.1) satisfies conditions (i)-(iv) of Section III.1. When the coarse graining is defined by sets of configuration space regions $\{\Delta_\alpha^k\}$ as discussed above, then (3.1) coincides with the sum-over-histories decoherence functional previously introduced in (II.2.2). However, more general partitions are possible.

The density matrix $\rho(q^i, q'^i)$. may be thought of as defining the initial condition of the closed system under consideration. Some initial conditions may be specified simply and elegantly as conditions on the class of fine-grained histories. For example,

$$\rho(q^i, q'^i) = \delta(q^i - Q^i)\delta(q'^i - Q^i)$$
(III.3.2)

corresponds to the condition that paths $q^i(t)$ and $q'^i(t)$ both begin at the particular configuration space point $Q^i$ at $t = t_0$. An initial condition that $\rho$ represents a pure momentum eigenstate

$$\rho(q^i, q'^i) = (2\pi\hbar)^{-n} \exp\left[ iP_i(q^i - q'^i)/\hbar \right]$$
(III.3.3)

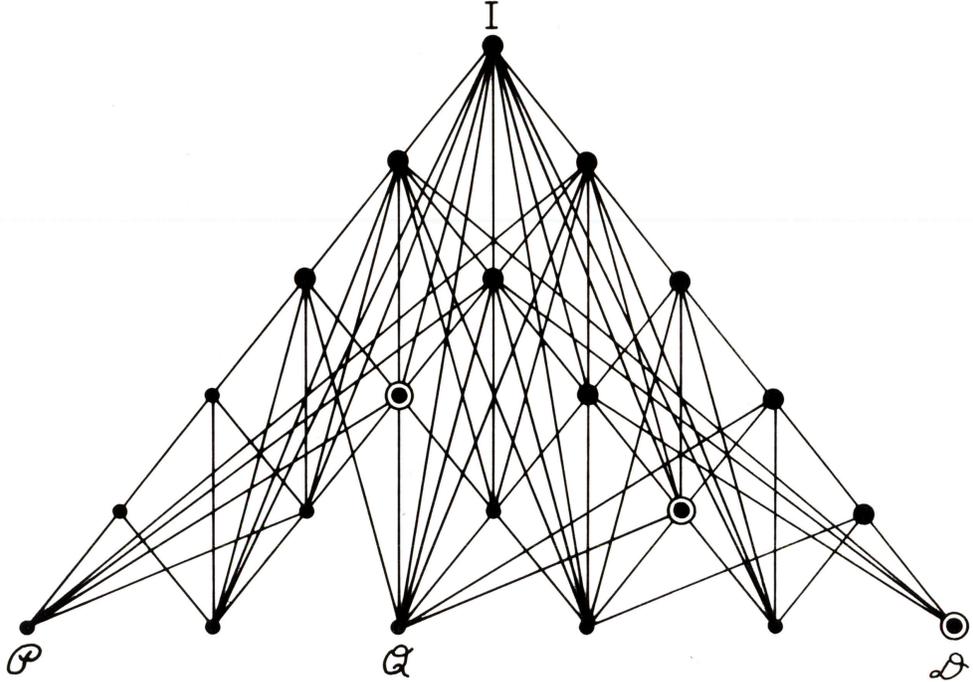

**Fig. 4:** *The schematic structure of the space of sets of possible histories in Hamiltonian quantum mechanics. Each dot in this diagram represents an exhaustive set of alternative histories for the universe. (This is not a picture of the branches defined by a given set!) Such sets, denoted by $\{[P_\alpha]\}$ in the text, correspond in the Heisenberg picture to time sequences $(P^1_{\alpha_1}(t_1),\ P^2_{\alpha_2}(t_2),\ \cdots\ P^n_{\alpha_n}(t_n))$ of sets of projection operators, such that at each time $t_k$ the alternatives $\alpha_k$ are an orthogonal and exhaustive set of possibilities for the universe. At the bottom of the diagram are the completely fine-grained sets of histories each arising from taking projections onto eigenstates of a complete set of observables for the universe at every time. For example, the set $\mathcal{Q}$ is the set in which all field variables at all points of space are specified at every time. This set is the starting point for Feynman's sum-over-histories formulation of quantum mechanics. $\mathcal{P}$ might be the completely fine-grained set in which all field momenta are specified at each time. $\mathcal{D}$ might be a degenerate set in which the same complete set of operators occurs at every time. But there are many other completely fine-grained sets of histories corresponding to all possible combinations of complete sets of observables that can be taken at every time.*

*The dots above the bottom row are coarse-grained sets of alternative histories. If two dots are connected by a path, the one above is a coarse graining of the one below — that is, the projections in the set above are sums of those in the set below. A line, therefore, corresponds to an operation of coarse graining. At the very top is the degenerate case in which complete sums are taken at every time, yielding no projections at all other than the unit operator! The space of sets of alternative histories is thus partially ordered by the operation of coarse graining.*

*The heavy dots denote the decoherent sets of alternative histories. Coarse grainings of decoherent sets remain decoherent.*

can be approximated through a condition on paths that defines momentum by time of flight. (See e.g. Feynman and Hibbs, 1965.)

When initial and final conditions are expressed as conditions on the fine-grained paths, $\mathcal{C}$, we may write compactly

$$D(h, h') = \int_{h, \mathcal{C}} \delta q \int_{h', \mathcal{C}} \delta q' e^{i\left(S[q^i(t)] - S[q'^i(t)]\right)/\hbar} \ , \qquad \text{(III.3.4)}$$

where the sum is over paths $q^i(t)$, $q'^i(t)$ meeting the initial condition, the final condition that their endpoints coincide, and lying in the partitions $h$ and $h'$ respectively.

The structure of the collection of sets of coarse-grained histories in sum-over-histories quantum mechanics is illustrated in Fig. 5. Because there is a *unique* fine grained set of histories, many fewer coarse grainings are possible in a sum-over-histories formulation than in a Hamiltonian one, and the space of sets of coarse-grained histories is a lattice rather than a semi-lattice.

### III.4. Differences and Equivalences between Hamiltonian and Sum-Over-Histories Quantum Mechanics for Theories with a Time.

From the perspective of generalized quantum theory the sum-over-histories quantum mechanics of Section III.3 is different from the Hamiltonian quantum mechanics of Section III.2. Even when the action of the former gives rise to the Hamiltonian of the latter, the two formulations differ in their notions of fine-grained histories, coarse graining and in the resulting space of coarse-grained sets of histories as Figs 4 and 5 clearly show. Yet, as is well known, the sum-over-histories formulation and the Hamiltonian formulation are equivalent for those particular coarse grainings in which the histories are partitioned according to exhaustive sets of configuration space regions, $\{\Delta_\alpha^k\}$, at various times $t_k$. More precisely the sum-over-histories expression for the decoherence functional, (3.1), is *equal* to the Hamiltonian expression, (2.1), when the latter is evaluated with projections onto the ranges of coördinates that occur in the former. Crucial to this equivalence, however, is the existence of a well defined physical time in which the paths are single-valued. To see this let us review the derivation of the equivalence in more detail.

When restricted to projections $P_{\Delta_\alpha}$ onto ranges $\Delta_\alpha$ of the $q$'s, the trace in (2.1) may be expanded as follows:

$$Tr\left[P_{\Delta_n}(t_n) \cdots P_{\Delta_1}(t_1) \rho P_{\Delta'_1}(t_1) \cdots P_{\Delta'_n}(t_n)\right]$$

$$= \int dq_f \int dq_0 \int dq'_0 < q_f t_f | P_{\Delta'_n}(t_n) \cdots P_{\Delta'_1}(t_1) | q_0 t_0 >$$

$$\times < q_0 t_0 | \rho | q'_0 t'_0 > < q'_0 t'_0 | P_{\Delta'_1}(t_1) \cdots P_{\Delta'_n}(t_n) | q_f t_f > \ . \qquad \text{(III.4.1)}$$

Here, to keep the notation manageable, the index "$i$" on the $q$'s has been suppressed, the index $k$ on $\Delta_\alpha^k$ has been suppressed, $|qt>$ has been written for the Heisenberg state that is an eigenvector of $q^i(t)$ with eigenvalue $q^i$, and $dq$ for the volume element on the space spanned by the $q^i$. We now demonstrate the following identity

$$< q_f t_f | P_{\Delta_n}(t_n) \cdots P_{\Delta_1}(t_1) | q_0 t_0 > = \int_{[q_0 \Delta_1 \cdots \Delta_n q_f]} \delta q(t) e^{iS[q^i(t)]/\hbar} \ , \qquad \text{(III.4.2)}$$

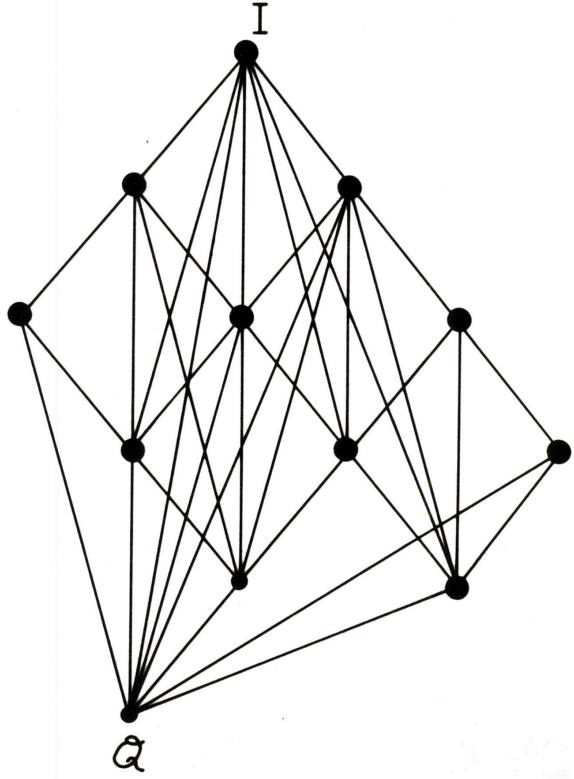

**Fig. 5.** *The schematic structure of the space of sets of histories in sum-over-histories quantum mechanics. The completely fine-grained histories arise from a single complete set of observables, say the set $\mathcal{Q}$ of field variables $q^i$ at each point in space and every time. The possible coarse-grained histories will then be a subset of those of Hamiltonian quantum mechanics illustrated in Fig. 4.*

where the sum is over all paths that begin at $q_0$ at $t_0$ pass through $\Delta_1, \cdots, \Delta_n$ at $t_1, \cdots, t_n$ respectively and end at $q_f$ at time $t_f$. To see how the argument goes, consider just one interval $\Delta_k$ at time $t_k$. The matrix element on the left of (4.2) may be further expanded as

$$< q_f t_f | P_{\Delta_k}(t_k) | q_0 t_0 > = \int_{\Delta_k} dq_k < q_f t_f | q_k t_k > < q_k t_k | q_0 t_0 > \quad . \qquad (\text{III.4.3})$$

Since the paths cross the surface of time $t_k$ at a single point $q_k$, the sum on the right of (4.2) may be factored as shown in Fig. 6,

$$\int_{[q_0 \Delta_k q_f]} \delta q \; e^{iS[q^i(t)]/\hbar} = \int_{\Delta_k} dq_k \left( \int_{[q_k q_f]} \delta q e^{iS[q^i(t)]/\hbar} \right) \left( \int_{[q_0 q_k]} \delta q e^{iS[q^i(t)]/\hbar} \right) \quad . \qquad (\text{III.4.4})$$

But, it is an elementary calculation to verify that

$$< q'' t'' | q' t' > = \int_{[q' q'']} \delta q e^{iS[q^i(t)]/\hbar} \qquad (\text{III.4.5})$$

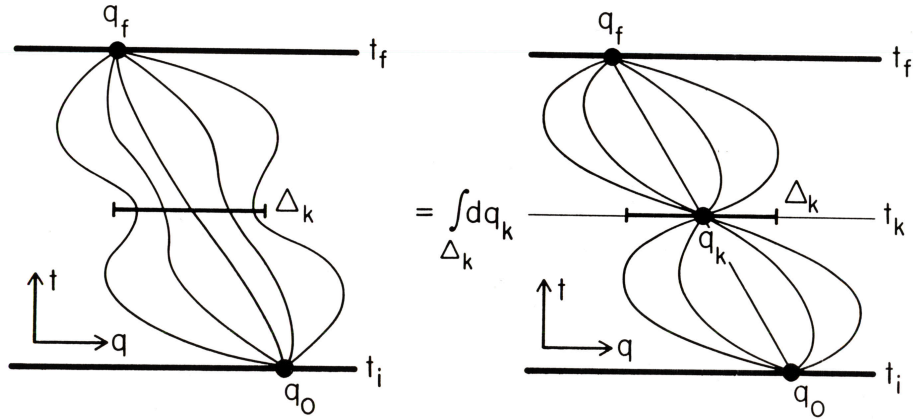

**Fig. 6:** *Factoring a sum over paths single-valued in time across a surface of constant time. Shown at left is the sum over paths defining the amplitude to start from $q_0$ at time $t_0$, proceed through interval $\Delta_k$ at time $t_k$, and wind up at $q_f$ at time $t_f$. If the histories are such that each path intersects each surface of constant time once and only once, then the sum on the left can be factored as indicated at right. The factored sum consists of a sum over paths before time $t_k$, a sum over paths after time $t_k$, followed by a sum over the values of $q_k$ at time $t_k$ inside the interval $\Delta_k$. The possibility of this factorization is what allows the Hamiltonian form of quantum mechanics to be recovered from a sum-over-histories formulation. The sum over paths before and after $t_k$ define wave functions on that time-slice and the integration over $q_k$ defines their inner product. The notion of state at a moment of time and the Hilbert space of such states is thus recovered.*

*If the sum on the left were over paths that were multiple valued in time, the factorization on the right would not be possible.*

and that inverting the time order on the right is the same as complex conjugation. Thus (4.3) is true and, by extension, also the equality (4.1).

The equivalence of the Hamiltonian and the sum-over-histories formulations of quantum mechanics on position coarse grainings is thus seen to be a consequence of the existence of surfaces in the extended configuration space $(t, q^\alpha)$ that the histories cross once and only once. We shall soon discuss cases where there are no such surfaces and no associated time.

Despite their equivalence on certain coarse-grained sets of alternative histories, Hamiltonian quantum mechanics and sum-over-histories quantum mechanics are different because their underlying sets of fine-grained histories are different. Are the more limited coarse grainings of sum-over-histories quantum mechanics adequate for physics? They are if all testable statements can be reduced to statements about configuration space variables — positions, fields of integer and half-interger spin, etc. Certainly this would seem sufficient to describe the coarse graining associated with any classical domain.

III.5. Classical Physics and the Classical Limit of Quantum Mechanics.

In a trivial way classical physics may be regarded as a generalized quantum mechanics. The three elements are:

1) *Fine-grained histories:* The fine-grained histories are paths in phase space, $(p_i(t), q^i(t))$, parametrized by the physical time.

2) *Coarse graining:* The most familiar type of coarse graining is specified by cells in phase space at discrete sequences of time. The paths are partitioned into classes defined by which cells they pass through.

3) *Decoherence Functional:* From the perspective of quantum theory, the distinctive features of classical physics are that the fine-grained histories are *exactly* decoherent and exactly correlated in time according to classical dynamical laws. A decoherence functional that captures these features may be constructed as follows: Let $z^i = (p_i, q^i)$ serve as a compact notation for a point in phase space. $z^i(t)$ is a phase space path. Let $z^i_{cl}(t; z^i_0)$ denote the path that is the classical evolution of the initial condition $z^i_0$ at time $t_0$. The path $z^i_{cl}(t) = (p^{cl}_i(t), q^i_{cl}(t))$ satisfies

$$\dot{p}^{cl}_i = -\frac{\partial H}{\partial q^i_{cl}} \quad , \quad \dot{q}^i_{cl} = \frac{\partial H}{\partial p^{cl}_i} \; , \tag{III.5.1}$$

where $H$ is the classical Hamiltonian, and satisfies the initial condition $z^i(t_0; z^i_0) = z^i_0$. Define a classical decoherence functional, $D_{cl}$, on pairs of fine-grained histories as

$$D_{cl}[z^i(t), \; z'^i(t)] = \delta[z^i(t) - z'^i(t)] \int d\mu(z^i_0)\delta[z^i(t) - z^i_{cl}(t; z_0)]f(z^i_0) \; . \tag{III.5.2}$$

Here $\delta[\cdot]$ denotes a functional $\delta$-function on the space of phase space paths, and $d\mu(z^i)$ is the usual Liouville measure, $\Pi[dp_i \; dq^i/(2\pi\hbar)]$. The function $f(z^i_0)$ is a real, positive normalized distribution function on phase space which gives the initial condition of the closed classical system. The first $\delta$-function in (5.2) enforces the exact decoherence of classical histories; the second guarantees correlation in time according to classical laws.

A coarse graining of the set of alternative fine-grained histories may be defined by giving exhaustive partitions of phase space into regions $\{R^k_\alpha\}$ at a sequence of times $t_k$, $k = 1, \cdots, n$. Here, $\alpha$ labels the region and $k$ the partition. As in Section II, we denote one history in the coarse-grained set corresponding to a particular sequence $\alpha_1, \cdots, \alpha_n$ by $[R_\alpha]$. The decoherence functional for the set of coarse-grained alternative classical histories is

$$D_{cl}\Big([R_\alpha], [R_{\alpha'}]\Big) = \int_{[R_\alpha]} \delta z \int_{[R_{\alpha'}]} \delta z' D_{cl}[z^i(t), z'^i(t)] \; , \tag{III.5.3}$$

where the integral is over pairs of phase space paths restricted by the appropriate regions and the integrand is (5.2). Easily one has

$$D_{cl}\Big([R_\alpha], [R_{\alpha'}]\Big) = \delta_{\alpha_1\alpha'_1} \cdots \delta_{\alpha_n\alpha'_n} \; p_{cl}(\alpha_1, \cdots, \alpha_n) \; , \tag{III.5.4}$$

where $p_{cl}(\alpha_1, \cdots, \alpha_n)$ is the classical probability to find the system in the sequence of phase space regions $[R_\alpha]$ given that it is initially distributed according to $f(z^i_0)$.

It is then also easy to see that (5.3) and (5.1) satisfy the conditions (i)-(iv) of Section III.1 for decoherence functionals.

In certain situations the decoherence functional of a quantum mechanics may be well approximated by a classical decoherence functional of the form (5.2). For example, in Hamiltonian quantum mechanics it may happen that for some coarse grained set of alternative histories $\{[P_\alpha]\}$

$$
\begin{aligned}
D\Big([P_\alpha], [P_{\alpha'}]\Big) &= Tr\left[P^n_{\alpha_n}(t_n)\cdots P^1_{\alpha_1}(t_1)\rho P^1_{\alpha'_1}(t_1)\cdots P^n_{\alpha'_n}(t_n)\right] \\
&\simeq D_{cl}\Big([R_\alpha], [R_{\alpha'}]\Big) \, ,
\end{aligned}
\tag{III.5.5}
$$

for some corresponding coarse graining of phase space $\{[R_\alpha]\}$ and distribution function $f$. One has then exhibited the classical limit of quantum mechanics.

Some coarse graining is needed for a relation like (5.5) to hold because otherwise the histories would not decohere. Moreover, a relation like (5.5) cannot be expected to hold for *every* coarse graining. Roughly, we expect that the projections $\{P_\alpha\}$ must correspond to phase space regions, for example, by projecting onto sufficiently crude intervals of configuration space and momentum space or onto coherent states corresponding to regions of phase space. (See, e.g. Hepp 1974, Omnès, 1989 for more on this.) Moreover, for a fixed coarse graining, a relation like (5.5) cannot not hold for every initial condition $\rho$. Only for particular coarse grainings and particular $\rho$ do we recover the classical limit of a quantum mechanics in the sense of (5.5)

III.6. Generalizations of Hamiltonian Quantum Mechanics.

As the preceeding example of classical physics illustrates, there are many examples of generalized quantum mechanics that do not coincide with Hamiltonian quantum mechanics. The requirements for a generalized quantum mechanics are weak. Fine-grained histories, a notion of coarse graining, and a decoherence functional are all that is needed. There are probably many such constructions.* It is thus important to search for further physical principles with which to winnow these possibilities. In this search there is also the scope to investigate whether the familiar Hamiltonian formulation of quantum mechanics might not itself be an approximation to some more general theoretical framework for quantum cosmology valid only for certain coarse grainings and particular initial conditions. If $D$ were the deoherence functional of the generalization then

$$
D(h, h') \simeq Tr\left[P^n_{\alpha_n}(t_n)\cdots P^1_{\alpha_1}(t_1)\rho P^1_{\alpha'_1}(t_1)\cdots P^n_{\alpha'_n}(t_n)\right]
\tag{III.6.1}
$$

only for certain $\{h\}$'s and corresponding strings of $P$'s and for a limited class of $\rho$'s. Thus, in cosmology it is possible to investigate which features of Hamiltonian quantum mechanics are fundamental and which are "excess baggage" that only appear to be fundamental because of our position late in a particular universe able to employ only limited coarse grainings.† In the next sections I shall argue that one such feature is the preferred time of Hamiltonian quantum mechanics.

---

* As further examples, one can imagine decoherence functionals based on purely Euclidean sums-over-histories (although then there is no decoherence and no probabilities) or decoherence functionals describing certain linear alternatives to quantum mechanics (e.g. Pearle, 1989).

## IV. TIME IN QUANTUM MECHANICS

Time plays a special role in the quantum mechanics of cosmology set forth in Section II. Every projection in a history was assumed to be characterized by *some time t*. It was possible to define exhaustive sets of alternatives for the universe *at one time*. The string of projections defining a history was *time ordered* in the fundamental formula. As a consequence the future was treated differently from the past in the predictive formalism and there was a quantum mechanical *arrow of time*.

In this section we shall ask whether such special roles for time in quantum mechanics might not be equally well seen as special features of our *particular* universe in generalized quantum mechanical frameworks for prediction in which such roles are not so singled out. In answering this question affirmatively we shall illustrate, in simple, take it or leave it, ways, routes towards generalization that will become essential for our discussion of quantum spacetime below. At the same time we shall illustrate how certain notions, in particular the notion of "state of the system at a moment of time", are inextricably linked to the preferred time in quantum mechanics.

### IV.1. Observables on Spacetime Regions

The measurable quantities in a field theory are not the values of a field at a spacetime point, $\phi(x)$. Rather they are the averages of fields over spacetime regions of the form

$$\phi(R) = \frac{1}{V(R)} \int_R d^4x \, \phi(x) \qquad (IV.1.1)$$

where $V(R)$ is the volume of the region $R$. A region of negligible temporal extent approximates a spatial field average "at one moment of time". However, in a different Lorentz frame such regions will have extension in time. It would be possible to restrict attention to the Lorentz invariant class of regions which have negligible extent in *some* timelike direction, but the natural Lorentz invariant class of observables for field theory consists of field operators averaged over general spacetime regions with extent in both space and time. These are the variables that occur in discussions of field measurements (Bohr and Rosenfeld 1933, 1950, DeWitt 1962). The projection operators onto ranges of values of such average fields are easily constructed from the operators (1.1). But what times should be assigned to sequences of such projection operators to construct the Hamiltonian decoherence functional for such coarse-grained histories according to (II.2.3)? There is no natural answer and therefore there is no natural Hamiltonian quantum mechanics. However, a generalized quantum mechanics for these observables can be constructed* for coarse-grained histories consisting of field averages over spacetime regions that are

---

† For more along these lines see Hartle (1990b).
* The author owes the essential ideas of this extension to R. Sorkin.

*causally consistent* in the following sense: The future of a spacetime region $R$ is the union of the future light cones and their interiors for each point in $R$. The past of $R$ is similarly defined. Two regions, $R''$ and $R'$, are said to be causally consistent if neither region intersects *both* the future and the past of the other. Thus, there is one member of the pair such that every point of it lies to the future or is spacelike separated from the other. A set of regions is said to be causally consistent if each pair is causally consistent.

A causally consistent set of regions can be partially time ordered. A region $R''$ lies to the future of $R'$ if there are some points in $R''$ that are the future of $R'$. Two regions that are entirely spacelike to each other are therefore not ordered and for this reason there is only a *partial* time ordering of a causally consistent set of spacetime regions.

With these definitions the three ingredients of a generalized quantum mechanics for observables defined from field averages over a set of causally consistent spacetime regions are as follows:

1) *Fine-Grained Histories:* The fine-grained histories are the four-dimensional field configurations, $\phi(x)$.

2) *Coarse Graining:* Coarse grainings are defined by specifying ranges of products of field averages defined on a causally consistent set of spacetime regions $R_i, i = 1, \cdots, n$. We denote the projections onto an exhaustive and exclusive set of ranges of a quantity constructed from field operators of the region $R_i$ by $P_\alpha^k(R_i)$ where $k$ denotes the particular quantity considered. A history consists of a sequence $[P_\alpha] = \left( P_{\alpha_1}^1(R_1), \cdots, P_{\alpha_n}^n(R_n) \right)$ of such "yes-no" alternatives. An exhaustive set of histories is obtained by allowing the $\alpha$'s to run over an exhaustive set of ranges.

Combining smaller ranges into bigger ones is one kind of coarse graining. Forming new sets of regions by taking unions of old preserving causal consistency is another example of coarse graining. In both cases the projections defining the coarser graining are sums of the projections defining the finer grained sets of histories.

3) *Decoherence Functional:* The decoherence functional of a set of coarse-grained histories is defined by

$$D\Big([P_\alpha], [P_{\alpha'}]\Big) = Tr\left[ P_{\alpha_n}^n(R_n) \cdots P_{\alpha_1}^1(R_1) \rho P_{\alpha_1'}^1(R_1) \cdots P_{\alpha_n'}^n(R_n) \right] , \qquad (IV.1.2)$$

where the projection operators are ordered according to the *partial* time order of the regions $R_i$. If two regions are not time ordered, then they are spacelike separated, the projections associated with them commute, and the value of (1.2) is unaffected by their order in the expression.

These three ingredients define a quantum mechanics with a weaker notion of time ordering than in the familiar Hamiltonian framework. The generalization coincides with Hamiltonian quantum mechanics when the regions have a negligible time extension in a particular Lorentz frame* but generalizes it to allow the

---

* An intermediate generalization could be based on a Lorentz invariant coarse graining that consisted entirely of averages over spatial regions with negligible extent in time provided all possible such regions were allowed and not just those associated with the constant time surfaces of a particular Lorentz frame. For a given set of such regions a family of non-intersecting spacelike hypersurfaces can be found such that each region lies in some spacelike surface. An equivalent Hamiltonian quantum mechanics can be formulated with a notion of state on these spacelike surfaces.

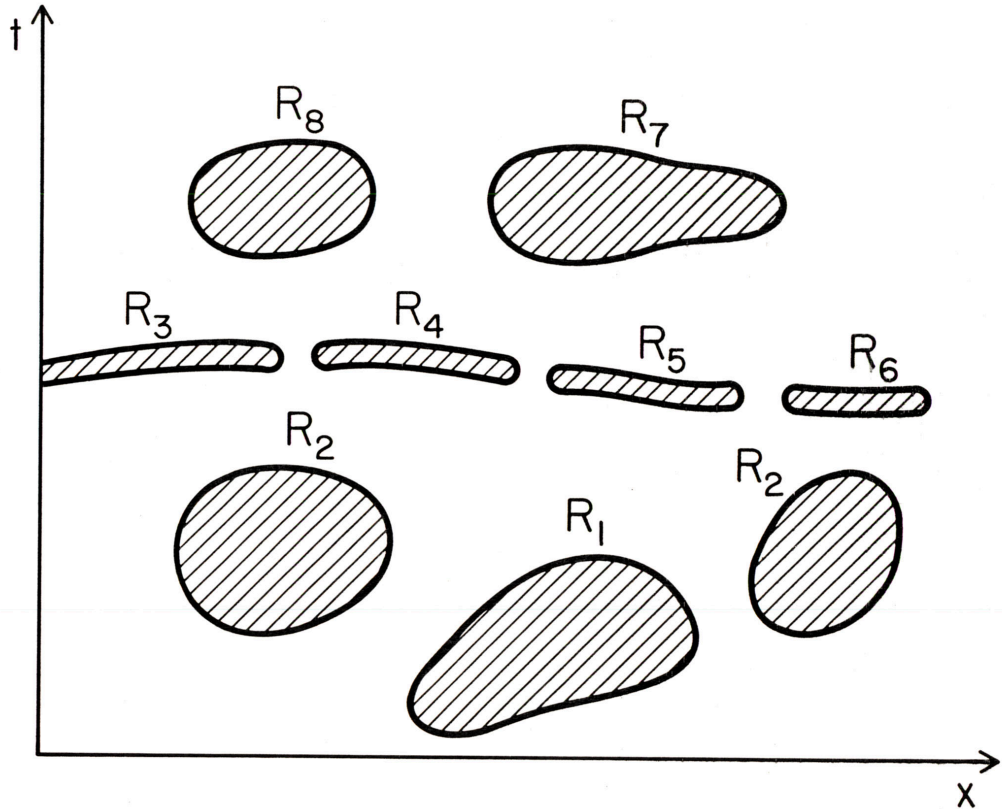

**Fig. 7:** *Causally consistent spacetime regions and their partial time order. A pair of spacetime regions is causally consistent if neither region contains both some points that are to the future and some points that are in the past of the other region. Put differently, there is one member of the pair such that every point in it lies to the future, or is spacelike separated from, every point in the other. A set of regions, such as the eight shown in this figure, are causally consistent if every pair is causally consistent.*

*Causally consistent regions can be partially time ordered. Region R is later than region $R'$ ($R > R'$) if there are points in R which are in the future of $R'$. In the figure, for example, $R_2 > R_1$, $R_8 > R_1$, $R_7 > R_4$, etc. ($R_2$ consists of two disconnected pieces) Completely spacelike separated regions are not time ordered as, for example, the pairs $(R_7, R_8)$, $(R_3, R_4)$, $(R_3, R_5)$, etc.*

*A generalized quantum mechanics can be constructed for coarse grainings of the completely fine-grained field history that consist of specifying ranges of values of field averages over causally consistent spacetime regions. The decoherence functional is given by the usual Heisenberg trace formula in which the operators respect the partial time order.*

discussion of alternatives that are not simply "at one moment of time". There is, however, a price for this generalization. It is no longer possible to follow (II.3.4) to construct an effective density matrix which gives an unambiguous notion of the state of the system on any spacelike surface that intersects one of the regions $R$.

Neither, therefore, is it possible to naturally describe evolution as a combination of unitarily evolving state vector and "collapse of the wave packet" for such times. It is not that it is impossible to define the notion of a unitarily evolving state vector. It is the "second law of evolution", the "reduction of the wave packet", that fails to hold in this generalized quantum mechanics when there are regions that are extended in both space and time. The Schrödinger picture does not exist for this generalized quantum mechanics. Despite this, the probabilities can be assigned to histories by the fundamental formula and the theory is predictive.

### IV.2. The Arrow of Time in Quantum Mechanics

There is an arrow of time in the quantum mechanics of Section II. The formalism treated the future differently from the past. Recall, for example, that we were able to predict the future from a state describing the universe at the present moment, but we could not retrodict the past. Mathematically such aspects of the arrow of time are consequences of the time ordering of the operators in the fundamental formula

$$D\left([P_\alpha],[P_{\alpha'}]\right) = Tr\left[P_{\alpha_n}^n(t_n)\cdots P_{\alpha_1}^1(t_1)\rho P_{\alpha_1'}^1(t_1)\cdots P_{\alpha_n'}^n(t_n)\right] , \qquad \text{(IV.2.1)}$$

$t_1 \le t_2 \le t_3 \le \cdots \le t_n$! This time ordering does not mean that quantum mechanics singles out an *absolute* direction in the parameter $t$. Field theory is TCP invariant. The TCP transformed projections

$$\tilde{P}_\alpha^k(-t) = (TCP)P_\alpha^k(t)(TCP)^{-1} \qquad \text{(IV.2.2)}$$

still evolve according to the Hamiltonian $H$. Thus (2.1) could equally well be written

$$D\left([P_\alpha],[P_{\alpha'}]\right) = Tr\left[\tilde{P}_{\alpha_n}^n(-t_n)\cdots \tilde{P}_{\alpha_1}^1(-t_1)\tilde{\rho}\ \tilde{P}_{\alpha_1'}^1(-t_1)\cdots \tilde{P}_{\alpha_n'}^n(-t_n)\right] \qquad \text{(IV.2.3)}$$

where $\tilde{\rho}$ is the TCP transformed $\rho$. In (2.3) the operators are anti-time-ordered. Either time ordering can, therefore, be used; the important thing is that on *one end* of the strings of $P$'s in (2.2) there is a knowable Heisenberg $\rho$ while at the other there is nothing. It is by convention that we call the end with the $\rho$ — the end that we know — the "past" and refer to an "initial" condition. It is a convention, however, that I shall stick to in the remainder of the lectures.

The future then is treated differently from the past in the quantum mechanics of Section II. A time asymmetry is built in. Of course, empirically the future *is* different from the past.* We know something of the past; we are ignorant of the future. But should this observed asymmetry be built in? Should it not be possible to consider quantum mechanically universes with less asymmetry between the future and the past? I will now argue that a slight generalization of the quantum mechanics of Section II would enable us to do so.

Imagine that we are members of a very advanced civilization interested in testing quantum cosmology in the laboratory. We learn how to isolate very large sections of the universe, thousands of Mpc on a side, and fix their quantum states at an initial moment of time at will. We do this for an ensemble of such systems selecting the initial states randomly. Allowing the systems to evolve to a later time we check on whether the system is in a particular state $|\Psi>$. If the answer is yes, we retain it in the ensemble; if not, we discard it.

---

* See, for example, the discussion in Penrose (1979).

What we have done, in effect, is to create an ensemble of "universes" of the type shown in Fig. 7 in which the future state is determined but the past state is random. Suppose we had investigated along the way whether the members of the ensemble contained observers and, if sufficiently advanced, what kind of quantum cosmology they would have induced. I do not know how to do the calculation that would predict what should be seen; although in principle it should be possible to do. One imagines that the prediction would be that most observers inside the region insulated from our initial conditions, would have induced a quantum mechanics of the kind we have been describing, asymmetric in time, but with a "past" and "future" that are oppositely ordered to ours. The observers would be "living backward in time".

Next, consider the case where we have not chosen the initial state of the ensemble randomly but according to some definite rule represented in quantum mechanics by a density matrix $\rho_f$. The probabilities of histories are, of course, easy to calculate for us. The joint probabilities for a history, $[P_\alpha]$, and the final state, $|\Psi>$, would arise from the decoherence functional

$$D\left([P_\alpha], [P_{\alpha'}]\right) = Tr\left[|\Psi><\Psi|P_{\alpha_n}^n(t_n)\cdots P_{\alpha_1}^1(t_1)\rho_f P_{\alpha_1'}^1(t_1)\cdots P_{\alpha_n'}^n(t_n)\right],$$
$$(IV.2.4)$$

which includes the projection selecting the final state. Probabilities for histories conditioned on this final state are these joint probabilities divided by a suitable normalization [cf. (II.3.3)]. One imagines that observers evolving in such a universe would have induced a correct formula for the decoherence functional giving these conditional probabilities. If they used conventions like ours they would have written it

$$D\left([P_\alpha], [P_{\alpha'}]\right) = Tr\left[\rho_f \hat{P}_{\alpha_n}^n(\tau_n)\cdots \hat{P}_{\alpha_1}^1(\tau_1)\rho_i \hat{P}_{\alpha_1'}^1(\tau_1)\cdots \hat{P}_{\alpha_n'}^n(\tau_n)\right]/Tr(\rho_i\rho_f)$$
$$(IV.2.5)$$

where $\rho_i = |\Psi><\Psi|$, the operators $\hat{P}_{\alpha_n}^n$ correspond exactly to the $P_{\alpha_n}^k$ when restricted to the spacetime region under investigation, but with times $\tau_n$ and labels $k$ ordered *back* from our future. In such a universe there would be both *initial* and *final* conditions. One could know something of both past and future. There would not be causality as we know it.

How do we know that we do not live in such a universe? The answer, I believe, is not determined *à priori*. Its an empirical question whether we can find out something about the future and the framework of quantum mechanics should be big enough to handle it if we can. The above framework is. It is a generalized quantum mechanics in the sense Section III.1. The fine-grained histories and notion of coarse grainings are the same as in Hamiltonian quantum mechanics. Only the decoherence functional is different; both initial *and* final conditions are possible. This quantum mechanics is applicable to our universe. The final condition that best fits our data is complete ignorance.

As was shown by Aharonov, Bergman, and Lebovitz (1964) over twenty years ago and as more recently discussed by Griffiths (1983) there is no arrow of time in this generalization of quantum mechanics. Because of the cyclic property of

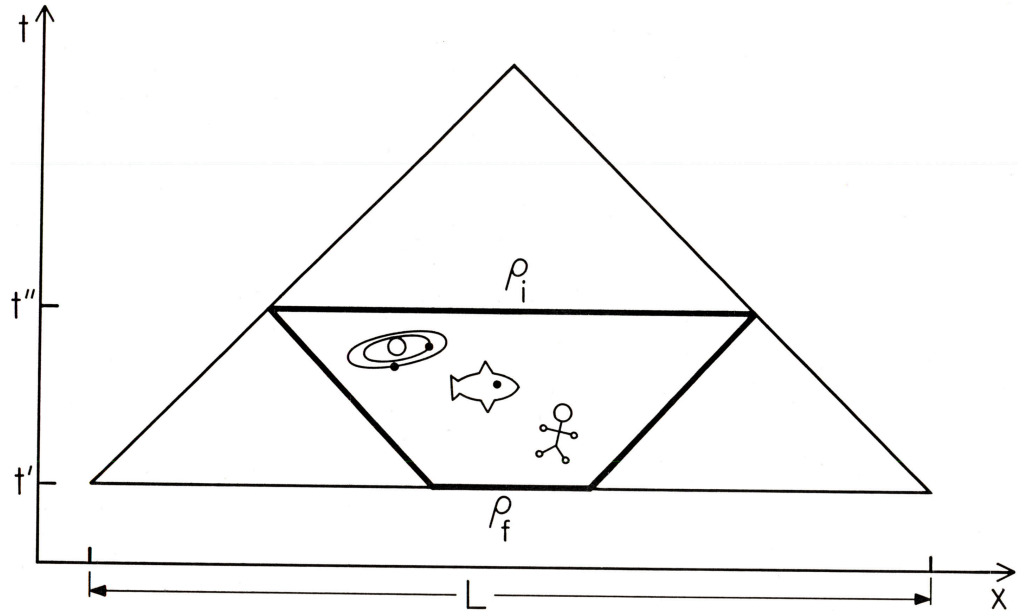

**Fig. 8:** *An experiment in quantum cosmology. The figure shows a thought experiment in quantum cosmology. An ensemble of very large regions of the spacetime of the late universe, one of which is illustrated here, is constructed by preparing the state of the system according to the statistics of a density matrix $\rho_f$ across a large spatial region of extent $L$ in the spacelike hypersurface $t = t'$. The region inside the large triangle (the future Cauchy development of the spatial region) is thus causally isolated from the initial condition of the larger universe. The ensemble of regions is refined by selecting the state on the surface $t = t''$ according to the statistics of a density matrix $\rho_i$. If $\rho_f \propto I$ (a random selection of the state at $t'$) the physics in the heavily outlined region should be indistinguishable from that in a universe with an initial condition $\rho_i$ in which the quantum mechanical arrow of time is reversed from its direction in the larger universe. The statistics of the evolution of IGUSes in such regions, and the physical theories they induce, are in principle predictions of the quantum cosmology of the larger universe and are subject to experimental test in such an ensemble! One expects that these IGUSes will, by induction, arrive at a quantum mechanics with a similar fundamental formula to that of the larger universe but with the quantum mechanical arrow of time reversed.*

*If the ensemble is constructed with a $\rho_f$ not proportional to $I$ then, one expects that the fundamental formula would have both initial and final conditions as in (IV.2.5). It is an empirical question whether or not we live in such a universe.*

the trace, it is time symmetric! Of course, particular $\rho_i$ and $\rho_f$ may be very different from one another producing an effective arrow of time for some physical phenomena, but there is no *independent* quantum mechanical arrow of time. The quantum mechanical formalism is time symmetric.

In this generalization there is no built in notion of causality and it is not possible to have a notion of "state at a moment of time" from which probabilities can be extracted either for events in the past or the future. ¿From this perspective the notion of the state of the system at a moment of time in Hamiltonian quantum mechanics is a consequence of our ignorance of the future.

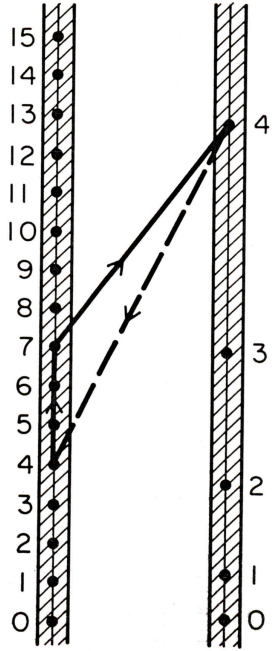

**Fig. 9:** *Closed timelike lines in a wormhole geometry. The figure shows a spacetime diagram (time upward) with two wormhole mouths (the shaded regions). The wormhole geometry is multiply connected so that it is possible to pass nearly simultaneously from points in one wormhole mouth to another. The wormhole mouth on the left remains at rest in an inertial frame. The one at right is initially at rest with respect to the first at $t = 0$ but then begins to rotate about it. The figure shows the corotating frame and the readings of a clock at the center of each wormhole mouth. As a consequence of time dilation in the rotating mouth, this spacetime has closed timelike curves of which one is shown. The dotted segment represents the nearly instantaneous passage through the wormhole throat.*

### IV.3. Topology in Time

As a third illustration of how closely our familiar formulation of quantum mechanics is tied to assumptions about spacetime, I would like to consider the question of constructing quantum mechanics on spacetimes whose time direction has a non-trivial topology.

An interesting example is the kind of wormhole spacetime discussed by Morris, Thorne, Yurtsver (1989) and others that is illustrated in Fig. 8. These are not the four-dimensional wormholes that are the subject of this school. They are handles on three-dimensional space. The topology of spacetime is $\mathbf{R} \times M^3$ with $M^3$ being multiply connected.

Imagine that before some time $t = t_s$ the wormhole mouths are at rest with respect to one another. At time $t_s$ they begin to rotate about one another and continue until a moment of time symmetry when they reverse their motion eventually coming to relative rest at time $t_e$. Before $t_s$ and after $t_e$ there are no closed timelike lines and it is possible to define surfaces of constant time that foliate those portions of spacetime. In between $t_s$ and $t_e$, however, because of time dilation in the rotating wormhole mouth, there are closed timelike lines, as Fig. 8 illustrates. By going through the wormhole throat it is effectively possible to go backward in time. Such wormhole spacetimes are time orientable but not causal.

It is clear that there is no straightforward Hamiltonian quantum mechanics in a wormhole spacetime between the surfaces $t_s$ and $t_e$. What would be the surfaces of the preferred time? How would unitary evolution of arbitrary states in the Hilbert space be defined in the presence of closed timelike lines?

A generalized quantum mechanics of the kind we have been discussing, however, may be constructed for this example using a sum-over-histories decoherence functional. The three ingredients would be the following*

1) *Fine-Grained Histories:* For the fine-grained histories we may take single-valued field configurations, $\phi(x)$, on the wormhole spacetime.

2) *Coarse Grainings:* The fine-grained histories may be partitioned according to their values on spacetime regions.† Select a set of spacetime regions, specify an exhaustive set of ranges for the average values of the field in these regions, and one has partitioned the four dimensional field configurations into classes, $\{h\}$, that have the various possible values of the average field. For example one might specify averaged spatial field configurations on an initial constant time surface with $t < t_s$ and on a final constant time surface with $t > t_e$. The resulting probabilities would be relevant for defining the $S$-matrix for scattering from the wormhole.

3) *Decoherence Functional:* A sum-over-histories decoherence functional is

$$D(h, h') = \int_h \delta\phi \int_{h'} \delta\phi' \ \delta[\phi_f(\mathbf{x}) - \phi'_f(\mathbf{x})]$$

$$\times \exp\left\{ i\Big( S[\phi(x)] - S[\phi'(x)] \Big)/\hbar \right\} \rho_0 \left[ \phi_0(\mathbf{x}), \ \phi'_0(\mathbf{x}) \right] \ . \qquad \text{(IV.3.1)}$$

The integrations are over field configurations between some initial constant time surface $t_0 < t_s$ and some final constant time surface $t_f > t_e$. $\phi_0(\mathbf{x})$ and $\phi'_0(\mathbf{x})$ are the spatial configurations on the initial surface; their integral is weighted by the density matrix $\rho_0$. $\phi_f(\mathbf{x})$ and $\phi'_f(\mathbf{x})$ are the spatial configurations on the final surface; their coincidence is enforced by the functional $\delta$-function. The integral over $\phi(x)$ is over the class of field configurations in the class $h$. For example, if $h$ specifies that the average value of the field in some region lies in a certain range, then the integral is only over $\phi(x)$ that have such average values. Formally, this decoherence functional satisfies conditions (i)-(iv) of Section III.1. The results of Friedman et al. (1990) on the existence of solutions to the classical initial value problem gives some hope that it may be well defined, at least for quadratic field theories.

With the generalized quantum mechanics based on the three elements described above probabilities can be assigned to coarse-grained sets of field histories

---

in the wormhole spacetime. These probabilities obey the standard probability sum rules. There is no equivalent Hamiltonian formulation of this quantum mechanics because this wormhole spacetime, with its closed timelike lines, provides no foliating family of spacelike surfaces to define the required preferred time. Nevertheless, the generalized theory is predictive. What has been lost in this generalization is any notion of "state at a moment of time" and of its unitary evolution in between the surfaces $t_e$ and $t_s$. This is not surprising for a region of spacetime that has no well defined notion of "at a moment of time".

### IV.4. The Generality of Sum Over Histories Quantum Mechanics.

What the examples of this Section argue is that the Hamiltonian formulation of quantum mechanics is closely tied to fixed spacetimes that admit a foliation by spacelike surfaces and that may have definite initial conditions but that must have ignorance of the future as a final condition. By contrast sum-over-histories quantum mechanics applies more generally to cases that do not have such a preferred time. Other examples would be interesting to investigate. A short list might include: field theories with interactions that are non-local in time (e.g. Bloch, 1952), field theory in identified flat spacetimes with interesting topology in time, particle path-integral quantum mechanics in spacetimes with interesting topology in time, theories with a discrete number of possible spacetimes, etc. However, in the next section I shall proceed directly to the case of quantum gravity where in general, there is *no* fixed spacetime. There I shall argue that the sum-over-histories formulation is the natural generally covariant way of constructing quantum mechanics.

## V. THE QUANTUM MECHANICS OF SPACETIME

### V.1. The Problem of Time

#### V.1.1. *General Covariance and Time in Hamiltonian Quantum Mechanics*

A consistent and manageable quantum theory of gravity is a central prerequisite for any quantum cosmology. Providing such a theory is the subject of intensive contemporary research in a variety of directions including string theory, generalized canonical quantum gravity, discrete spacetime models, low dimensional models, and non-perturbative approaches to the quantization of Einstein's theory. In all of these approaches, spacetime, whether a fundamental quantity or not, is a dynamical quantum variable rather than fixed and given as in all our previous discussion. Therefore, each of the theories, however they may differ on their assumptions concerning fundamental fields or their approach to the divergences of the theory, must confront the second count on which the usual framework of Hamiltonian quantum mechanics is insufficiently general for quantum cosmology. This concerns the "problem of time".*

The discussion of the preceeding sections shows that time plays a special and peculiar role in Hamiltonian quantum mechanics. What are the physical grounds for singling out one variable to play such a special role in the predictive formalism? They arise, I believe, from the fact that as observed on all directly accessible scales, over the whole of the accessible universe, spacetime does appear to have a fixed, classical geometry. It is this background geometry that supplies an unambiguous notion of time for quantum mechanics. In the spacetime of non-relativistic physics there is a preferred family of spacelike surfaces of constant Newtonian time that define the preferred time of non-relativistic quantum mechanics. In the spacetimes of special relativistic physics there are many foliating families of spacelike surfaces. There is thus an issue as to which defines the preferred time of quantum mechanics. Causality, however, implies that the quantum mechanics constructed from one choice is unitarily equivalent to that for any other. All choices give equivalent results.

---

* For classic reviews of the problem of time see Wheeler (1979) and Kuchař (1981).

In the quantum gravity of cosmological spacetimes there are no fixed backgrounds in general. In particular, in the early universe we expect quantum fluctuations of spacetime. What then supplies the preferred time required by Hamiltonian quantum mechanics? Certainly it is not the classical theory of spacetime — Einstein's general relativity. That theory is generally covariant* and no one family of spacelike surfaces is preferred over any other. Further in the absence of a fixed background to define a notion of causality there is no evidence that the quantum mechanics constructed from two different choices of preferred spacelike surfaces are unitarily equivalent.† There is thus a conflict between the framework of Hamiltonian quantum mechanics and covariant theories of spacetime such as general relativity or string theory. This is the problem of time.

The traditional route out has been to keep quantum mechanics "as is" and give up on spacetime. Perhaps there *is* a preferred family of spacelike surfaces in quantum mechanics. General covariance is thereby broken at the quantum level and the beautiful synthesis of Einstein and Minkowski can emerge only in the classical limit.‡

Perhaps there are other variables, now hidden, that would play the role of time in a Hamiltonian formulation of quantum gravity.‡ Perhaps in a theory in which spacetime is not a fundamental variable such a preferred time would be distinguished naturally. However, to formulate a Hamiltonian quantum mechanics with a time variable other than a family of spacelike surfaces in spacetime is, most honestly, a generalization of familiar quantum mechanics. All such generalizations have the obligation to show how the familiar formulation with spacelike surfaces as the preferred time variable emerges in appropriate limits.

---

* By a generally covariant theory I do not mean one which can be expressed in a form that is invariant under general coördinate transformations. Rather, as usual in general relativity, I mean a theory in which gravitational phenomena are described by the spacetime metric alone. Invariance can be accomplished for any theory by introducing sufficiently many tensor fields, say, those specifying a preferred family of spacelike surfaces. It is the absence of such fields — general covariance — that is the orgin of the problem of time in relativity.

† See, e.g. Isham and Kuchař (1985) and the references therein.

‡ This is the approach taken in what is usually called canonical quantum gravity. A preferred time variable is identified from among the variables describing three-geometry. The Wheeler-DeWitt equation is reorganized as a "Schrödinger" equation in this time variable and an inner product introduced in the remaining variables. For reviews see again Kuchař (1981) and the discussion in Ashtekar and Stachel (1991). By canonical quantum gravity we shall mean this kind of schema. We do not mean merely implementing the constraints as operator equations on superspace and neither do we mean the generalized canonical approach of Ashtekar and others (Ashtekar, 1988).

‡ See, e.g. the variety of ideas in Unruh and Wald (1989), Brown and York (1990), Henneaux and Teitelboim (1989), and Hoyle and Hoyle (1963).

Each of the ideas described above could be right. However, in these lectures I would like to explore another way to resolve the problem of time. This is the idea that Hamiltonian quantum mechanics with its preferred time is an insufficiently general framework for a generally covariant quantum theory of spacetime. I would like to pursue the idea that quantum theories of spacetime should be formulated in a generalized quantum mechanics in the sense of Section III that *is* generally covariant. In such a formulation Hamiltonian quantum mechanics would be but an approximation appropriate to those special initial conditions and those special coarse grainings in which spacetime *is* approximately classical and fixed and a notion of spacelike surface is defined. Further, I would like to sketch a particular sum-over-histories generalized quantum mechanics that is general enough to incorporate *both* generally covariant theories without a preferred time and also theories that have such a variable.* In this way a preferred time and its associated Hamiltonian quantum mechanics become a special property of certain theories, or an approximate property of the universe's initial condition, and not a prerequisite for the application of quantum mechanics to the geometry of the universe.

### V.1.2. *The "Marvelous Moment"*

As a response to the problem of time the idea is sometimes advanced that there is no fundamental notion of time. That it is sufficient for prediction to calculate probabilities for observation from a single wave function on a single spacelike surface. That any notion of time is to be recovered from a study on that surface of the probabilities for the correlations between the indicators of clocks and other variables. That our impression of the past is but an illusion more properly viewed as correlations between records existing at the "marvelous moment" now.† That quantum general relativity is most properly viewed as a theory of space rather than a theory of quantum spacetime.

However, to abandon spacetime just because there is no natural candidate for the ordering parameter in Hamiltonian quantum mechanics seems, to me, an overreaction. While it is no doubt true that many interesting probabilities, especially in cosmology, are for observations that are more or less on one spacelike surface they do not exhaust those predicted by familiar quantum mechanics nor those that are important.

As we saw in Section II.3, the objects of predictions in history are most honestly correlations between present records. However, these correlations can be said to describe history only when theory implies a further correlation between records in the *present* and events in the *past*. Our ability to calculate such probabilities is central to our ability to organize and explain *present* data. It would be an impossibly complex task to predict from one wave function of the universe analysed solely in the present that similar dinosaur bones should be located in similar geologic strata. The reason is that such correlations are not thereby distinguished from other correlations that exist in the present. The simple explanation for the observed correlation that similar dinosaur bones are located in similar strata is that dinosaurs *did* roam the earth many millions of years ago and their bones are persistent records of this epoch. It is by calculating probabilities between the past and different records today that correlations are predicted between these records. It is by such calculations that the probability for error in present records is estimated. Such calculations cannot be carried out solely on one spacelike surface.

---

* The ideas in this section were sketched in Hartle (1986b) and have been developed in Hartle (1988ab, 1989b). For related sum-over-histories approaches to a covariant quantum mechanics of spacetime see Teitelboim (1983abc) and Sorkin (1989).

† For modern versions of this idea see e.g. Page and Wootters (1983), Wootters (1984, 1986). The idea is an old one, see e.g. Augustine (399).

Many other interesting probabilities involve several spacelike surfaces. Just to describe in an objective way the subjective experience of the passage of time requires such probabilities. The correlations that distinguish classical spacetime are between a sequence of spacelike surfaces. Similarly, the correlations satisfied by other classical dynamical laws involve many spacelike surfaces. A physical system can be said to behave as a good clock when the probability is high that the position of its indicator is correlated with the location of *successive* spacelike surfaces in spacetime.

It may be that the search for unity between gravitation and other interactions will lead us to abandon as fundamental, time, space, or both. In this case a revision of the predictive framework of quantum mechanics seems inevitable. A distinction needs to be drawn, however, between such motivations and those arising from the preferred status of time in the familiar framework. Before we invoke the conflict between that familiar framework and covariant spacetime as reason to abandon one of the most powerful organizing concepts of our experience, it may be of interest to see whether the familiar framework of quantum mechanics might be generalized a bit to apply to theories of spacetime.

## V.2. A Quantum Mechanics for Spacetime

### V.2.1. *What we Need*

What we need is a generalized quantum mechanics in the sense of Section III that supplies probabilities for correlations between different spacelike surfaces, that does not prefer one set of spacelike surfaces to another, *and* that reduces approximately to Hamiltonian quantum mechanics when spacetime is classical in the late universe as a consequence of its initial condition. Such a generalized quantum mechanics is specified by the three elements: the set of fine-grained histories, a notion of coarse graining, and a decoherence functional. In this Section I shall describe an example of such a generalized quantum mechanics of spacetime. It is a sum-over-histories generalized quantum mechanics that assumes that spacetime is fundamental. This means, in particular, that the fine-grained histories are cosmological four geometries with matter field configurations upon them.

There is considerable speculation that spacetime is not fundamental. Even in a theory where it is not, however, it must be possible to construct a coarse graining that defines spacetime geometry on scales modestly below the Planck scale and to construct an effective quantum theory of such coarse-grained histories. In such a theory one might expect the following schematic sequence of approximations to the decoherence functional:

$$D_{\text{fundamental}}(h, h') \simeq D_{\text{spacetime}}(h, h') \simeq D_{\text{Hamiltonian}}(h, h') \ . \qquad (\text{V.2.1})$$

The first approximate equality describes how a spacetime theory could be an ef-fective limit of the more fundamental theory. It would be expected to hold only for coarse grainings that define spacetime and matter fields on scales of about the Planck length and above. The second approximate equality, of the kind discussed in Section III.6, describes how Hamiltonian quantum mechanics with its preferred time is recovered from a generalized quantum mechanics of spacetime. It would be expected to hold only for coarse graining defining classical spacetimes in the late universe with quantum matter fields on them.

Thus, the generalized sum-over-histories quantum mechanics to be described may be thought of either as a model for the kind of framework necessary in a more fundamental theory or as a representation of the effective theory of spacetime such a theory is expected to provide.

### V.2.2. *Sum-Over-Histories Quantum Mechanics for Theories Without a Time*

I begin by describing generalized sum-over-histories quantum mechanics for theories with no preferred time. The three elements of this formulation are supplied as follows:

1) *Fine-grained histories:*   The fine-grained histories are paths, $Q^\alpha(\tau)$, in a configuration space that includes the physical time if there is one. The value of a parameter $\tau$ along the path is not specified as part of a fine-grained history; only the path itself is specified. The parameter $\tau$ is a label, useful in constructing a sum-over-paths, but not itself assigned a probability in general. Thus, in the Hamiltonian quantum mechanics discussed in Section II, $Q^\alpha = (t, q^i)$. There, because the histories are single-valued in $t$, the time $t$ *could* be used as the parameter $\tau$. No such time would exist, for example, for the quantum mechanics on the spacetimes of Section IV.3 that are multiply connected in time.

2) *Coarse-grained histories:*   One type of coarse graining is defined by giving regions in the configuration space and partitioning the paths according to whether they pass through them or not (Fig. 10). Thus, with two regions $R_1$ and $R_2$ there are the set of four exhaustive alternative classes of histories:

   $h_1$: The paths that go through both regions $R_1$ and $R_2$ at least once.

   $h_2$: The paths that go through $R_1$ at least once but not $R_2$.

   $h_3$: The paths that go through $R_2$ at least once but not $R_1$.

   $h_4$: The paths that go through neither $R_1$ nor $R_2$.

A coarser graining might be

   $\tilde{h}_1$: The paths that go through $R_1$ or $R_2$ at least once.

   $\tilde{h}_2$: The paths that go through neither $R_1$ nor $R_2$.

A finer graining might specify the exact number of times a path crosses a region. For the case of the quantum mechanics of Section II these regions were associated with a precise time, but in the discussion of the quantum mechanics of field averages they have extent in time.

3) *Decoherence functional:*   In a sum-over-histories quantum mechanics this

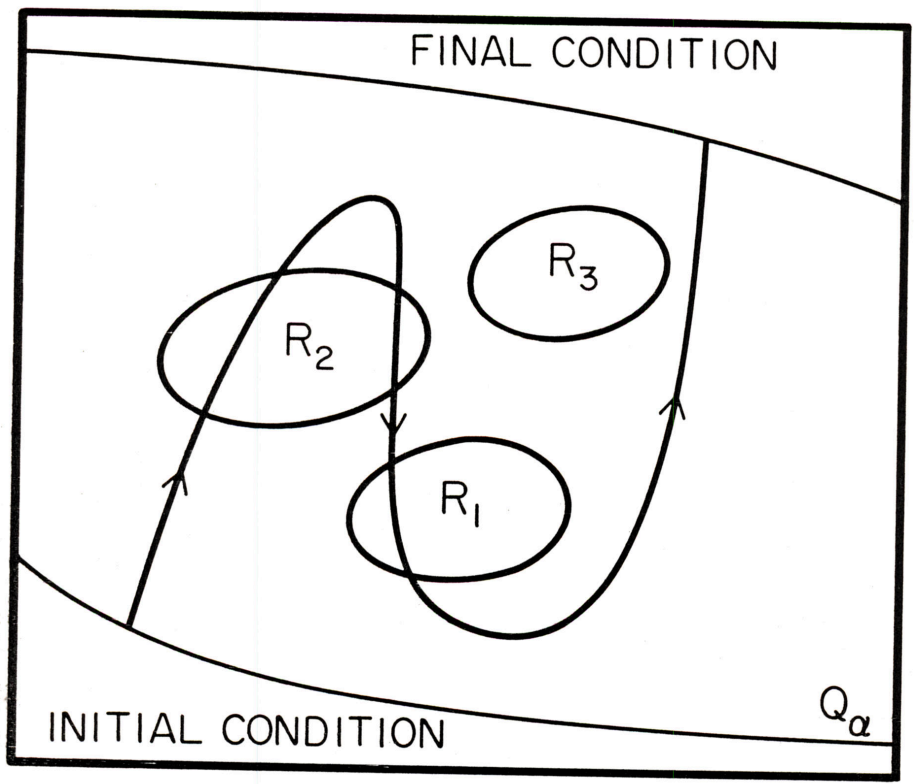

**Fig. 10:** *Coarse-grained histories for a sum-over-histories quantum mechanics for theories without a preferred time variable. The figure is a schematic illustration of a configuration space $\{Q_\alpha\}$ that includes any physical time. The fine-grained histories are all possible paths thus space satisfying prescribed initial and final conditions. A coarse-grained set of histories is a partition of the set of all paths into exhaustive and exclusive classes. The behavior of histories with respect to a set of regions can be used to define such a partition. The illustrated history is a member of the class that passes through regions $R_1$ and $R_2$ at least once but not through $R_3$.*

has the form

$$D(h_i, h_j) = \int_{h_i, \mathcal{C}} \delta Q^\alpha \int_{h_j, \mathcal{C}} \delta Q'^\alpha e^{i\left(S[Q^\alpha(\tau)] - S[Q'^\alpha(\tau)]\right)/\hbar} \ . \qquad (\text{V}.2.2)$$

The sum over $Q^\alpha(\tau)$ is over all paths in the class $h_i$ that satisfy certain initial and final conditions. The sum over $Q'^\alpha(\tau)$ is similar but in the class defined by $h_j$. In sum-over-histories quantum mechanics these conditions, $\mathcal{C}$, include the initial condition of the universe (the initial $\rho$) and ignorance of the future or other final condition.

We cannot expect such a quantum mechanics to have an equivalent Hamiltonian formulation unless there are surfaces in the space of the $Q^\alpha$ that the paths cross once and only once (Fig. 11). This was the lesson of Section III.4. Thus, we cannot expect to recover a notion of "state at a moment of time" (there is no time) or its unitary evolution. We have the predictibility without states and con-servation of probability (represented by the probability sum rules) without unitary evolution.*

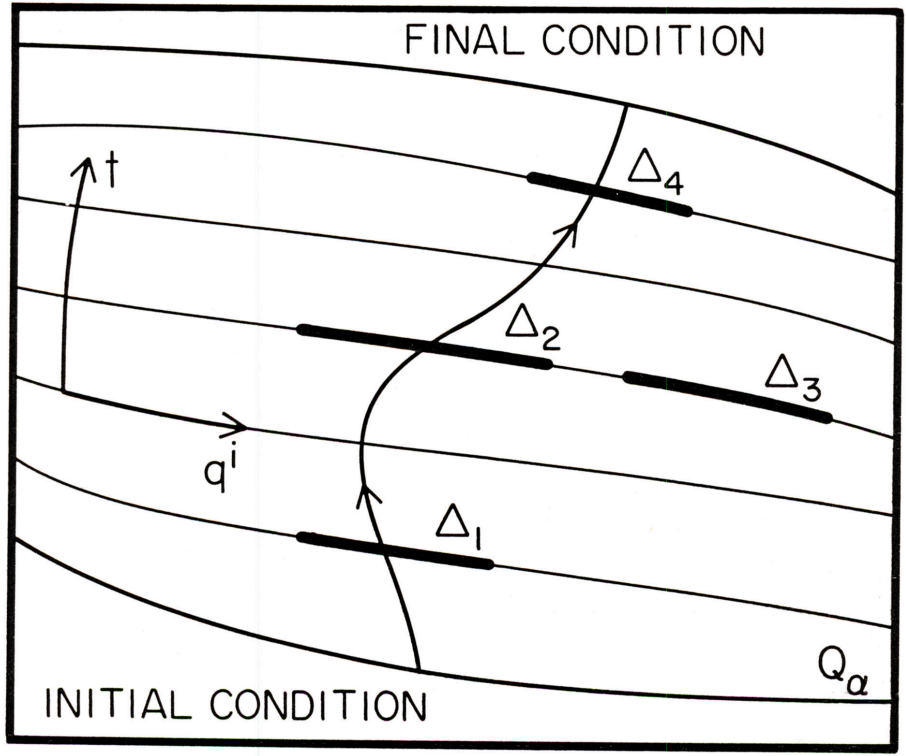

**Fig. 11:** *Hamiltonian quantum mechanics as a special case of general-ized sum-over-histories quantum mechanics. If the fine-grained histories have the property that there is a set of surfaces that the paths cross once and only once, then it is possible to construct an equivalent Hamilto-nian formulation for coarse grainings that are regions limited to these surfaces. As described in Section III.4 the sums over histories can be factored across such surfaces and the factorization used to define unitar-ily evolving states and their inner product on the preferred surfaces.*

V.2.3. Sum-Over-Spacetime-Histories Quantum Mechanics

---

\* One might have thought one could factor the sum over histories across *any* surface into a sum on one side of the surface times a sum on the other followed by a sum over the various ways the paths could cross the surface thereby defining states and then inner products on these surfaces. For sums-over-histories defined on a spacetime lattice this is indeed possible. However, as the lattice spacing becomes small the sum defining some particular way of crossing the surface (say, a fixed number of intersection points) vanishes. This is because the dominant paths are non-differentiable and the expected number of crossings is infinite. As a consequence no useful Hamiltonian quantum mechanics is recovered in the continuum limit. (Hartle, 1988a)

How does the sum-over-histories quantum mechanics described above look specifically for four-dimensional cosmological spacetimes? I shall first sketch the story in terms of words and pictures. More technical details of how to do the relevant sums over histories are sketched in Section V.3.

The fine-grained histories in cosmology are all cosmological four-geometries with matter field configurations upon them. An example is the standard classical Friedman evolution of a closed universe from the big bang to a big crunch. Of course, there are many other possible non-classical histories that must be assigned amplitudes quantum mechanically. These histories may be thought of as successions of three-dimensional geometries — an expanding and contracting three sphere in the case of the Friedmann universe. They may, therefore, be thought of as paths in the space of three-geometries and three-dimensional matter field configurations (Fig. 12). This can be made explicit by writing the metric in a gauge in which $g_{00} = -1$ and $g_{0i} = 0$:

$$ds^2 = -dt^2 + h_{ij}(\mathbf{x}, t)dx^i dx^j \ . \tag{V.2.3}$$

The three-metric $h_{ij}(\mathbf{x}, t)$ describes the three-geometry and $t$ defines the succession of them. Fine-grained histories may thus be thought of as curves in the superspace of three-metrics, $h_{ij}(\mathbf{x})$, and spatial matter field configurations $\chi(\mathbf{x})$, parameterized by the proper time $t$. Put differently, the $Q^\alpha$ of the general discussion above are $(t, h_{ij}(\mathbf{x}), \chi(\mathbf{x}))$.

Implicit in such a description is a limitation on the topology of the cosmological histories to be $\mathbf{R} \times M^3$ where $M^3$ is some compact three-manifold. This, as we shall see in Section V.2.4, is not an essential limitation. The sums over geometries that define the decoherence functional could be extended to include geometries that bifurcate into separate universes, include wormholes, etc. However, to keep the discussion manageable let us, for the moment, make this simplifying assumption on topology.

There are many kinds of coarse grainings of these fine-grained histories that might be of interest. Of course, the coarse grainings must be diffeomorphism invariant. That is the set of possible four-metrics describing possible cosmological histories must be partitioned into diffeomorphism invariant classes. Perhaps the coarse grainings which are most analogous to those used in familiar quantum mechanics are those associated with a set of non-overlapping regions in superspace, $\{R_i\}$, that are invariant under three-dimensional diffeomorphisms. A history which passes through a region $R$ once, has a sequence of foliating, spacelike, three-surfaces whose spatial geometries and matter field configurations are fixed within the error allowed by the region $R$. A history that passes through $R$ twice would have two such regions, and so forth.

Coarse graining by superspace regions that are thin in one direction can capture the canonical notion of "at a moment of time". Consider, for example, a region of superspace which lies along a constant value of some scalar field, $\chi$, to be used as a "clock" and allows negligible variation in the value of that scalar field. A history which passes through such region once has one spacelike surface on which $\chi$ has the specified value. A history which passes through such a region twice has two such regions, etc. A history passing through two such regions with different values of $\chi$, is a history which has two spacelike surfaces with different "times", $\chi$.

By using the word "time" in this discussion I do not mean to imply that any of these variables is the preferred time of Hamiltonian quantum mechanics. I shall argue below that they are not. It is through such variables, however, that the canonical notion of configurations on several different spacelike surfaces is defined. More general regions can be considered that make use of the proper time in the full configuration space $\{t, h_{ij}(\mathbf{x}), \chi(\mathbf{x})\}$. For example, one could consider coarse grainings in which the proper time $t$ was specified between spacelike surfaces labeled by $\chi$.

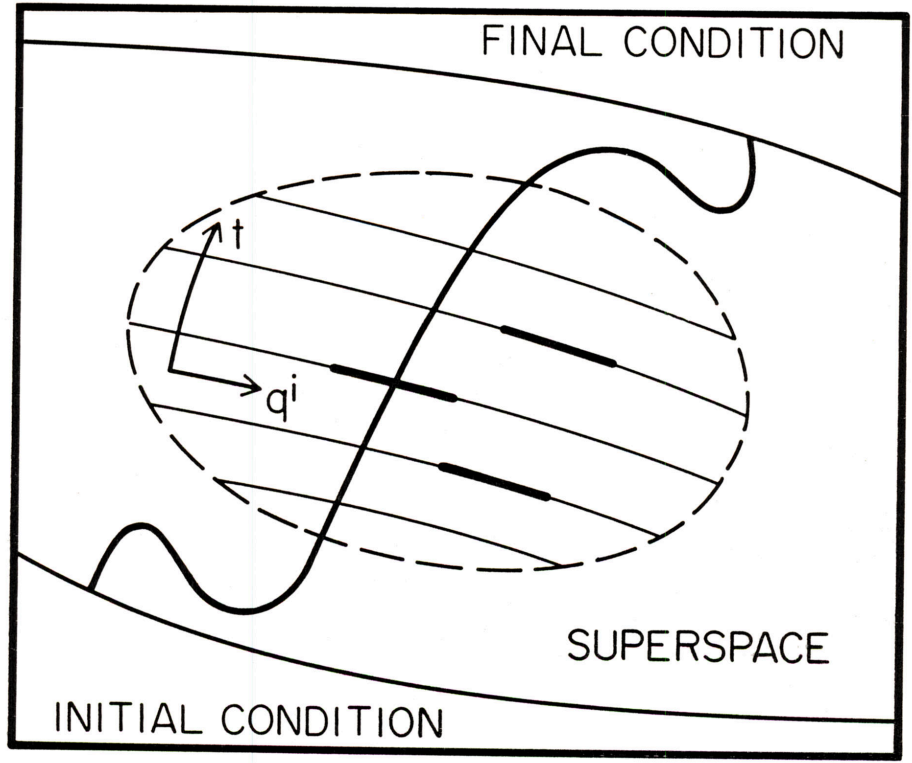

**Fig. 12:** *Superspace. A cosmological history is a four dimensional cosmological spacetime with matter fields upon it. A two dimensional representation of such a history is shown in the upper left of this figure proceeding from a big bang to a big crunch. In the gaussian gauge of (2.3) a cosmological history can be thought of as a succession of three-dimensional geometries and spatial matter field configuration. Superspace is the space of such three-dimensional geometries and matter field configurations. A "point" in superspace is particular three-geometry and spatial matter field configuration. The succession of three-geometries and matter fields that make up a four-geometry and field history, therefore, trace out a path in superspace.*

Given a set of exclusive regions in the configuration space, the set of fine-grained histories can be partitioned into exhaustive and exclusive classes by their behavior with respect to these regions, exactly as in the general discussion of the preceding section. Exhaustive sets of coarse-grained sets of alternative histories histories thus can be defined. As a simple example, consider the coarse-grainings specified by one region $R$ in superspace that is thin in the $\chi$ direction. The fine-grained histories can be partitioned into the classes:

$h_0$ : histories that never pass through $R$,

$h_1$ : histories that pass through $R$ at least once.

These classes might be described as follows: $h_1$ is the class of histories that have at least one spacelike surface with the specified $\chi$, the specified range of $h_{ij}(\mathbf{x})$, and other matter fields, and *any* value of $t$. $h_0$ is the class with no such spacelike surfaces.

A more refined coarse graining would be

$\tilde{h}_0$ : histories that never pass through $R$,

$\tilde{h}_1$ : histories that pass through $R$ once,

$\tilde{h}_2$ : histories that pass through $R$ twice,

$\tilde{h}_3$ : histories that pass through $R$ three times, etc.

In such a coarse graining the *number* of spacelike surfaces with given three-geometry and spatial field configuration in the four-geometry is specified.

The above discussion of coarse graining is incomplete on at least two important physical issues that are topics for further research.

1) *Which Coarse Grainings Make Sense?* We know from non-relativistic quantum mechanics that it is possible to specify coarse grainings in words that make no sense when examined in the full light of the mathematics of functional integrals (e.g. Hartle, 1988a). Consider, for example, partitioning the fine-grained paths of a non-relativistic particle according to how many times they cross a segment of timelike surface. Such a partition can be defined in a lattice approximation to the sums over histories. In the continuum limit, however, the amplitude for any fixed, finite number of crossings vanishes. This is because the paths are non-differentiable and the expected number of crossings is infinite. Which of the coarse grainings of geometries sketched above fail to make sense in this manner? We need a more explicit and manageable definition of a sum over geometries to find out.

2) *Which Coarse Grainings do **We** Use?* Our observations fall far short of determining anything like the three-geometry on a spacelike surface or the spatial matter field configurations there. We deal with much coarser grainings of the universe that are heavily branch dependent in the sense of Section II.8. How are they most honestly described as partitions of the fine-grained histories discussed above?

With these caveats in mind, we can pass on to the third element of a generalized quantum mechanics for spacetime — the decoherence functional. For sum-over-histories quantum mechanics the decoherence functional is naturally defined on a set of coarse-grained histories $\{h_i\}$ as

$$D(h_i, h_j) = \int_{h_i, \mathcal{C}} \delta g \delta \phi \int_{h_j, \mathcal{C}} \delta g' \delta \phi' e^{i(S[g, \phi] - S[g', \phi'])/\hbar} \ . \qquad (\mathrm{V}.2.4)$$

Here, $S$ is the action for gravity and matter fields. The integral is over four-dimensional metrics, $g$, and field configurations, $\phi$, that lie in the partition $h_j$. The integral over $g'$ and $\phi'$ is similarly defined with respect to the partition $h_j$. It is assumed that the initial and final conditions on the histories are incorporated in the sum over histories as conditions, $\mathcal{C}$, on the fine-grained histories in a way analogous to that discussed in Section III.3. These conditions may involve both $h_i$ and $h_j$ as in the conditions that enforce the coincidence of the final endpoints in (II.2.2). It is because of such conditions that (2.4) does not factor. I shall not discuss the details of these conditions further because the exact forms representing, say, the "no boundary" initial condition and final ignorance are still open questions. I assume that they exist.

In addition to the conditions, $\mathcal{C}$, several other ingredients of the functional integral need to be supplied. These include specification of the parametrization of the sum-over-geometries, the measure, the gauge fixing machinery, and the contour of integration. A few aspects of these problems are discussed in the next subsection. Until they are specified equations like (2.4) must be regarded as schematic forms.

Using this decoherence functional, decoherent sets of histories can be identified and probabilities assigned to them that approximately obey all the rules of probability theory. There is, in general, no possible choice of time variable such that this quantum mechanics of spacetime can be put into Hamiltonian form. For that to be the case, we would need a time function on superspace whose constant time surfaces the paths cross once and only once as in Fig. 11. There is none. Put differently, there is no purely geometrical quantity that uniquely labels a spacelike hypersurface. The volume of the universe, for example, may single out just a few surfaces in a classical cosmological history, but in quantum mechanics we must consider all possible histories. A non-classical history may have arbitrarily many surfaces of a given volume.

While we do not recover a Hamiltonian formulation precisely and generally, we may recover it *approximately* in restricted domains of superspace, for special coarse grainings, and for particular initial conditions. Suppose, for example, that the initial condition were such that for coarse grainings defined by sufficiently unrestrictive regions of superspace, $R$, in a regime of three geometries much larger than the Planck scale, only a *single* spacetime geometry, $\hat{g}$, contributed to the sum defining the decoherence functional (Fig. 13). Then if

$$S[g, \phi] = S_E[g] + S_M[g, \phi] \ , \tag{V.2.5}$$

we would have approximately

$$D\left(h_i, h_j\right) \simeq \int_{h_i, \mathcal{C}} \delta\phi \int_{h_j, \mathcal{C}} \delta\phi' \ e^{i(S[\hat{g}, \phi] - S[\hat{g}, \phi'])/\hbar} \ . \tag{V.2.6}$$

The remaining sum over $\phi(x)$ defines the quantum mechanics of a field theory in the background spacetime $\hat{g}$. Any family of spacelike surfaces in this background picks out a unique field configuration since the sum is over fields that are single-valued on spacetime. The field paths are then single-valued in any time defined by a foliating family of spacelike surfaces, there is a notion of causality, and we recover an ordinary field theory in the background spacetime $\hat{g}$. Because the field histories are single valued in time the sums over fields may be factored across any spacelike surface of the geometry $\hat{g}$, as in Section III.4, and an equivalent Hamiltonian formulation of this quantum mechanics recovered.

Typical proposals for theories of the initial condition do not single out a single classical history for the late universe. It is hard, for example, to see how a simple initial condition could summarize all the complexity we see in our particular history. Rather typical proposals, such as the "no boundary" proposal, predict an *ensemble* of possible decohering background, classical spacetimes. This will be discussed in more detail in Section VI. In each member of the ensemble an approximate

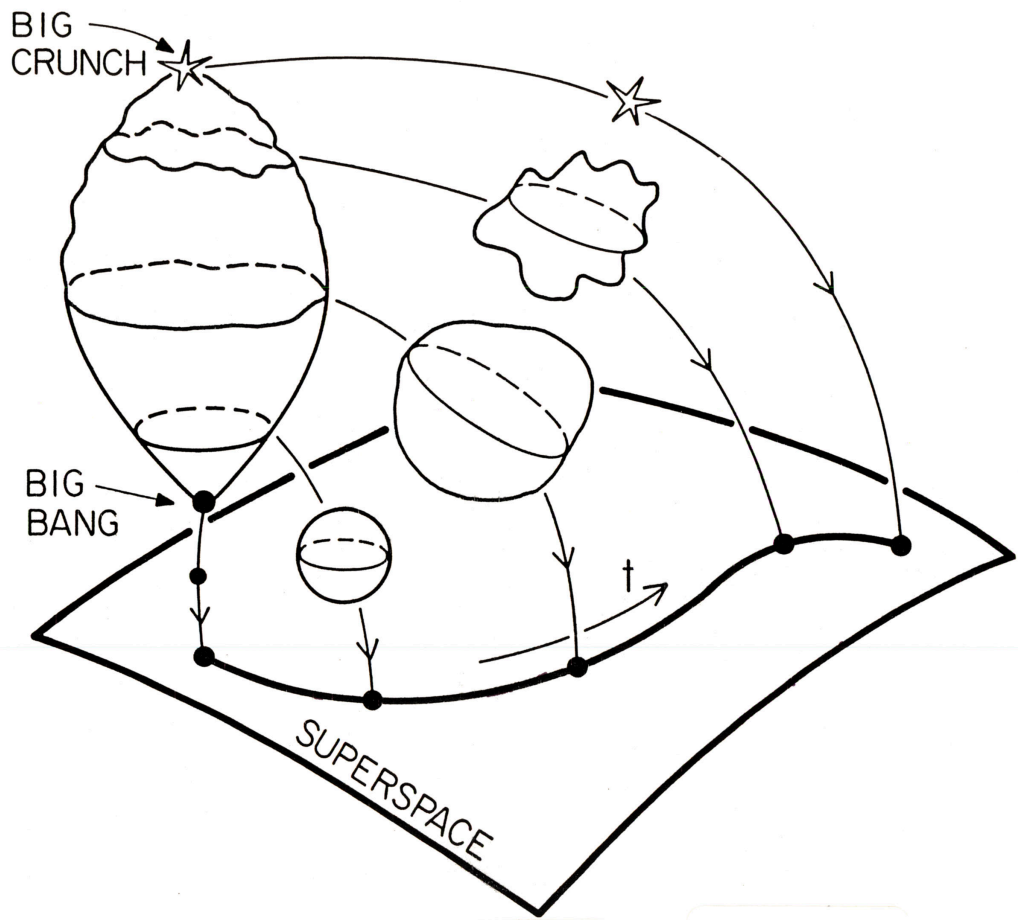

**Fig. 13:** *Recovery of Hamiltonian quantum mechanics in the late universe. The figure shows a schematic representation of the superspace of three-geometries and spatial matter field configurations. The large three-geometries of the late universe are contained in the region surrounded by the dotted line. For some initial and final conditions it may be true that, for coarse grainings that fix spacetime geometry only on scales well above the Planck length, only a class of decoherent classical spacetimes contribute to the sum-over-histories defining the decoherence functional. The remaining sum over matter fields is then over histories that assume one and only one spatial field configuration on any of the spacelike surfaces of the classical geometry. That sum then defines a decoherence functional for the matter fields that does have an equivalent Hamiltonian formulation. The possible preferred times are the preferred time directions of the classical spacetimes. In this way Hamiltonian quantum mechanics could be an emergent, approximate feature of the late universe appropriate to those initial conditions and coarse grainings that imply approximately classical spacetime there.*

Hamiltonian quantum mechanics is the associated background spacetime could be constructed. If an initial condition does not predict decoherent quasiclassical spacetime on familiar scales it is simply inconsistent with observation in a manifest way.

It would be in this way that the familiar Hamiltonian framework of quantum mechanics emerges as an approximation appropriate to the classical spacetime about our special position in the late universe as a consequence of its particular initial condition.

### V.2.4. *Extensions and Contractions*

The extension of the quantum mechanics of spacetime described in the preceding subsections to topologies other than $\mathbf{R} \times M^3$ presents no issues of principle. The fine-grained histories consist of the variety of manifolds allowed in the defining sums over histories with metrics and matter fields upon them. Coarse grainings are partitions of these fine-grained histories into exclusive classes. The decoherence functional would be constructed as in (2.4) with additional sum over manifolds in the allowed class, for example as

$$D(h_i, h_j) = \sum\nolimits_{MM'} \nu(M)\nu(M') \int_{h_i, \mathcal{C}(M)} \delta g \, \delta \phi \int_{h_j, \mathcal{C}(M)} \delta g' \, \delta \phi'$$

$$\times \exp\left\{ i\Big( S[g, \phi, M] - S[g', \phi', M'] \Big)/\hbar \right\}, \qquad \text{(V.2.7)}$$

where $\nu(M)$ is a positive weight on the class of manifolds summed over.* The further elements which need to be specified when topology is not fixed include the weight $\nu(M)$ and the dependence of the action, $S$, and conditions, $\mathcal{C}$, on the manifolds. A more serious problem is to identify the physically meaningful coarse grainings.

As the discussion in Sections V.2.2 and III.4 show, the predictions of sum-over-histories quantum mechanics can coincide with a Hamiltonian formulation for particular coarse grainings when histories are single-valued in a physical time. Thus, a sum-over-histories formulation is general enough to deal *both* with theories that have such a physical time and those that do not. Were there a physical time variable waiting to be discovered in spacetime theories, it might be still instructive to begin with a sum-over-histories point of view. Alternatively, a preferred time can be introduced into the theory by imposing restrictions on the class of fine-grained histories.

To formulate a canonical quantum gravity in sum-over-histories terms first identify the extended configuration space $\{Q^\alpha\}$ that includes the physical time. Then, within that space identify surfaces of constant time and restrict the fine-grained histories to paths that are single-valued in that time. Finally construct the decoherence functional according to (2.2) for suitable action and measure.

Consider by way of example, the popular choice of the trace of the extrinsic curvature, $K$, as a canonical time variable (Kuchař, 1972, York 1972). We could consider a superspace consisting of $K$ and the conformal three-metric $\tilde{h}_{ij}(\mathbf{x})$. The fine-grained histories would then be paths in this superspace that are single-valued in $K$. Appropriate coarse grainings would define alternative values of $\tilde{h}_{ij}(\mathbf{x})$ and matter fields at given values of $K$. This choice of fine-grained histories is not generally covariant in the sense that term is used here because a general metric may have many surfaces of a given value of $K$. The canonical fined-grained histories are, therefore, a non-covariant restriction of the general set. Nevertheless the resulting quantum mechanics may still be of interest to investigate, and the sum-over-histories formulation may provide interesting an alternative tool with which to do so.

---

\* It is assumed that the sum is over a class of manifolds that is classifiable. This is not the case for the class of all four-manifolds. For some discussion of this issue see Hartle (1985b), Geroch and Hartle (1986) and the references therein.

## V.3. The Construction of Sums Over Spacetime Histories

I would now like to turn to the more technical issue of what one means by the sums over histories defining the decoherence functional. The particular method I shall employ, although still completely formal, will, through increased concreteness, shed some light on the diffeomorphism invariance of the theory and its connection with the role of a preferred time.[*] For true concreteness we should consider lattice techniques for computing sums over geometries using, say, the methods of the Regge calculus.[**]

We were not able to construct a Hamiltonian quantum mechanics of spacetime that preserved the general covariance of the classical theory because Hamiltonian quantum mechanics required a preferred family of spacelike surfaces. One set of spacelike surfaces is as good as any other. However, there is a standard trick in quantum mechanics that is often useful in constructing amplitudes for theories with such symmetries. We extend the description of the histories with auxiliary labels that break the symmetry. We calculate amplitudes as always, but we make the rule that physically acceptable coarse grainings ignore these labels so that we sum *amplitudes* over them before squaring to calculate probabilities.

The most familiar example is in the theory of identical particles. We can begin a discussion of $N$ identical particles by introducing $N$ coördinates, $\vec{X}_i$, $i = 1, \cdots, N$ where the label $i$ distinguishes one particle from another. A general wave function $\psi(\vec{X}_1, \cdots, \vec{X}_N)$ characterizes a state in which one particle is distinguishable from another. However, when we sum that wave function over different possible values of the label for each argument we symmetrize it. The symmetry reflects the indistinguishability of the particles and the unobservability of the label. Other examples are the use of gauge variant fields rather than gauge invariant ones to describe gauge theories and the use of an unobservable proper time label to describe the relativistic particle.

In the language of decoherence functionals, the labels are treated differently from other variables in the final condition representing future ignorance. For example, in the sum-over-histories decoherence functional of eq.(II.2.2) the final $\delta$-function enforcing the coincidence of the paths at the final time does not involve the lavels. The final values of the label variables are summed *separately* in the expression for the decoherence functional. Put differently but equivalently, the inner product defining the overlap between different branches in (II.4.4) is in the Hilbert space of physical variables. If it is represented in terms of wave functions on an extended configuration space, including the label variables, these labels must be integrated out of each wave function separately before constructing the overlap on the physical variables.

---

[*] The material in this subsection was reported in Hartle (1989b).

[**] See, e.g. Hamber and Williams (1985, 1986ab), Hamber (1986), Hartle (1985a, 1986a, 1989a).

We can do the same thing with time. We introduce labels for a preferred family of spacelike surfaces. We construct quantum amplitudes by the familiar rules using these surfaces as the preferred time. Since the labels are unobservable, we ignore these labels in all coarse grainings. By introducing auxiliary labels we break general covariance. By integrating over them we restore it. I will now work out this idea in some detail focussing, for convenience, on amplitudes rather than decoherence functionals. (It's then one sum-over-histories to write rather then two.)

Isham and Kuchař (1985) have given a convenient formalism for the additional labels needed to describe a preferred family of spacelike surfaces in spacetime. A foliating family of spacelike surfaces is described by four functions, $X^\mu(\tau, x^i)$, that specify on which spacelike surface a point of spacetime lies and where it lies in the surface. We may think of these four functions as four additional scalar fields on the spacetime or more physically as the readings $X^\mu = (T, X^m)$ of ideal clocks and rods that give a system of coordinates for spacetime. If, for definiteness, we further assume that the trajectories of the clocks at fixed $X^m$ are orthogonal by the surfaces of constant $T$, then the spacetime metric in these special coordinates is

$$ds^2 = -dT^2 + s_{mn}(\mathbf{X}, T)dX^m dX^n \quad . \tag{V.3.1}$$

The time $T$ of the preferred coordinates supplies the preferred time of quantum mechanics. The geometrical quantum variables are the components of the $s_{mn}$. Thus we write for a state

$$\psi = \psi[T, s_{mn}(\mathbf{X}), \chi(\mathbf{X})] \quad . \tag{V.3.2}$$

This evolves by the Schrödinger equation

$$-i\frac{\partial \psi}{\partial T} + H\psi = 0 \tag{V.3.3}$$

where $H$ is the total Hamiltonian. For Einstein gravity we could construct $H$ by the standard canonical procedure from the action

$$\ell^2 S_E = 2 \int_{\partial M} d^3 x \sqrt{h} K + \int_M d^4 x \sqrt{-g}(R - 2\Lambda) \tag{V.3.4}$$

after (3.1) has been substituted into it. ($\ell = (16\pi G)^{\frac{1}{2}}$ is the Planck length in the units with $\hbar = c = 1$ that are used throughout.) The inner product is

$$(\psi, \psi') = \int \delta s_{mn} \delta \chi \; \bar{\psi}(T, s_{mn}, \chi) \psi'(T, s_{mn}, \chi) \quad . \tag{V.3.5}$$

In relativity, however, we should be able to state our results invariantly using any spacelike surface, not just the special surfaces of constant $T$. Indeed, it is important to do so to express the invariance of the theory under coordinate transformations. Isham and Kuchař (1985) tell us how to do it. In a general system of coordinates $x^\mu = (\tau, x^i)$ the metric (3.1) is

$$ds^2 = \left[ -\nabla_\alpha T \nabla_\beta T + s_{mn} \nabla_\alpha X^m \nabla_\beta X^n \right] dx^\alpha dx^\beta. \tag{V.3.6}$$

Substitute this into the action (3.4). One obtains an action that is a functional of the $X^\mu(x)$ and $s_{mn}(x)$. Equivalently it may be thought of as a functional of the embedding functions and the three-metric $h_{ij}$ on a surface of constant general coordinate $\tau$. This is because $s_{mn}$ and $h_{ij}$ are related on that surface by

$$h_{ij} = s_{mn}D_iX^mD_jX^n + D_iTD_jT \ , \qquad (\text{V.3.7})$$

$D_i$ being the derivative in the surface.

The action that results is for a parametrized theory in which the coördinates $X^\mu$ have been elevated to the status of dynamical variables coupled to curvature. The theory is invariant under diffeomorphisms because the coordinates $(\tau, x^i)$ were arbitrary. It is not, however, generally covariant because gravitational phenomena are now described by four scalar fields $X^\mu$ in addition to the metric. As a consequence of diffeomorphism invariance there are four constraints. Classically they can be written

$$n^\mu P_\mu + \mathcal{H}(\pi^{ij}, h_{ij}, \pi_\chi, \chi) = 0 \ , \qquad (\text{V.3.8}a)$$

and

$$(D_iX^\mu)P_\mu + \mathcal{H}_i(\pi^{ij}, h_{ij}, \pi_\chi, \chi) = 0 \ . \qquad (\text{V.3.8}b)$$

Here, $\mathcal{H}$ and $\mathcal{H}_i$ are the familiar Hamiltonian and momentum constraints of the classical theory – functions of the three-metric, $h_{ij}$, its conjugate momentum, $\pi^{ij}$, the scalar field, $\chi$, and its conjugate momentum, $\pi_\chi$. $P_\mu$ is the momentum conjugate to $X^\mu$. $n^\mu[X^\mu, h_{ij}]$ is the unit normal to the constant $T$ hypersurfaces. It can be expressed in terms of the $X^\mu$ and $h_{ij}$ alone because $n_\mu \propto \epsilon_{\mu\nu\sigma\tau}(D_1X^\nu D_2X^\sigma D_3X^\tau)$.

Quantum mechanically the states are described by wave functions

$$\psi = \psi\big[X^\mu(\mathbf{x}), h_{ij}(\mathbf{x}), \chi(\mathbf{x})\big] \qquad (\text{V.3.9})$$

that satisfy operator forms of the constraints (3.8)

$$in^\mu\frac{\delta\psi}{\delta X^\mu(\mathbf{x})} = \mathcal{H}\left(-i\frac{\delta}{\delta h_{ij}(\mathbf{x})}, h_{ij}(\mathbf{x}), -i\frac{\delta}{\delta\chi(\mathbf{x})}, \chi(\mathbf{x})\right)\Psi \ , \qquad (\text{V.3.10}a)$$

$$iD_iX^\mu\frac{\delta\psi}{\delta X^\mu(\mathbf{x})} = \mathcal{H}_i\left(-i\frac{\delta}{\delta h_{ij}(\mathbf{x})}, h_{ij}(\mathbf{x}), -i\frac{\delta}{\delta\chi(\mathbf{x})}, \chi(\mathbf{x})\right)\Psi \ . \qquad (\text{V.3.10}b)$$

Eq. (3.10a) is the covariant form of the Schrödinger equation (3.3) (i.e. the Tomonaga-Schwinger equation). Eq. (3.3) follows from (3.10a) by considering only variations in $\Psi$ that uniformly advance a surface of constant $T$. The additional constraints (3.10b) ensure that $\Psi$ is independent of the choice of spatial coordinates, $x^i$.

So far, the formalism is an exotic version of quantum mechanics but familiar in all of its basic aspects. Let us now turn to the calculation of probabilities. For simplicity suppose that the universe is in a pure state characterized by a wave function $\Psi[X^\mu, h_{ij}, \chi]$. The crucial decision in the calculation of probabilities is the status of the variables $X^\mu(\mathbf{x})$. If, as here, they are unobservable labels then amplitudes should be summed over them and then squared to yield probabilities for prediction. Thus, for example, the amplitude $\Psi[h_{ij}, \chi]$ that a single spacelike surface in the geometry has a metric $h_{ij}(\mathbf{x})$ and a matter field configuration $\chi(\mathbf{x})$ is

$$\Psi\big[h_{ij}(\mathbf{x}), \chi(\mathbf{x})\big] = \int_{\text{foliations}} \delta X^\mu(\mathbf{x}) \psi\big[X^\mu(\mathbf{x}), h_{ij}(\mathbf{x}), \chi(\mathbf{x})\big] \ . \qquad \text{(V.3.11)}$$

The integration is over all foliations of spacetime. In particular, the integration is over time, $T$, — both positive and negative values.

In constructing the sum (3.11) general covariance is restored, for if the $X^\alpha$ are integrated over diffeomorphism invariant range, the constraints (3.10) imply* for $\Psi$

$$\mathcal{H}(\mathbf{x})\Psi = 0, \quad \mathcal{H}_i(\mathbf{x})\Psi = 0. \qquad \text{(V.3.12)}$$

Another way to see that covariance is restored is to show that the amplitude so defined can be restated as a sum over geometries and field configurations and nothing else. Eqs.(3.10) are formally like the Schrödinger equation of ordinary quantum mechanics. In a familiar way, $\Psi$ could be constructed as a sum over $h_{ij}(x)$, $\phi(x)$, and $X^\mu(x)$ that matches the initial condition and the arguments of $\Psi$. The sum over $X^\mu(\mathbf{x})$ in (3.11) is just that needed in addition to $h_{ij}(\mathbf{x})$ to sum over all four-geometries. ¿From the form of the metric (3.6), one sees that by summing over all the $X^\mu(x)$ one is, in effect, summing over all the components of the metric — the $g_{00}$, $g_{0i}$ parts as well as the $h_{ij}$.

There is a good deal of gauge invariance in such a sum that must be fixed. Under an infinitesimal diffeomorphism, $x^\alpha \to x^\alpha + \xi^\alpha(x)$, the scalar $X^\mu$ transform as

$$X^\mu(x) \to X^\mu(x) + \xi^\alpha(x)\nabla_\alpha X^\mu(x) \ . \qquad \text{(V.3.13)}$$

A diffeomorphism that maps a region of the manifold into itself must not affect the ranges of the coördinates $x^\alpha$; $\xi^\alpha$ must, therefore, vanish on the boundaries. Thus, the values of $X^\mu(\mathbf{x})$ on the final surface cannot be transformed away. Put differently by a diffeomorphism we can arrange for $g_{00} = -1$ and $g_{0i} = 0$ in between two spacelike surfaces. These are, in fact, the coördinates of (3.1). However, the specification of the surface of interest in such coördinates, $X^\mu(\mathbf{x})$, carries physical information — the location and orientation of the surface with respect to the initial condition. The integration over foliations thus includes, in particular, an integration over the physical time that separates the surface from the initial condition *over both positive and negative values*. Including both positive and negative values is necessary to ensure the constraints (3.13) which express diffeomorphism invariance. (See, e.g. Teitelboim, 1983a, Halliwell and Hartle, 1990.)

### V.4. Some Open Questions

The slight improvement in concreteness afforded by the methods of the last section over the sketchy development of the preceeding one should not obscure the fact that there are many unresolved issues, both mathematical and physical, in this approach to a quantum kinematics of cosmology. Starting with the more technical issues, there is the issue of the measure and gauge fixing in the defining sums over geometries. A measure can be induced from the formal "Liouville" measure in the "Schrödinger" quantum mechanics thus supplying a kind of unitarity in the label time $X^\mu(\mathbf{x})$ But, how exactly is the sum-over-foliations to be carried out? It is not a simple functional integral because arbitrary functions $X^\mu(\mathbf{x})$ will not describe an imbedding that is non-self-intersecting. What exactly is the connection of the present discussion with Teitelboim's (1983abc) propagator approach and how exactly is the machinery for diffeomorphism fixing to be implemented?

---

* For more details see Halliwell and Hartle (1990).

Turning to more physical questions, there is the issue of how to implement specific proposals for the initial condition especially the "no boundary" proposal? What is the correct representation of final ignorance? What are the physically appropriate and mathematically sensible coarse grainings to be allowed? What are the details of the recovery of the limit of Hamiltonian quantum mechanics when spacetime in classical? In particular, what is the exact connection between the observables of familiar quantum mechanics in fixed backgrounds and the regions of superspace? Are such regions enough? Are more complicated partitions of histories useful? How does transformation theory emerge? What is the objective mechanism for the decoherence of spacetime? That is, what are the sets of decohering histories with high classicality? How are the semiclassical rules shortly to be described to be justified in quantum cosmology?

Finally, how is such a program to be carried out in more fundamental theories of quantum gravity such as string theory or in theories where spacetime is perhaps not even a basic variable?

In view of these remaining issues, perhaps the main message that I would hope the reader would take away from this part of these lectures is just the following: The familiar structure of quantum mechanics is closely entwined with the assumption of a fixed background spacetime. If we are to have a generally covariant quantum theory of spacetime we may have to generalize this framework. Feynman's sum-over-histories framework may supply such a generalization in which Hamiltonian quantum mechanics emerges as an approximation appropriate to our specific initial conditions and our special place in the late universe.

## VI. PRACTICAL QUANTUM COSMOLOGY

### VI.1. The Semiclassical Regime

The aim of the quantum mechanics of cosmology described in the previous sections is to extract predictions for correlations among observations of the universe today from a theory of its dynamics and its initial condition. By and large, however, the framework has not been applied in detail to produce predictions from anything that might be called a theory of the initial condition. The historical reason is that theories of the initial condition, and the quantum mechanical framework for extracting predictions from these theories, are developing together. Earlier work has, for the most part, focused on predictions of the ensemble of possible classical spacetimes of the late universe. This is the prediction that can be confronted most directly with observation. Practical prescriptions for extracting predictions of classical spacetime from the wave function of the universe were developed by analogy with non-relativistic quantum mechanics. By and large, these prescriptions *assume* the decoherence of alternative classical spacetimes and identify the form of wave functions from which one can expect classical correlations in time. These are (as of mid 1990) the rules for practical quantum cosmology. It seems appropriate to review them, if only to pose the problem of justifying them in terms of a more precise quantum framework such as that of Section V.

### VI.2. The Semiclassical Approximation to the Quantum Mechanics of a

Non-Relativistic Particle.

Let us recall how the semiclassical approximation works in non-relativistic quantum mechanics. Suppose we are given at $t = 0$, an initial wave function $\psi(X)$. Then the amplitude to arrive at position $X$ at time $t$ having passed through position intervals $\Delta_1 \cdots \Delta_n$ at time $t_1, \cdots, t_n$ is (see, e.g. Caves, 1986, 1987, Stachel, 1986):

$$< Xt|P_{\Delta_n}(t_n) \cdots P_{\Delta_1}(t_1)|\psi > = \int dX_0 \int_{[\Delta_\alpha]} \delta X \; e^{iS[X(t)]/\hbar} \psi(X_0) \quad . \quad \text{(VI.2.1)}$$

The sum is over all the paths that start at $X_0$ at $t = 0$, pass through the intervals $\Delta_1, \cdots, \Delta_n$ at the appointed times, and wind up at $X$ at time $t$ (Fig. 14).

We recover classical dynamics when this path integral can be done by the method of steepest descents. For then, only when $\Delta_1, \cdots, \Delta_n$ lined up so that a classical path from $X_0$ to $X$ passes through them will the amplitude (2.1) be non-vanishing. Classical correlations are thus predicted. Classical correlations are, however, as we know from Section II, only one aspect of classical behavior. The other is decoherence. However, here I am assuming, as in Section II.10, that the particle has been localized in intervals $\Delta_1, \cdots, \Delta_n$ by a "measurement" so that decoherence is accomplished by the interactions of the localizing apparatus with the rest of the universe.

Whether a steepest descents approximation is appropriate for the path integral depends on the intervals $\Delta_1, \cdots, \Delta_n$, the times $t_1, \cdots, t_n$, and the initial wave function $\psi(X)$. The $\Delta_1, \cdots, \Delta_n$ must be large enough and the times $t_1, \cdots, t_n$ separated enough to permit the destructive interference of the non-classical paths by which the steepest descents approximation operates. But, $\psi(X)$ must be right as well. There are a number of standard forms for $\psi(X)$ for which the steepest descents approximation can be seen to be valid. For example, if $\psi(X)$ describes a wave packet with position and momentum defined to an accuracy consistent with the uncertainly principle, and the time intervals between the $t_k$ are short compared with the time over which it spreads, and the $\Delta_\alpha$ are greater than its initial width, then only a single path will contribute significantly to the integral — that classical path with the initial position and momentum of the wave packet. Another case is when $\psi(X)$ corresponds to *two* initially separated wave packets. Then, *two* different classical paths contribute to the steepest descents approximation to (2.1) corresponding to the two sets of initial data. A unique classical trajectory is not predicted but rather one of two possible classical evolutions each with some probability. That is, given one of the $\Delta$'s a classical correlation of the rest is predicted. More precisely, the *conditional* probability for a particular classical trajectory *given* that the particle passed through an interval that lies on one and not on the other is a number near unity.

In general, therefore, a rather detailed examination of $\psi(X)$ is needed to determine if there are classical correlations predicted and what they are. However, there is a simple case where these predictions can be read off immediately. This is when the wave function $\psi(X, t)$ is well approximated by a linear combination of terms like

$$\psi(X, t) \approx \Delta(X, t) \; e^{\pm iS(X,t)/\hbar} \quad \text{(VI.2.2)}$$

where $\Delta(X, t)$ is a real slowly varying function of $X$ and $S/\hbar$ is a real, rapidly varying function of $X$. Eq.(2.2) thus separates $\psi$ into a slowly varying prefactor and a rapidly varying exponential. It follows from the Schrödinger equation in these

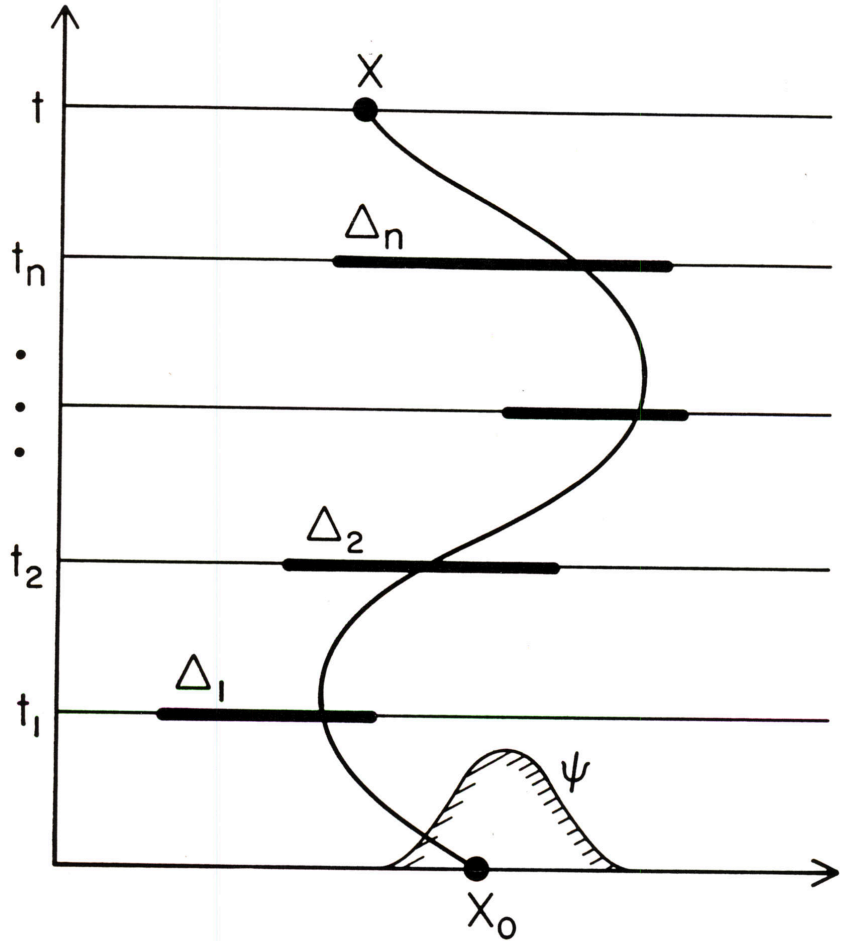

**Fig. 14:** *The semiclassical approximation to the quantum mechanics of a non-relativistic particle. Suppose at time t = 0 the particle is in a state described is a wave function $\psi(X_0)$. Its subsequent evolution exhibits classical correlations in time if successive determinations of position are correlated according to classical laws, that is, if the amplitude for non-classically correlated positions is near zero. The existence of such classical correlations is, therefore, a property not only of the initial condition but also the coarse graining used to analyse the subsequent motion. Classical correlations are properties of coarse-grained sets of histories of the particle. The amplitude for the particle to pass through intervals $\Delta_1, \Delta_2, \cdots, \Delta_n$ at times $t_1, \cdots, t_n$ and arrive at X at t is the sum of $\exp(iS)$ over all paths to X(t) that pass through the intervals, weighted by the initial wave function. For suitably spaced intervals in time, suitably large intervals $\Delta_i$, and suitable initial wave function $\psi$, this sum may be well approximated by the method of steepest descents. In that case, only when the intervals $\Delta_i$ are aligned about a classical path will there be a significant contribution to this sum. Classical correlations are thus recovered. How many classical paths contribute depends on the initial condition $\psi(X_0)$. If, as illustrated here, it is a wave packet whose center follows a particular classical history then only that particular path will contribute significantly. By contrast, if $\psi$ is proportional to $\exp[iS(X_0)]$ for some classical action $S(X_0)$, then all classical paths that satisfy $m\dot{X} = \partial S/\partial X$ will contribute. Then the prediction is of an ensemble of classical histories, each one correlated accords the classical equations of motion.*

circumstances that $S$ is a classical action approximately satisfying the Hamilton-Jacobi equation

$$-\frac{\partial S}{\partial t} + H\left(\frac{\partial S}{\partial X}, X\right) = 0 \ , \qquad (\text{VI.2.3})$$

where $H$ is the Hamiltonian:

$$H = \frac{P^2}{2M} + V(X) \quad . \qquad (\text{VI.2.4})$$

The forms (2.2) are called *semiclassical* approximations. When the semiclassical approximation (2.2) is inserted in (2.1), the functional integral and the integral over $X_0$ are integrals of a slowly varying prefactor with a rapidly varying exponent. This is immediately of the form for which the steepest descents approximation will be valid for suitable intervals $\Delta_1, \cdots, \Delta_n$ and times $t_1, \cdots, t_n$. Like the two wave packet example above, a unique classical trajectory is not predicted. The wave function (2.2) is not peaked about some *particular* initial data. In fact, by the slowly varying assumption for $\Delta$, it treats many $X_0$'s equally. However, the wave function (2.2) *does* lead to the classical connection between position and momentum implied by the action $S$. If $\psi$ is integrated over a wave packet of appropriate width, centered about $X_0$, then, because

$$S(X) = S(X_0) + \left(\frac{\partial S}{\partial X}\right)_{X_0}(X - X_0) + \cdots \qquad (\text{VI.2.5})$$

only momenta satisfying the classical relation

$$P = \frac{\partial S}{\partial X} \qquad (\text{VI.2.6})$$

will contribute significantly. The width of the packet must be wide enough to allow for rapid variation of $\exp(iS/\hbar)$ but not so wide that higher terms in the expansion (2.5) are important. Thus, for suitable subsequent intervals $\Delta_i$ and times, $t_1, \cdots, t_n$ a semiclassical wave function predicts not one classical trajectory, neither all of them, but just those for which the initial momenta are related by (2.6) for the particular classical action $S$. It thus predicts an ensemble of classical trajectories, each differing from the other by the constant needed to integrate (2.6).

The prefactor $\Delta$ is also of significance. $|\Delta(X_0, 0)|^2$ is probability of an initial $X_0$. Given that subsequent values of $X$ are correlated by the classical trajectory with this initial $X$ and the initial momentum (2.6), $|\Delta(X_0, 0)|^2$ may be thought of as the probability of a particular classical trajectory crossing the surface $t = 0$, although the variation over trajectories is necessarily weak. The order $\hbar$ implication of the Schrödinger equation is that

$$\frac{\partial |\Delta|^2}{\partial t} + \nabla \cdot \left(|\Delta|^2 \frac{\nabla S}{M}\right) = 0 \qquad (\text{VI.2.7})$$

so that the probability density $|\Delta|^2$ is conserved along the trajectories.

VI.3. Semiclassical Prediction in Quantum Cosmology

The conclusion of the above analysis in particle quantum mechanics is that when the wave function is well approximated by a semiclassical form that is a linear combination of terms like (2.2), then it predicts that sufficiently crude observations of position and momentum will be correlated along one of the classical trajectories in the ensemble determined by the action $S$ with a relative weight proportional to the prefactor squared. In quantum cosmology this conclusion is generalized to superspace to provide a direct prescription for extracting semiclassical predictions from the wave function of the universe.

This generalization has been widely discussed in the literature and I shall only briefly review it here. For more details and references to earlier literature, the reader might consult the papers of Halliwell (1987) and Padmanabhan and Singh (1990).

The wave function of the universe is the amplitude $\Psi[h_{ij}(\mathbf{x}), \chi(\mathbf{x})]$ introduced in (V.3.11) that the universe has a spacelike surface with three-metric $h_{ij}(\mathbf{x})$ and three-dimensional matter field configurations $\chi(\mathbf{x})$. This wave function is fixed by the initial condition. For example, in the "no boundary" proposal, it is prescribed directly as a certain Euclidean functional integral as described in the lectures of Halliwell in this volume. To illustrate a possible semiclassical approximation to $\Psi$ consider a linear combination of terms of the form.

$$\Psi[h_{ij}(\mathbf{x}), \chi(\mathbf{x})] = \Delta[h_{ij}(\mathbf{x})]e^{\pm iS_0[h_{ij}(\mathbf{x})]/\ell^2}\psi[h_{ij}(\mathbf{x}), \chi(\mathbf{x})] \ , \qquad \text{(VI.3.1)}$$

where $\Delta$ and $\psi$ are slowly varying functionals of $h_{ij}(\mathbf{x})$ and $S_0[h_{ij}(\mathbf{x})]/\ell^2$ is a real classical action for gravity alone. That is, $S_0[h_{ij}(\mathbf{x})]$ is the action

$$S_0 = 2\int_{\partial M} d^3x \sqrt{h}K + \int_M d^4x\sqrt{-g}(R - 2\Lambda) \qquad \text{(VI.3.2)}$$

evaluated along a particular extremum having the metric $h_{ij}(\mathbf{x})$ on $\partial M$. (Here, as usual $\ell = (16\pi G)^{1/2}$ is the Planck length and $\Lambda$ is the cosmological constant.) Which extrema contribute to (3.1) is determined by the initial condition. We shall return to the conditions that separate $\Delta$ from $\psi$ in a moment.

The action $S_0[h_{ij}(\mathbf{x})]/\ell^2$ satisfies the classical constraints (Peres, 1962, Gerlach, 1969)

$$\ell^2 G_{ijk\ell}(\mathbf{x})\pi^{ij}(\mathbf{x})\pi^{k\ell}(\mathbf{x}) + \ell^{-2}h^{\frac{1}{2}}\left(2\Lambda - {}^3R(\mathbf{x})\right) = 0 \ , \qquad \text{(VI.3.3a)}$$

$$D_j\pi^{ij}(\mathbf{x}) = 0 \ , \qquad \text{(VI.3.3b)}$$

where the momentum $\pi^{ij}(\mathbf{x})$ conjugate to $h_{ij}(\mathbf{x})$ is

$$\ell^2\pi^{ij}(\mathbf{x}) = \frac{\delta S_0}{\delta h_{ij}(\mathbf{x})} \ . \qquad \text{(VI.3.4)}$$

In these equations $D_i$ is the derivative constructed from the metric $h_{ij}(\mathbf{x})$, ${}^3R(\mathbf{x})$ is that metric's scalar curvature and

$$G_{ijk\ell} = \tfrac{1}{2}h^{-\frac{1}{2}}\left(h_{ik}h_{j\ell} + h_{i\ell}h_{jk} - h_{ij}h_{k\ell}\right) \ . \qquad \text{(VI.3.5)}$$

The gradient (3.4) defines a vector field on superspace and its integral curves are the classical spacetimes that give rise to the action $S_0$. For example, if we work in the guage where four-metrics have the form

$$ds^2 = -d\tau^2 + h_{ij}(\tau, \mathbf{x}) dx^i dx^j \ , \tag{VI.3.6}$$

then eq.(3.4) becomes

$$\tfrac{1}{2} \frac{dh_{ij}}{d\tau} = G_{ijk\ell} \frac{\delta S_0}{\delta h_{k\ell}} \quad . \tag{VI.3.7}$$

Integrating (3.7) we recover a four-metric (3.6) that satisfies the Einstein equation. The values of $\psi$ along such an integral curve define $\psi$ as a function of $\tau$

$$\psi = \psi \left[ h_{ij}(\tau, \mathbf{x}), \chi(\mathbf{x}) \right] = \psi \left[ \tau, \chi(\mathbf{x}) \right] \quad . \tag{VI.3.8}$$

The wave function $\Psi[h_{ij}(\mathbf{x}), \chi(\mathbf{x})]$ must satisfy the operator form of the constraints (V.3.12) that implement the underlying gravitational dynamics. The three momentum constraints, $\mathcal{H}_i \Psi = 0$, guarantee that $\Psi$ is independent of the choice of coördinates in the spacelike surface. The fourth constraint may be written out formally as

$$\mathcal{H}(\mathbf{x}) \Psi = \left[ -\ell^2 \nabla_{\mathbf{x}}^2 + \ell^{-2} h^{\frac{1}{2}} (2\Lambda - ^3 R) + h^{\frac{1}{2}} \hat{T}_{nn}(\chi, -i\delta/\delta\chi) \right] \Psi = 0 \ . \tag{VI.3.9}$$

Here,

$$\nabla_{\mathbf{x}}^2 = G_{ijk\ell}(\mathbf{x}) \frac{\delta^2}{\delta h_{ij}(\mathbf{x}) \delta h_{k\ell}(\mathbf{x})} + \begin{pmatrix} \text{linear derivative} \\ \text{terms depending} \\ \text{on factor ordering} \end{pmatrix} \tag{VI.3.10}$$

and $\hat{T}_{nn}$ is the stress-energy of the matter field projected into the spacelike surface (the Hamiltonian density) expressed as a function of the matter field $\chi(\mathbf{x})$ and the operator $-i\delta/\delta\chi(\mathbf{x})$ corresponding to its conjugate momentum. This fourth constraint is called the Wheeler-DeWitt equation (DeWitt, 1967, Wheeler, 1968). The implications of the Wheeler-DeWitt equation (3.9) for that part of the semiclassical approximation that varies slowly with three-metric may be found by inserting the approximation (3.1) into (3.9), using the Hamilton-Jacobi equation (3.3a), and neglecting second derivatives of slowly varying terms with respect to the three-metric. The result is an equation for $\Delta\psi$ that can be organized in the following form:

$$-i\psi \left[ (\nabla_{\mathbf{x}}^2 S_0)\Delta + 2G_{ijk\ell} \frac{\delta S_0}{\delta h_{ij}} \frac{\delta \Delta}{\delta h_{k\ell}} \right] + \Delta \left[ -i2G_{ijk\ell} \frac{\delta S_0}{\delta h_{ij}} \frac{\delta \psi}{\delta h_{k\ell}} + h^{\frac{1}{2}} \hat{T}_{nn} \psi \right] = 0 \ . \tag{VI.3.11}$$

We now impose the condition that the two terms in (3.11) vanish separately. This defines the decomposition of the slowly varying part, $\Delta\psi$, into $\Delta$ and $\psi$.

The condition on the $\psi$ resulting from (3.11) may be rewritten using (3.7) and (3.8) as

$$i \frac{\partial \psi}{\partial \tau} = h^{\frac{1}{2}} \hat{T}_{nn} \left( \chi, -i \frac{\delta}{\delta\chi} \right) \psi \ . \tag{VI.3.12}$$

This is the Schrödinger equation in the field representation for a quantum matter field $\chi$ executing dynamics in a background geometry of the form (3.6).

The condition on $\Delta$ arising from (3.11) implies the following relation

$$G_{ijk\ell} \frac{\delta}{\delta h_{ij}} \left( |\Delta|^2 \frac{\delta S_0}{\delta h_{k\ell}} \right) = 0 \ . \tag{VI.3.13}$$

This is the equation of conservation of the current $|\Delta|^2(\delta S_0/\delta h_{ij})$ in superspace. It is the analog of the similar relation (2.7) in non-relativistic quantum mechanics. Indeed, in view of (3.7), this is just the statement that the "density in superspace", $|\Delta|^2$, is conserved along classical trajectories, the integral curves of (3.7).

The Schrödinger equation (3.12), the conservation law (3.13), and the analogy with non-relativistic quantum mechanics suggest the following rule for extracting predictions from a semiclassical approximation: Wave functions of the form (3.1) predict that sufficiently coarse-grained histories of spacetime variables will be correlated as on one of the classical spacetimes in the ensemble defined by $S_0[h_{ij}(\mathbf{x})]$ through (3.7). The various possible classical histories occur with a probability weight proportional to $|\Delta|^2$ reflecting the specific initial condition.* The ensemble includes the spacetimes predicted by each contributing action if several different terms of the form (3.1) contribute to the semiclassical approximation. For coarse-grained histories involving matter fields as well as spacetime variables, the quantum mechanics of the matter fields will be that of field theory in one of the classical spacetimes in the ensemble. In short, a semiclassical wave function that is a superposition of terms like (3.1) predicts classical spacetime and quantum matter field theory in that classical curved spacetime.

Many other semiclassical approximations are possible besides the one based on the form (3.1) and the action (3.2). For example, an approximate form in which both spacetime and some matter variables behave classically would involve an action defining the rapidly varying part of the wave function which depended on both kinds of variables. One can consider ensembles of classical geometries driven by expectation values of matter fields in which the constraints (3.3) contain such terms as sources. Systematic approaches to obtaining such approximate wave functions by expanding the solutions to the Wheeler-DeWitt equation in powers of the inverse Planck length have been extensively discussed.† Indeed, it is essential

---

* The conserved current can be used to define a probability density on classical trajectories. More precisely, this is the relative probability that a classical trajectory crosses one part of a surface in superspace orthogonal to these trajectories rather than crossing another part. It is the conservation of current expressed by eq.(3.13) that makes the interpretation consistent and natural as has been stressed by Vilenkin (1988). This is consistent with thinking of $|\Psi|^2$ as a probability density on superspace because a flux through a hypersurface is related to the integrated superspace probability density over a volume behind the hypersurface with a thickness proportional to the local velocity. In a semiclassical approximation these notions are all defined and both ways of calculating a probability density on classical trajectories agree. Thanks are due to D. Page and A. Vilenkin for a discussion on this point.

† See, e.g. Halliwell (1987) and Padmanabhan and Singh (1990) for discussion and further references

to consider approximations with both matter and geometry behaving classically since the late universe is certainly not a solution of the vacuum Einstein equation. The defining feature of all these semiclassical approximations is the separation of the wave function into a part rapidly varying in certain variables governed by a classical action and a more slowly varying part. There are different approximations depending on what variables are distinguished in this way.

Of course, there is no necessity for a theory of the initial condition to imply a wave function that can be well approximated by any semiclassical form like (3.1), but it is likely that successful theories will. The reason is that a successful theory must predict classically behaving spacetime on scales above the Planck length in the late universe. If it does not it is simply inconsistent with observation in a manifest way. A wave packet, not of the form (3.1), could still imply classical behavior and indeed a *particular* classical spacetime. But a single classical history with all the complexity of the present classical universe is unlikely to be predicted by a *simple* theory of initial condition. An ensemble of possibilities not strongly preferred one to the other in most features seems more natural. Present theories do, by and large, predict the classical behavior of spacetime through semiclassical forms like (3.1).

As a simple application of this discussion which is relevant for this school, consider the prediction of the value of the cosmological constant. We determine the present effective value of the cosmological constant by fitting the observed data on the expansion of the universe with solutions of Einstein's equation. A probabilistically distributed cosmological constant would be a consequence of an initial condition that predicted an ensemble of classical universes, some with one value of the effective $\Lambda$, measured by the expansion of the universe, some with another. The associated wave function of the universe would be well approximated by a semiclassical approximation. An example is the form:

$$
\begin{aligned}
\Psi\left[h_{ij}(\mathbf{x}), \chi(\mathbf{x})\right] \approx \int d\Lambda \Delta\left[h_{ij}(\mathbf{x}), \Lambda\right] \cos\left(S_0[h_{ij}(\mathbf{x}), \Lambda]/\ell^2\right) \\
\times \psi\left(h_{ij}(\mathbf{x}), \Lambda, \chi(\mathbf{x})\right) \quad . \qquad \text{(VI.3.14)}
\end{aligned}
$$

Universes with different $\Lambda$ would, therefore, be predicted with a relative weight proportional to $|\Delta|^2$. It is argued elsewhere in this volume that, in the no boundary proposal, a sum over wormhole topologies can be replaced at low energies by such an integral over an effective $\Lambda$ and further that $\Delta$ will be essentially arbitrarily sharply peaked about $\Lambda = 0$. The point to stress here is that, since the cosmological constant is determined from the large scale motion of the classical universe, it is through a prediction of the ensemble of classical possibilities and their relative weights that we most honestly have a prediction of its value.

The above prescription for extracting semiclassical predictions from a wave function for the universe is perhaps compelling on the basis of its analogy with non-relativistic quantum mechanics. However, it should be possible to justify this prescription on a more fundamental basis than this analogy as, for example, it is justified in non-relativistic quantum mechanics in Section VI.2. Specifically, it should be possible to supply a quantitative answer to the questions: How good an approximation does a semiclassical form like (3.1) supply to the genuine probabilities of the theory? How small are the probabilities for *non*-classical behavior of a set of alternative histories that are coarse-grained according to the above prescription?

As in the non-relativistic quantum mechanics of Section VI.2, answers to such questions require the probabilities for sets of histories involving coarse-grained alternatives on more than one spacelike surface. A wave function of the universe on a single spacelike surface will not be sufficient. What is needed first is an analysis of the decoherence of coarse-grained spacetime histories in the context of a generalized quantum mechanics like that of Section V. Suggestive calculations of spacetime decoherence phenomena have been made by several authors (e.g. Zeh, 1986, 1988, Kiefer, 1987, Fukuyama and Morikawa, 1989, Halliwell, 1989, and Padmanabhan, 1989) but not for sets of alternative *histories*. Then a calculation of the probabilities for patterns of classical correlation in such sets of decoherent histories needs to be carried out for particular theories of the initial condition. While the basic ingredients of such a demonstration can be found in Section V — sum-over-histories decoherence functional, coarse graining by regions in superspace, steepest descents approximation to the integrals defining probabilities, etc. — filling in this sketch represents an important class of outstanding problems.

ACKNOWLEDGEMENTS

      The author is indebted to many physicists over many years for discussions of the issues addressed in these lectures. Special thanks are due to M. Gell-Mann. Most of the material in Section II is based on joint work with him and the author is grateful for his permission to reproduce here significant parts of the paper (Gell-Mann and Hartle, 1990) in which it was reported. These ideas were influential in the later part of the lectures as have been many discussions with J. Halliwell, K. Kuchař, and R. Sorkin. On particular points the author has benefited from conversations with A. Ashtekar, S. Coleman, R. Griffiths, G. Horowitz, J. Lebovitz, R. Omnès, D. Page, R. Penrose, and A. Vilenkin. Preparation of these lectures was supported in part by NSF grants PHY 85-06686 and PHY 90-08502. The author is grateful for the hospitality of the Institut des Hautes Études Scientifique where the lectures were completed.

# REFERENCES

For a subject as large as this one it would be an enormous task to cite the literature in any historically complete way. I have attempted to cite only papers that I feel will be *directly* useful to the points raised in the text. These are not always the earliest nor are they always the latest. In particular I have not attempted to review or to cite papers where similar problems are discussed from different points of view.

APPENDIX: BUZZWORDS

Only a casual inspection of the literature reveals that many interpreters of quantum mechanics who agree completely on the algorithms for quantum mechanical prediction, disagree, often passionately, on the words with which they describe these algorithms. This is the "words problem" of quantum mechanics. The agreement on the algorithms for prediction suggests that such disagreements may have as much to do with people as they do with physics. This does not mean that such issues are unimportant because such diverging attitudes may motivate different directions for further research. However, it is important to distinguish such motivation from properties of the theory as it now exists.

A few "buzzwords" characterize the words problem for quantum mechanics. They are phrases like "reduction of the wave packet", "many worlds", "non-locality", "state", etc. These are words that evoke or challenge some of the core assumptions that guide physicists in their work. To avoid confusion among the variety of preconceived meanings commonly held for such terms, they have been avoided in the preceding discussion. Now, in this appendix, it seems appropriate to return to a brief discussion of the author's attitudes and preferences concerning these words (*circa* mid-1990). These comments are collected together in this appendix to stress that they are not essential to the preceding discussion and to emphasize that they represent nothing further than the author's own preferences and opinions in these matters. The text's discussion of the quantum mechanical process of prediction for closed systems is self-contained as far as it goes and the material in this appendix may be dispensed with. Alternatively, the reader may choose different words with which to surround the discussion and different attitudes to it. In this spirit no attempt has been made to describe, discuss, confront, or refer to other discussions of these words.

1. *State.* In classical physics there is a description of a system at a moment of time that is all that is necessary to both predict the future *and* retrodict the past. The most closely analogous notion in quantum mechanics is the effective density matrix, $\rho_{\text{eff}}(t)$, of eq. (II.3.4) expressed either in the Heisenberg picture, as there, or in the Schrödinger picture. However, this quantum mechanical notion of "state at a moment of time", has a very different character from the classical analog. The future may be predicted from $\rho_{\text{eff}}$ *alone* but to retrodict the past requires, *in addition*, a knowledge of the initial condition. (See Section II.3.) The quantum mechanical notion of state is, therefore, already considerably weaker in its power to summarize probabilities than its classical analog for deterministic theories.

It is important to distinguish the notions of "state at a moment of time" represented by $\rho_{\text{eff}}(t)$ in the discussion in Section II from the initial condition of the system represented by the initial Heisenberg $\rho$. Both are commonly referred to as the "state of the system". However, the conclusion of the discussion in Sections III and IV is that while an initial condition, or its equivalent, is an essential feature of the quantum mechanical process of prediction, a notion of "state at a moment of time" is not. As Section II shows, the familiar theory may be organized without this notion. When, as in quantum theories of spacetime, such as that in Section V, there is no well defined notion of time it is unlikely that it is possible to introduce a notion of "state at a moment of time".

2. *Reduction of the Wave Packet.* Two senses of this phrase can be distinguished. The first concerns the updating of probabilities by an *IGUS* on acquisition of information. The second concerns the evolution in time of the effective density matrix, $\rho_{\text{eff}}(t)$, corresponding to the notion of "state at a moment of time". I shall consider these senses separately, first for the quantum mechanics of closed systems. Then I shall discuss the evolution of the effective density matrices, $\rho_{s,\text{eff}}(t)$, of sub-systems under observation in the Copenhagen approximation.

Much has been made of the normalization of joint probabilities that occurs in the calculation of the conditional probabilities for prediction [eq. (II.3.2)] or retrodiction [eq. (II.3.3)]. An *IGUS* utilizing these formulae would update the conditional probabilities of interest as new information is acquired (or perhaps lost). There is, however, nothing specifically quantum mechanical about such updating; it occurs in any statistical theory. In a sequence of horse races the joint probabilities for a sequence of eight races is naturally converted, after the winners of the first three are known, into conditional probabilities for the outcomes of the remaining five races by exactly this process. All probabilities are available to the *IGUS*, but, as new information is acquired, new conditional probabilities become relevant for prediction and retrodiction.

For those quantum mechanics of closed systems that permit the construction, according to (II.3.4), of an effective density matrix, $\rho_{\text{eff}}(t)$, to summarize present information for future prediction, the process of the reassessment of probabilities described above can be mirrored in its "evolution" according to the following rule: The effective density matrix, $\rho_{\text{eff}}(t)$, is constant in the Heisenberg picture between two successive times when data is acquired, $t_k$ and $t_{k+1}$. When new information is acquired at $t_{k+1}$, $\rho_{\text{eff}}(t)$ changes by the action of a new projection on each side of (II.3.4) and division by a new normalizing factor. One could say that "the state of the system is reduced"* at $t_{k+1}$. It might be clearer to say that a new set of conditional probabilities has become appropriate for future predictions and therefore a new $\rho_{\text{eff}}(t)$ is relevant.

It should be clear that in the quantum mechanics of a closed system this "second law of evolution" for $\rho_{\text{eff}}(t)$ has no special, fundamental status in the theory and no particular association with a measurement situation or any physical process. It is simply a convenient way of organizing the time sequence of probabilities that are of interest to a particular *IGUS*. Indeed, as the development of Section II shows, it is possible to formulate the quantum mechanics of a closed system without ever mentioning "measurement", "an effective density matrix", its "reduction" or its "evolution". Further, as in the framework for quantum spacetime discussed in Section V, there may be quantum mechanical theories where it is not possible to introduce an effective density matrix at all, much less discuss its "evolution" or "reduction".

It is in the ideal measurement model of Section II.10, upon which the Copenhagen approximation to quantum mechanics is based, that we can connect the reassessment of probabilities with the "reduction of the wave packet" on *measurement*. There, as can be seen from (II.10.8), [cf. (II.3.4)-(II.3.5)] the effective density matrix.

$$\rho_{s,\text{eff}}(t_k) = \frac{s_{\alpha_k}^k(t_k) \cdots s_{\alpha_1}^1(t_1) \rho_s s_{\alpha_1}^1(t_1) \cdots s_{\alpha_k}^k(t_k)}{tr\left[s_{\alpha_k}^k(t_k) \cdots s_{\alpha_1}^1(t_1) \rho_s s_{\alpha_1}^1(t_1) \cdots s_{\alpha_k}^k(t_k)\right]} \tag{A.1}$$

summarizes present information for future prediction of the subsystem under observation. *Every projection operator in (A.1) is part of a measurement situation* in this idealized model. That is, in the larger universe of apparatus and subsystem each projection is exactly correlated with an exactly decohering record variable. Thus, it is possible to say that $\rho_{s,\text{eff}}(t)$ is constant in between measurements (in this Heisenberg picture), but is "reduced" *at a measurement*.

---

* Typically it is not reduced very much! The $P$'s of a coarse graining of a typical *IGUS* fix almost none of the variables of the whole universe and therefore correspond to very large subspaces of its Hilbert space. Most variables are still available, untouched, for future projections.

Two remarks may be useful concerning the "reduction of the wave packet" in the Copenhagen approximation. First, again, the quantum mechanics of a subsystem under observation may be formulated directly in terms of probabilities for histories [eq. (II.10.6)] without an effective density matrix or its reduction. To introduce these notions is, therefore, to some extent a choice of words. Second, and more importantly, the association of the "reduction" with "measurement" is a special property of the ideal measurement model. This has suggested to some that there is a physical mechanism behind the reduction of the wave packet. However, in the more general situations in which a closed system is considered, there is no necessary association of "reduction" with a measurement situation.

Do the Everett class of interpretations eliminate the "reduction of the wave packet"? Some have said so. (Everett 1957, DeWitt 1970). The argument is crudely that only probabilities for correlations at one moment of time — the "marvelous moment now" — are of interest. For these $\rho_{\text{eff}} = \rho$ and no further reduction need be contemplated. However, in general, probabilities for histories involving more than one time are of interest and for these sequences of projections are necessary. (See Section V.1.2.) Then, the Everett interpretation *can*, if one so chooses, be formulated in terms of a $\rho_{\text{eff}}(t)$ that is "reduced". On the other hand, the "reduction of the wave packet" is not a necessary element of a quantum mechanics of cosmology. If one chooses, it need never be mentioned. It is, thus, no less necessary or more necessary in the Everett class of formulations than it is in the Copenhagen approximation to it. Its a matter of words. In a generalized quantum mechanics these words may not even be possible.

3.*The Measurement Problem*. Quantum mechanics does not predict a particular history for a closed system; it predicts the probabilities of a *set of alternative* histories. This is the case even when the histories constitute a quasiclassical domain and refer to the "macroscopic" description of objects consisting of many particles. Some describe this state of affairs as the "quantum measurement problem" or even the "quantum measurement paradox". However, such words can be confusing because there is no evidence that quantum mechanics is logically inconsistent, no evidence that it is inconsistent with experiment, and no evidence of known phenomena that could not be described in quantum mechanical terms.

If there is a "quantum measurement problem", therefore, nothing said in this exposition of quantum mechanics will resolve it. It is not a problem *within* quantum mechanics; rather it seems to be a problem that certain researchers have *with* quantum mechanics. Some find quantum mechanics unsatisfactory by some standard for physical theory beyond consistency with experiment. The intuition of others suggests that in domains where the predictions of quantum mechanics have not yet been fully tested an experimental inconsistency will emerge and a different theory will be needed. For example, perhaps the interference between "macroscopically" different configurations predicted by quantum mechanics will not be observed. (See, e.g. Leggett, 1980, Tesche, 1990). What is needed to meet such standards, or to resolve such experimental inconsistencies, should they develop, is not further research on quantum mechanics itself, but rather a new and conceptually different theoretical framework. It would be of great interest to have serious and compelling alternative theories if only to suggest decisive experimental tests of quantum mechanics.

4. *Many Worlds.* Quantum mechanics describes *sets of alternative* histories of the universe and within a given set one cannot assign "reality" simultaneously to different alternatives because they are contradictory. Everett (1957), DeWitt (1970) and others have described this situation, not incorrectly, but in a way that has confused some, by saying that all the alternative histories are "equally real". What is meant is that quantum mechanics prefers no alternative over another except through its probability.

The author prefers the term "many histories" to "many worlds" as less confusing and less inflammatory. However, either set of words, and no doubt others as well, may be used to describe this theoretical framework without affecting its predictive content.

5. *Non-Locality.* In an EPR or EPRB situation a choice of measurements, say $\sigma_x$ or $\sigma_y$ for a given electron, is correlated with the behavior of $\sigma_x$ or $\sigma_y$ respectively for another electron because the two together are in a singlet spin state even though widely separated. A situation in which an *IGUS* measures the $x$-component of the spin decoheres from one in which the $y$-component is chosen, but in each case there is also a correlation between the information obtained about one spin and the information obtained about the other. This behavior is called "non-local" by some authors. However, it is straightforward to show very generally using techniques of the present formulation that it involves no non-locality in the sense of quantum field theory and no signaling outside the light cone. (For alternative demonstrations cf. Ghirardi, Rimini, and Weber, 1980, Jordan, 1983.)

Consider the ideal measurement model discussed in Section II.10. In a particular Lorentz frame, let $\{s^1_{\alpha_1}(t_1)\}$ correspond to a set of alternatives at time $t_1$ but localized in space. For example, the projection operators $s^1_{\alpha_1}(t_1)$ might be projections onto ranges of field averages at time $t_1$ over a certain spatial region $R_1$. Let $\{s^2_{\alpha_2}(t_2)\}$ be another set of alternatives at a later time $t_2$ defined for a region $R_2$ *every point of which is spacelike separated from every point of* $R_1$. Let the initial density matrix be $\rho_s$. These alternatives are assumed to decohere because they are measured as described in Section II.10.

If no measurement is carried out at time $t_1$, the probability of finding alternative $\alpha_2$ at the later time $t_2$ is

$$p_{\text{no meas}}(\alpha_2) = tr\left[s^2_{\alpha_2}(t_2)\rho_s\right] \ . \tag{A.2}$$

If a measurement is carried out at time $t_1$, but the results are not known (because they cannot be independently signaled from $R_1$ to $R_2$ faster than the speed of light) then probability of finding alternative $\alpha_2$ is

$$p_{\text{meas}}(\alpha_2) = \sum_{\alpha_1} p(\alpha_2, \alpha_1)$$
$$= \sum_{\alpha_1} tr\left[s^2_{\alpha_2}(t_2)s^1_{\alpha_1}(t_1)\rho_s s^1_{\alpha_1}(t_1)s^2_{\alpha_2}(t_2)\right] \ . \tag{A.3}$$

In general (A.3) and (A.2) will not be equal because of interference. This is consistent because they correspond to two physically distinct situations: In the situation described (A.2) no measurement was made at time $t_1$. A measurement was made in that described by (A.3). However, in the case of spacelike separated regions $R_2$ and $R_1$, the local operators $s^k_{\alpha_2}(t_2)$ and $s^k_{\alpha_1}(t_1)$ *commute* by relativistic causality. The operators $s^1_{\alpha_1}(t_1)$ in (A.3) can therefore be moved to the outside of the trace, moved from one side of $\rho_s$ to the other by the trace's cyclic property, and eliminated using $(s^1_{\alpha_1})^2 = s^1_{\alpha_1}$ and $\sum_{\alpha_1} s^1_{\alpha_1} = 1$. Thus, the relativistic causality of the underlying fields implies

$$p_{\text{meas}}(\alpha_2) = p_{\text{no meas}}(\alpha_2), \qquad (A.4)$$

so that by a local analysis of the second measurement one cannot tell whether the first was even carried out, much less gain any information about its outcome if it was.

6. *Reality*. Quantum mechanics prefers no one set of histories to another except by such criteria as decoherence and classicality. Quantum mechanics prefers no one history to another in a given set of alternative decohering histories except by probability. Thus, the only element of the theory that might conceivably lay claim to the title of a unique, absolute, independent "reality" is the collection of all *sets* of alternative coarse-grained histories of the universe, or what is essentially the same thing, its initial condition.*† Yet, to use the word "reality" in this way is contentious, for this notion has no relation to the familiar "reality" of our impressions. What are these impressions and how are they described quantum mechanically? The familiar sense of reality arises, it seems, from the agreement among many and varied collections of *IGUS*es on the values of the quasiclassical variables in a quasiclassical domain and the experience that this agreement is largely independent of circumstance, position, and time. In quantum mechanics this agreement would be described as follows: A coarse graining can be associated with each *IGUS* which includes certain quasiclassical projection operators that the *IGUS* can perceive and projection operators (not necessarily quasiclassical) that describe the *IGUS*'s memory in which these perceptions are registered. To have a good memory means that there is a nearly full correlation between the operators describing the *IGUS*'s memory and the quasiclassical operators of the quasiclassical domain. Perception is thus a particular type of measurement situation. Agreement among several *IGUS*es means that there is a correlation between the various memories and common projection operators of a quasiclassical domain. The correlations will not be perfect. There may be fluctuations and, indeed, situations where there is a correlation between an *IGUS*es memory and some *other* part of its memory rather than the appropriate quasiclassical variable describe symptons of schizophrenia commonly described as "loss of contact with reality". Despite such anomalies, the agreement that exists would seem to be the source of our impression of an independent "reality".

The focus by *IGUS*es on the quasiclassical operators of a quasiclassical domain can be explained by understanding evolution of *IGUS*es in the universe. That is the only way of understanding why *IGUS*es employ the coarse grainings they do.

---

* This is worse than "all the alternative histories (worlds) are equally real". It would imply that "all the alternative *sets* of decohering histories are equally real".

† As Bohm (1952), deBroglie (1956), Bell (1981), and others have demonstrated, it is possible to use words to describe quantum mechanics that themselves specify a "reality". However, the predictions of quantum mechanics appear to be unaffected by this choice. If that is the case, then such issues as the existence of quasiclassical domains or the description of the reality of familiar experience remain as issues in the alternative descriptions.

If, as a consequence of the initial conditions of the universe and the dynamics of the fundamental fields, there is an essentially unique quasiclassical domain, then it is plausible that $IGUS$es evolved to exploit this possibility that our particular universe presents. (See Section II.12.) The coarse grainings describing what $IGUS$es perceive are then *all coarser grainings of the coarse graining defining the essentially unique quasiclassical domain.* $IGUS$es agree because they are perceiving the same quasiclassical projection operators. Thus, although quantum mechanics prefers no one set of histories to another, or one history in a given set to another, $IGUS$es may have evolved to do so.

Thus, if an essentially unique set of decohering alternative histories with high classicality is an emergent feature of our universe it would seem reasonable to associate the term "reality" in its familiar sense with that set of histories or with the individual history in the set correlated with our present memory. Reality would then be an *approximate* notion contingent on the approximate standard for decoherence, the initial condition of the universe, and the dynamics of the elementary fields. Universes for which no quasiclassical domains were emergent would have no such notion of "reality". The evolution, perceptions, and behavior of $IGUS$es in a universe for which there is more than one quasiclassical domain are open and very interesting questions. Thus, a central question for serious theoretical research in quantum cosmology is whether our universe exhibits more than one quasiclassical domain and, if so, the consequences of this fact for the evolution and behavior of $IGUS$es and the evolution of their notions of "reality".